\documentclass[12pt]{article}
\pdfoutput=1
\usepackage{amsfonts}
\usepackage{amsmath}
\usepackage{url}
\usepackage{hyperref}
\usepackage{bbold}
\usepackage{slashed}
\usepackage{tikz}
\usepackage{graphicx}
\usepackage{epstopdf}
\usepackage{subfigure}
\usepackage{pgfplots}
\usepackage{amssymb}
\usepackage{color}
\usepackage{mathrsfs}
\usepackage{fancybox}
\usepackage{lipsum}
\usepackage{eurosym}
\usepackage{tcolorbox}
\usepackage{physics}
\usepackage{tensor}

\numberwithin{equation}{section}

\newcommand{\slr}{\text{SL}(2,\mathbb{R})}
\newcommand{\sltr}{\text{SL}(2,\mathbb{R})}
\newcommand{\dpi}{\mathcal{D}}
\newcommand{\mj}{\mathcal{J}}

\newcommand{\mo}{\mathcal{O}}

  \newcommand{\sj}[6]{ \begin{Bmatrix}
   #1 & #2 & #3 \\
   #4 & #5 & #6 
  \end{Bmatrix}}
\newcommand{\Poincare}{Poincar\'e }
\newcommand{\sinc}{\text{sinc}}
\newcommand{\average}[1]{\left\langle #1 \right\rangle}

\newcommand{\airy}{\text{Ai}}
\newcommand{\tlc}{t_\text{LC}}

\begin{document}
\begin{titlepage}

\setcounter{page}{1} \baselineskip=15.5pt \thispagestyle{empty}

\vfil

${}$
\vspace{1cm}

\begin{center}
\def\thefootnote{\fnsymbol{footnote}}
\begin{changemargin}{0.05cm}{0.05cm} 
\begin{center}
{\Large \bf Clocks and Rods in Jackiw-Teitelboim Quantum Gravity 
}
\end{center} 
\end{changemargin}

~\\[1cm]
{Andreas Blommaert\footnote{\href{mailto:andreas.blommaert@ugent.be}{\protect\path{andreas.blommaert@ugent.be}}}, Thomas G. Mertens\footnote{\href{mailto:thomas.mertens@ugent.be}{\protect\path{thomas.mertens@ugent.be}}} and Henri Verschelde\footnote{\href{mailto:henri.verschelde@ugent.be}{\protect\path{henri.verschelde@ugent.be}}}}
\\[0.3cm]
\vspace{0.7cm}
{\normalsize { \sl Department of Physics and Astronomy,
\\[1.0mm]
Ghent University, Krijgslaan, 281-S9, 9000 Gent, Belgium}} \\
\vspace{0.5cm}

\end{center}


 \vspace{0.2cm}
\begin{changemargin}{01cm}{1cm} 
{\small  \noindent 
\begin{center} 
\textbf{Abstract}
\end{center} 
We specify bulk coordinates in Jackiw-Teitelboim (JT) gravity using a boundary-intrinsic radar definition. This allows us to study and calculate exactly diff-invariant bulk correlation functions of matter-coupled JT gravity, which are found to satisfy microcausality. We observe that quantum gravity effects dominate near-horizon matter correlation functions. This shows that quantum matter in classical curved spacetime is not a sensible model for near-horizon matter-coupled JT gravity. This is how JT gravity, given our choice of bulk frame, evades an information paradox. This echoes into the quantum expectation value of the near-horizon metric, whose analysis is extended from the disk model to the recently proposed topological completion of JT gravity \cite{sss2}. Due to quantum effects, at distances of order the Planck length to the horizon, a dramatic breakdown of Rindler geometry is observed. 
}

\end{changemargin}
 \vspace{0.3cm}
\vfil
\begin{flushleft}
\today
\end{flushleft}

\end{titlepage}

\newpage
\tableofcontents
\vspace{0.5cm}
\noindent\makebox[\linewidth]{\rule{\textwidth}{0.4pt}}
\vspace{1cm}

\addtolength{\abovedisplayskip}{.5mm}
\addtolength{\belowdisplayskip}{.5mm}

\def\plus{\raisebox{.5pt}{\tiny$+$\smpc}}

\addtolength{\parskip}{.6mm}
\def\spc{\hspace{1pt}}

\def\nspc{{\hspace{-2pt}}}
\def\ff{\rm\smpc f\smpc} 
\def\fff{\mbox{Y}}
\def\ww{{\rm w}}
\def\smpc{{\hspace{.5pt}}}

\def\zz{{\spc \rm z}}
\def\xx{{\rm x\smpc}}
\def\xxi{\mbox{\footnotesize \spc $\xi$}}
\def\jj{{\rm j}}
 \addtolength{\baselineskip}{-.1mm}

\renewcommand{\Large}{\large}

\setcounter{tocdepth}{2}
\addtolength{\baselineskip}{0mm}
\addtolength{\parskip}{.4mm}
\addtolength{\abovedisplayskip}{1mm}
\addtolength{\belowdisplayskip}{1mm}

\setcounter{footnote}{0}


\section{Introduction and Summary}
The goal of this work is to explore the bulk of Jackiw-Teitelboim (JT) quantum gravity, starting from the boundary observer's data. This program has attracted a lot of attention throughout the past years, especially in AdS$_3$/CFT$_2$ where constructions within CFT$_2$ are endowed with a gravitational interpretation in an emergent asymptotically AdS$_3$ space. 
\\~\\
JT gravity is a model of 2d quantum gravity and is the spherically symmetric sector of 3d gravity. More explicitly its action is \cite{Jackiw:1984je,Teitelboim:1983ux}:
\begin{equation}
\label{JTaction}
S_{\text{JT}}[g,\Phi] = \frac{1}{16\pi G}\int d^2x \sqrt{-g}\Phi \left(R-\Lambda\right) + S_{\text{GH}},
\end{equation}
in terms of the metric $g$ and the dilaton field $\Phi$.\footnote{We focus on negative cosmological constant $\Lambda = -2/L^2$, and we set $L=1$ from here on.} It is the universal low-energy limit of SYK-like models \cite{KitaevTalks,Sachdev:1992fk,Polchinski:2016xgd,Jevicki:2016bwu,Maldacena:2016hyu,Jevicki:2016ito,Cotler:2016fpe,wittenstanford,Turiaci:2017zwd,Gross:2017hcz,Gross:2017aos,Das:2017pif,Das:2017wae,Berkooz:2018qkz,gurari,Berkooz:2018jqr}, and has attracted a great deal of attention lately \cite{Almheiri:2014cka,jensen,malstanyang,ads2,mandal,Nayak:2018qej,shockwaves,pets,Moitra:2018jqs,paper4}.
Arguably it is the simplest model of quantum gravity that is nontrivial, meaning it has black holes and a nonzero Hamiltonian. It would then seem to make sense to first investigate questions like bulk reconstruction in this simplified NAdS$_2$/NCFT$_1$ setup. The idea is that the calculations are now more tractable than for example in AdS$_3$/CFT$_2$, whilst the theory is still complex enough to result in nontrivial lessons about quantum gravity.
\\~\\
On a technical level, our exploration of the bulk of this model will cover the following topics:
\begin{itemize}
    \item A boundary-intrinsic definition of a bulk frame.
	\item The exact calculation of local bulk correlation functions of a matter-coupled JT gravity.
    \item The exact calculation of geometric observables in quantum gravity: the metric and the geodesic distance between two bulk points.
\end{itemize}
Let us be more specific.
\\~\\
The first step towards defining local bulk observables in any theory of quantum gravity is to find a suitable diffeomorphism invariant definition of a bulk frame. This can be achieved by anchoring bulk points to the boundary by means of some geometric observable. Indeed, diffeomorphism gauge invariance is explicitly broken on the boundary, so boundary coordinates are physical. In this 2d set up, the physical coordinate is just the proper time of the boundary observer, see for example \cite{malstanyang,ads2}. Any sensible bulk frame should therefore have a definition in terms of the boundary observers proper time. More generically any bulk observable should come with a prescription for a boundary intrinsic procedure to obtain said observable.
\\
One could imagine various such definitions of bulk frames. A popular one is to anchor a bulk point to the boundary by fixing the geodesic distance between said bulk point and several reference points on the boundary, see \cite{Donnelly:2015hta,Almheiri:2017fbd} in general and \cite{ref3,Lewkowycz:2016ukf,ref4,ref5,ref6,Chen:2017dnl,Chen:2018qzm} for AdS$_3$/CFT$_2$ in particular. In 2d we require two such points. We will use a variant of this and anchor bulk points to the boundary via incoming and outgoing lightrays, or null geodesics. We imagine placing a fictitious mirror at the location of the bulk point, and define the spatial bulk coordinate of a point to be one half the time it takes a light ray emitted by the boundary observer to reflect off said mirror and find its way back to to the observer's detector. This procedure is implementing the old radar definition of constructing a coordinate frame. This light ray definition of defining bulk points was investigated holographically in generic dimensions in \cite{Engelhardt:2016wgb}.
\\
We wish to emphasize an important point here. Although anchoring bulk points to the boundary via lightrays seems to be a very natural thing to do, it remains a \emph{choice}. Whether or not one would want to make this particular choice is then up for debate. Other possibilities exist, and could be worth investigating. We come back to this in section \ref{s:otherframes}.
\\~\\
Given a solvable model of quantum gravity, one natural thing to do would seem to be to ask about the quantum gravity expectation values of diff-invariant geometric observables such as the local metric.\footnote{We will be interested in the metric tensor $ds^2=g(dx,dx)$ as a map from the manifold $\mathcal{M}$ to $\mathbb{R}$. This is a scalar under diffeomorphisms unlike the actual rank $(0,2)$ tensor field $g$.} This requires the computation of quantum gravity path integrals of the type 
\begin{equation}
    \average{\mathcal{O}}\equiv \int [\dpi g]\, \mathcal{O}(g) e^{-S[g]}.\label{qgpathintegral}
\end{equation}
In a generic theory of quantum gravity there are several major obstructions. 
\begin{itemize}
    \item In more than three dimensions, pure gravity is non-renormalizable and does not make sense as a UV-complete theory. One should then embed Einstein Hilbert gravity in a non-local completion such as string theory and study its correlation functions. 
    \\
    Two dimensions is special though: the theory \emph{is} renormalizable, so \eqref{qgpathintegral} is a sensible object to study. This also implies bulk locality might a priori be possible even at Planckian scales. The flipside of course is that the lessons on locality from our 2d model are not expected to be universally true in higher dimensions, except possibly in three dimensions. We leave this to future work. 
    \item Secondly, off-shell metrics appearing in the path integral can generically be quite exotic. Consequently there exists in general no closed expressions for the geometric observables $\mathcal{O}(g)$. For example there is generically no easy functional relation between $g$ and the geodesic distance between two points.
    \\
    This is where JT gravity is unique. All off-shell metrics in the JT path integral are known, and we can calculate geometric observables in them. For example, we will see that the geodesic distance between two bulk points is basically a bulk two-point function, and that the metric is given by a Schwarzian boundary two-point function. Both of these can be calculated exactly.\footnote{The boundary two point function has been calculated numerous times in the literature \cite{altland,altland2,schwarzian,Mertens:2018fds,paper3,kitaevsuh,belo,zhenbin}.}
    \item Finally there is the technical hurdle of actually being able to calculate the resulting path integral \eqref{qgpathintegral}. For example, in 3d gravity the calculation of the partition function $\mathcal{O}=\mathbf{1}$ is still very much an open problem \cite{wittenpartition,maloneywitten,jensenads3}, let alone some complicated correlation function. 
\end{itemize}

\noindent Next to purely geometic observables, we will also consider matter-coupled JT gravity, where the matter sector supplies the operators $\mathcal{O}(g)$ and the correlators \eqref{qgpathintegral} are of the form:
\begin{equation}
\int \left[\dpi g\right] \left[\dpi \Phi\right]\Big(\left[\dpi \phi\right]\phi_1(g \cdot x_1) \hdots \phi_n(g \cdot x_n) e^{-S_{\text{mat}}[\phi,g]}\Big) e^{-S_{\text{JT}}[g,\Phi]}.\label{main}
\end{equation}
The most important point here is that the locations of the operator insertions are bulk points as defined via our radar definition, and as such depend explicitly on the bulk metric $g$, as emphasized by the notation $g\cdot x$.\footnote{Let us give a preliminary introduction to this phenomenon within our specific context. We can think of the metrics in the JT gravity path integral as different patches of the AdS$_2$ \Poincare coordinate patch with coordinates $X$. Naively when considering matter-coupled JT gravity one might think of correlators of operators like $\phi(X)$. One might then be led to think that the matter correlator will return some fixed answer independent of the shape of the patch, and factor out of the gravitational path integral. However, any suitable boundary intrinsic definition of bulk points, such as our radar definition, implies that the location of the bulk point $g \cdot x$ in \Poincare coordinates $X$ depends on the shape of this patch. The bulk point effectively becomes fuzzy, and this will result in nontrivial quantum gravitational effects.} \\
We will be interested primarily in this work in understanding the validity of the semi-classical approximation of \eqref{main}. 
When can we describe correlators of the type \eqref{main} in terms of quantum matter on a classical background $g_0$:
\begin{equation}
\int \left[\dpi \phi\right]\phi_1(g_0 \cdot x_1) \hdots \phi_n(g_0 \cdot x_n) e^{-S_{\text{mat}}[\phi,g_0]}\,\, ?
\end{equation}
Here, $g_0$ is the saddle of JT gravity with no operator insertions, and one should keep in mind the explicit dependence on $g_0$ in the coordinate locations.  \\
Conversely, when is it \emph{not} valid to use the semiclassical approximation? This is when backreaction becomes important. In JT gravity, the model remains exactly solvable including matter backreaction.
\\~\\
We will find (perhaps unsurprisingly) that both at late times as well as in the very-near-horizon region, such quantum gravity effects become important. Though intuitive, this has implications for the information paradox in JT gravity. One of the assumptions of the Hawking calculation is that quantum matter on classical curved spacetime is a valid approximation to quantum gravity in the near-horizon region. We find here via very explicit calculation an example of a matter-coupled quantum gravity theory for which this does not hold. We comment on this in section \ref{s:information}.
The conclusion is that JT quantum gravity seems currently the only sensible theory of quantum gravity in which we can actually calculate path integrals of the type \eqref{qgpathintegral}. This is the topic of section \ref{sect:geometry}.
\\~\\
An important remark is that the quantum gravity path integral \eqref{qgpathintegral} requires an additional input of information: we need to specify what kind of metrics should contribute.
\\
At the very least we should constrain the asymptotics of the metric near the holographic boundary. On top of this, there is the question of which Euclidean topologies may contribute. The boundary of Euclidean AdS$_2$ is a circle. This can be filled in by a disk, but also by higher genus Riemann surfaces. Whether or not these higher genus contributions are included in the path integral \eqref{qgpathintegral} is a \emph{choice} that defines the quantum gravity theory we are working with. This makes contact with the old question on whether topology-changing dynamics is present in quantum gravity or not. 
\\
Throughout most of this work we will focus on the JT disk theory dual to Schwarzian quantum mechanics:
\begin{equation}
\label{SSch}
S[f] = -C\int dt \, \left\{f,t\right\}, \qquad \left\{f,t\right\} \equiv \frac{f'''}{f'} - \frac{3}{2}\left(\frac{f''}{f'}\right)^2
\end{equation}
with $\left\{f,t\right\}$ the Schwarzian derivative of $f$. Here, the field $f$ accounts for all physically inequivalent frames on the disk that satisfy $R=-2$ \cite{malstanyang,ads2}. This is a manifestly unitary model with a continuous density of states $\rho_\infty(E)=\theta(E)\sinh 2\pi \sqrt{E}$.
\\
When we investigate the near-horizon metric in sections \ref{s:metric} and \ref{s:information}, and the near-horizon matter correlation functions in section \ref{sect:otherprimaries}, the specifics of this density of states turn out to be crucial. More microscopic models of quantum gravity, such as the SYK model, have a spectrum $\rho_L(E)$ with more fine structure than the universal Schwarzian density of states $\rho_\infty (E)$.\footnote{The latter is recovered in the parametric limit $L\to\infty$.} This immediately raises the question how near-horizon physics is affected by this additional fine structure.
\\
To this end we consider in section \ref{sect:topo} and appendices \ref{app:othermanifolds} and \ref{app:latetimermt} the topologically complete JT model, where higher genus Euclidean contributions are included. The resulting theory of quantum gravity was recently shown to have a non-perturbative completion as a matrix integral \cite{sss2,sss,talks}. The theory should be thought of as a statistical average over individual unitary realisations, much like SYK. Each of the individual realizations has a discrete density of states. Taking the statistical average results in a continuous spectral density $\rho_L(E)$.
\\~\\
This work is structured as follows. 
\\
In {\bf section \ref{sect:mcjt}}, we couple JT gravity to bulk CFT matter. This is the main section of this work. We calculate and discuss the local bulk two-point function. The focus is on understanding how and when quantum gravity effects cause significant modifications from semi-classical CFT correlators. We find that quantum gravity effects dominate near-horizon physics. In addition, we compare correlators in pure and thermal states, to understand when a form of the Eigenvalue Thermalization Hypothesis (ETH) holds, and how the theory goes away from it.\footnote{ETH was studied within this class of models in \cite{Kourkoulou:2017zaj,Eberlein:2017wah,Sonner:2017hxc, shockwaves}.} 
\\
In {\bf section \ref{s:localoperator}}, we interpret the results from section \ref{sect:mcjt} as defining local bulk operators in quantum gravity. This interpretation is strengthened by an investigation of the analytic structure. 
\\
In {\bf section \ref{sect:geometry}}, we discuss the exact quantum expectation value of the metric operator as measured by a boundary observer. He observes a breakdown of semiclassical gravity close to the horizon, in line with the general lessons from section \ref{sect:mcjt}. In particular, at distances of order the Planck length $\ell_P$ from the horizon, he finds an answer that is not the Rindler metric, but instead has singular behavior.  
\\
In {\bf section \ref{sect:topo}}, we consider the gravitational matrix integral discussed in \cite{sss2}. Our goal is to understand how finite $L$ matrix integral effects affect the metric. We find that they take over at distances of order $\ell_P/L$ from the horizon. In this new parametric region, the breakdown of semiclassical gravity is even more dramatic than the effects observed in the JT disk model. In all we are led to believe that the statement \emph{quantum gravity effects dominate near horizon physics} is rather universal.
\\
In {\bf section \ref{s:information}}, we discuss implications for the information paradox. In particular we highlight how the importance of non-perturbative effects in the gravitational coupling $C$ for near-horizon physics, follows from the late time power-law decay of boundary correlation functions.\footnote{See \cite{maldacena2001,Dyson:2002nt,Goheer:2002vf,Barbon:2003aq} for the original arguments and \cite{Fitzpatrick:2016ive,kaplan1606,kaplan1609,kaplan1703num,anoushartmansonner,dyergurari} for the specific case of AdS$_3$/CFT$_2$ .} Similar conclusions hold for finite $L$ matrix integral effects. This disproves one of the assumptions that make up the information paradox. Therefore the paradox is averted. At least that is, within JT gravity, and given our choice of bulk frame.

\subsection{Defining a Bulk Frame}
\label{s:building}
\begin{figure}[h]
\centering
\includegraphics[width=0.5\textwidth]{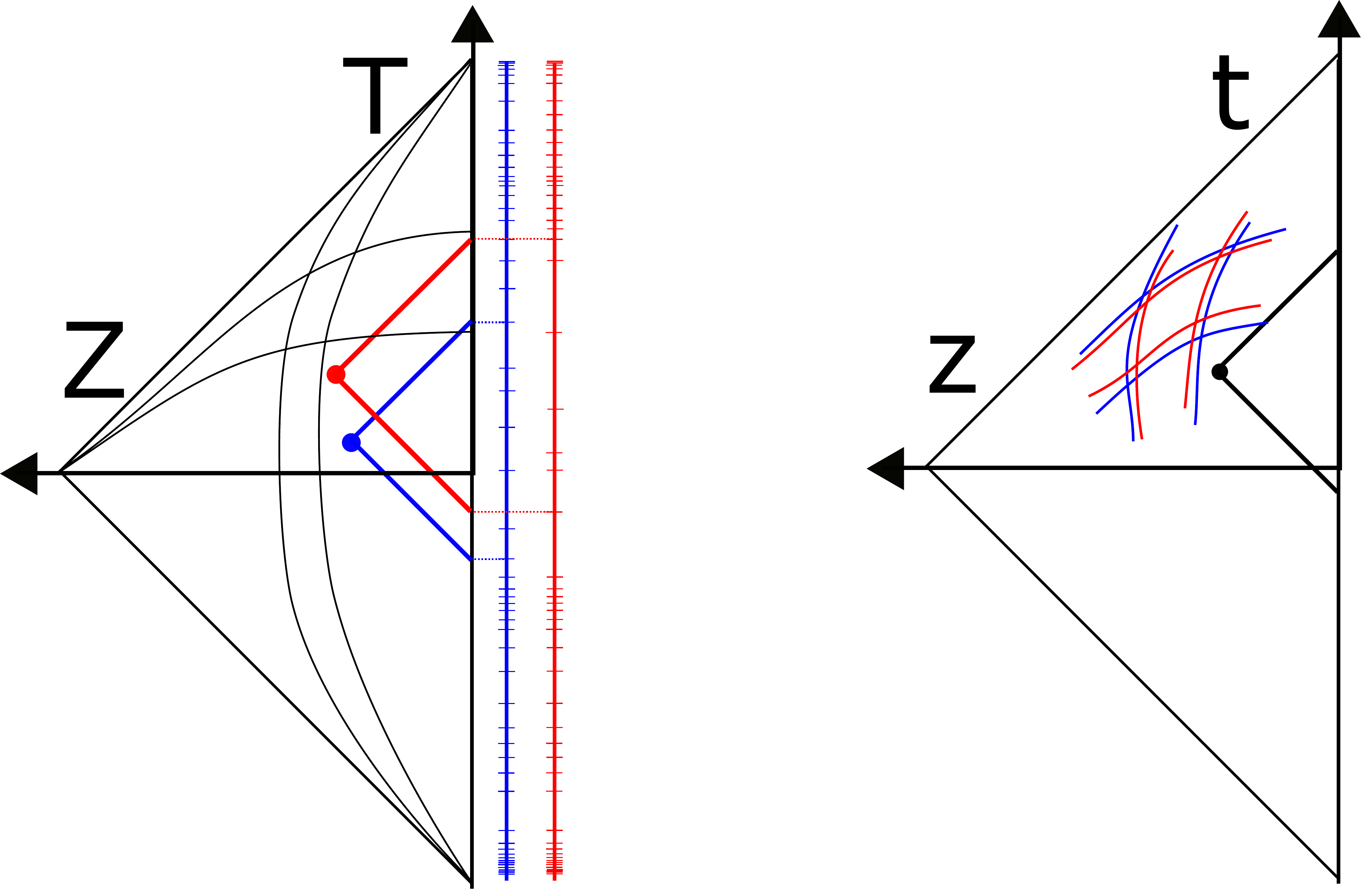}
\caption{Left: Building the bulk frame from the Poincar\'e perspective $(T,Z)$. For two reparametrizations (blue and red), we construct the bulk point given two boundary observer time coordinates $t_1$ and $t_2$. The ticks in the clock-ticking pattern are depicted. The bulk location is fuzzy in the Poincar\'e geometry, but the metric is fixed. Right: Building the bulk frame from the observer's perspective $(t,z)$. The bulk location is fixed, but the metric fluctuates.}
\label{fig:bulkpoints}
\end{figure}

Performing the path integral over $\Phi$ in \eqref{JTaction} localizes the path integral on bulk metrics that are locally AdS$_2$. In particular this means all such metrics can be thought of as cut-outs of Poincar\'e AdS$_2$ described by the coordinates:
\begin{equation}
ds^2=\frac{1}{Z^2}(dZ^2-dT^2).\label{poincare}
\end{equation}
The boundary observer can be though of as living on a wiggly boundary curve $Z(t) = \epsilon \dot{f}(t)$. His proper time $t$ is then related to the \Poincare time $T$ by the map $T=f(t)$. This field $f$ is the only physical degree of freedom in the system \cite{malstanyang,ads2,jensen}, and it is weighed with a Schwarzian action \eqref{SSch}.
\\~\\
From the perspective of the boundary observer (i.e. in his frame), the path integral over $f$ is interpreted as a path integral over inequivalent bulk metrics, as in Figure \ref{fig:bulkpoints}.
\\
He constructs the bulk frame as explained above, by shooting light rays into the bulk and collecting them back. As such, he associates coordinates $v=t-z$ and $u=t+z$ to every bulk point, where $v=t_1$ is the time on his clock at which he sends the signal and $u=t_2$ the time at which he receives the signal.\footnote{We have implicitly chosen conformal gauge for the bulk metric 
\begin{equation}
    ds^2=e^{\omega(z,t)}(dz^2-dt^2).
\end{equation}}
\\
We can describe this same experiment in the \Poincare frame with identical outcome: a lightray is sent from the boundary at $T_1=f(t_1)$, reflects at the fictitious mirror at the location of the bulk point, and impinges on the boundary at \Poincare time $T_2=f(t_2)$. Via this experiment we associate coordinates $V=T_1$ and $U=T_2$ to this bulk point. Combining all elements we find an experimentally determined, or boundary-intrinsic map of the \Poincare frame ($U,V)$ to the new bulk coordinates ($u,v$):
\begin{equation}
    U=f(u),\qquad V=f(v)\label{map}.
\end{equation}
Via this map we arrive at the following bulk metric:
\begin{equation}
\boxed{ds^2(f)=\frac{\dot{f}(u)\dot{f}(v)}{(f(u)-f(v))^2}(d z^2-d t^2),}\label{bulkmetric}
\end{equation}
with a location-dependent conformal scaling factor, uniquely determined by the physical field $f(t)$.
\\
As an example consider the thermal boundary reparameterization $f(t) = \tanh \frac{\pi}{\beta}t$ which is a solution to the Schwarzian equation of motion, following from \eqref{SSch}. We find the thermal AdS$_2$ bulk metric:
\begin{equation}
\label{thermal}
ds^2(\beta)=\frac{4\pi^2}{\beta^2}\frac{dz^2-dt^2}{\sinh^2 \frac{2\pi}{\beta} z}.
\end{equation}
Notice that our definition of a bulk frame is similar to that of \cite{Donnelly:2015hta}, where a preferred frame (our $t$-coordinate) was specified in terms of a platform and a geodesic distance measured perpendicularly into the bulk from this platform.
\\
Let us re-emphasize that our construction is boundary-intrinsic: it is described entirely in terms of local operations performed by the boundary observer at fixed times on his own clock. This (holographic) feature is believed to be fundamental in any construction of quantum gravity, and is actually already contained in the semi-classical description of JT gravity \cite{Almheiri:2014cka, ads2}, and the boundary correlators in the Schwarzian theory \cite{schwarzian}.\footnote{Indeed, the boundary bilocal operator insertions are specified at proper times $t_i$, with the time coordinate $t$ running from $-\infty$ to $+\infty$. It would seem to make little sense to attempt to define an operator in e.g. the Poincar\'e time coordinate as it might be behind the horizon (i.e. later than $t=+\infty$).} 

\subsection{Quantum Fluctuations}
Within this construction, a diff-invariant local (scalar) bulk operator at the bulk point $(u,v)$ is specified as $\mathcal{O}(f(u),f(v))$, where the $u,v$ labels are times on the boundary observer's clock. We have in effect used our bulk frame \eqref{bulkmetric} to define the gravitational dressing of bulk operators into diff-invariant observables.
\\
As a consequence, the location of this operator in the \Poincare frame depends explicitly on $f(t)$. The path integral over metrics can be thought of as some statistical averaging, such that the operator $\mo(U,V)$ is fuzzy and smeared out in the \Poincare frame $(U,V)$. This is represented in Figure \ref{fig:bulkpoints} (left). 
\\
This fuzziness is important for what follows: it represents the coupling of the bulk matter degrees of freedom to the boundary graviton $f$. As a result, the matter correlator does not factor out of the gravitational path integral \eqref{main}, and we are left with a path integral over inequivalent frames $f(t)$.  For every fixed $f(t)$, the matter correlator in \eqref{main} is that of a matter field in AdS$_2$. However, when we path-integrate over the degrees of freedom $f(t)$, this structure will not be preserved. \\
Semiclassical gravitational physics (quantum matter on curved space time) is obtained when the metric path integral collapses to its saddle via localization. This does not happen when probing late time or near-horizon physics, where quantum fluctuations of the metric become important. Indeed, we will find that the theory is always effectively strongly coupled in the near-horizon region.

\subsection{Other Bulk Frames}\label{s:otherframes}
An important point is that all the calculations that follow hinge on defining observables using this particular radar definition of bulk coordinates.
\\
Alternative boundary-intrinsic definitions of bulk points construct different observables, and hence imply a different coupling of bulk matter to the boundary graviton, and in general could result in different physics. There is no obvious guiding tool here which selects one such definition as being preferred.
\\~\\
We introduce an alternative definition of bulk points using spatial geodesics in appendix \ref{app:geodesic}. It would be interesting to investigate how matter correlators behave in this set of coordinate frames, though we see no easy route towards an explicit calculation. \\
It would be interesting to understand if there are any physical constraints that can guide us to a specific choice. Our specific choice of frame is directly related to static observers' observations and constructions done in the bulk by static observers. E.g. building upon the computations we perform in section \ref{sect:mcjt}, it directly leads to a quantum gravitational generalization of the Unruh effect for an eternal black hole \cite{Mertens:2019bvy}. But how does one define a frame and associated observables relevant for other observers, e.g. infalling-observer physics?
\\~\\
For now, we will restrict ourselves to making one particular choice and work out \eqref{main} within that context. In the remainder of this paper, we thereto stick to the radar definition, and use the bulk frame \eqref{bulkmetric}.

\section{Matter-coupled JT Gravity}
\label{sect:mcjt}
In this section we couple bulk matter to JT quantum gravity \eqref{JTaction}. We will assume the matter sector $S_{\text{mat}}[\phi,g]$ does not couple to the dilaton $\Phi$ and is a 2d CFT. We consider matter correlation functions of the form:
\begin{equation}
\int \left[\dpi g\right] \left[\dpi \Phi\right]\Big(\left[\dpi \phi\right]\phi_1 \hdots \phi_n e^{-S_{\text{mat}}[\phi,g]}\Big) e^{-S_{\text{JT}}[g,\Phi]}.
\end{equation}
Performing the path integral over $\Phi$, and fixing the bulk diffeomorphism invariance as discussed above, we localize to a path integral over the metrics \eqref{bulkmetric}:
\begin{equation}
\int \left[\dpi f\right] \Big(\left[\dpi \phi\right]\phi_1 \hdots \phi_n e^{-S_{\text{mat}}[\phi,f]}\Big)e^{-S[f]}.    
\end{equation}
We first perform the path integral over the matter sector, before integrating over gravity:
\begin{equation}
\mathcal{O}(f) \equiv \left\langle \phi_1 \hdots \phi_n\right\rangle_{\text{mat}} = Z_\text{mat}[f]^{-1}\int \left[\dpi \phi\right] \phi_1 \hdots \phi_n e^{-S_{\text{mat}}[\phi,f]}.\label{matter}
\end{equation}
Since all off-shell frames are local conformal transformations of the Poincar\'e frame, we know the general expression for $\mathcal{O}(f)$.\footnote{An example is a CFT two-point function of a weight $h$ primary field:
\begin{equation}
    G_f(u_1,u_2)=\dot{f}(u_1)^h\dot{f}(u_2)^h G(f(u_1),f(u_2))=\left(\frac{\dot{f}(u_1)\dot{f}(u_2)}{(f(u_1)-f(u_2) )^2}\right)^{h}.\label{twopointf}
\end{equation}
}
In writing \eqref{matter}, we have chosen to normalize the CFT correlation function \emph{before} evaluating the gravitational path integral. A priori, this seems relevant as the 2d CFT matter partition function $Z_{\text{mat}}[f]$ depends explicitly on the metric via the conformal anomaly.\footnote{Explicitly:
\begin{equation}
\label{anomaly}
Z_{\text{mat}}[f] = \exp \left(- \frac{c}{92 \pi} \int d^2x\int d^2x' R(x) G(x,x') R(x')\right),
\end{equation}
where $\int R \to \int R + 2 \oint K$ for a manifold with boundary as we study here, and $G(x,x')$ the bulk-to-bulk Green's function.} However, it was shown in appendix C of \cite{zhenbin} that in this particular set-up, the Polyakov-Liouville action, describing the conformal anomaly, reduces to a boundary Schwarzian action again, but with a prefactor that is subdominant to $C$ from the gravitational sector. Hence this choice is immaterial for any of the results.\footnote{Our choice has a similar flavor to it as section 3 of \cite{Kabat:1992tb}. We also remark that the geometric observables discussed in section \ref{sect:geometry} must be calculated with this normalization. We remark that for pure boundary correlators \eqref{bdybdy}, this choice is not present. They arise from integrating out a 1d boundary matter CFT, which has no conformal anomaly to begin with.}
\\~\\
In the end we are left with a Schwarzian path integral:
\begin{equation}
\boxed{\left\langle \mathcal{O}\right\rangle \equiv \int \left[\dpi f\right]\mathcal{O}(f)e^{-S[f]}.}\label{full}
\end{equation}
Whether or not we can push the calculation to the end, depends only on our ability to calculate the Schwarzian correlation function of some possibly involved $\sltr$ invariant operator $\mo(f)$. 
\\~\\
In the remainder of this section we focus on the bulk two-point function. More generic correlation functions are left to future work.

\subsection{Scalar Bulk Two-Point Function}
\label{s:bulkpropagator}
\begin{figure}[h]
\centering
\includegraphics[width=0.65\textwidth]{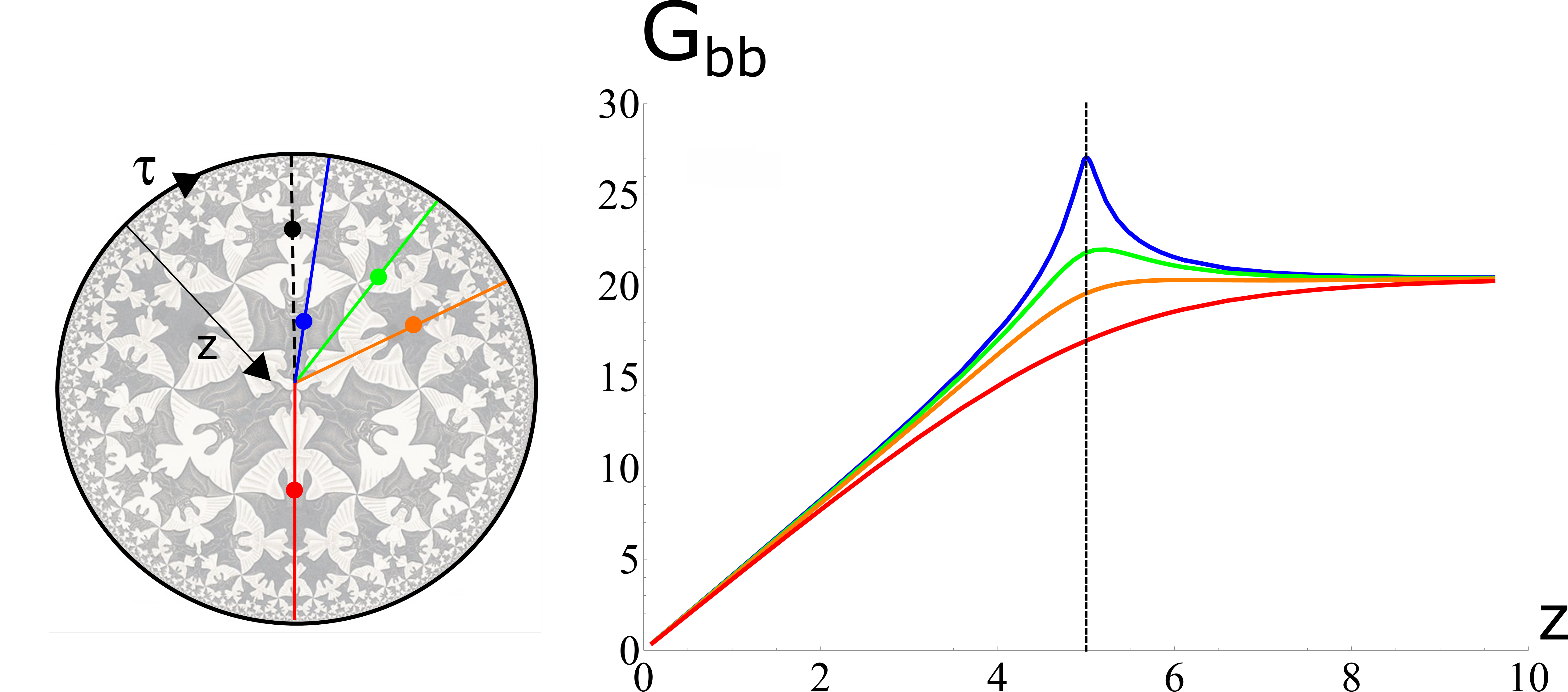}
\caption{Euclidean bulk-to-bulk two point function \eqref{bulkbulk} at finite temperature ($\beta=2\pi$, $C=1/2$), with one bulk point at ($t=0,z'=5$) and the second bulk point at ($it,z$), as a function of $z$. Blue: $it=0.1$, Green: $it=0.5$, Orange: $it=1$, Red: $it=\pi$.}
\label{fig:bbfiniteT}
\end{figure}

As our first example, we will consider specifically a free massless scalar $\phi$ coupled to JT gravity with total action:
\begin{equation}
\frac{1}{2}\int d^2x \sqrt{-g}\, g^{\mu\nu}\partial_\mu \phi \partial_\nu \phi + S_{\text{JT}}[g,\Phi].
\end{equation}
The goal is to compute the bulk two-point function $\left\langle \phi(p)\phi(q)\right\rangle$ within this full gravitational theory. We will make comments on the massive case further on.
\\~\\
As above, the matter path integral results in the Green's function of the field $\phi$ in the generically off-shell background $f$:
\begin{equation}
G_f(p,q)=\int \left[\dpi \phi\right] \phi(p)\phi(q) e^{-S_{\text{mat}}[\phi,f]}.
\end{equation}
The quantum gravity two-point function of the field $\phi$ is then:
\begin{equation}
\average{G(p,q)}=\int \left[\dpi f\right]G_f(p,q)e^{-S[f]},\label{qgpi}
\end{equation}
following \eqref{full}. The Lorentzian CFT two point function on the \Poincare half-plane \eqref{poincare} is well-known:\footnote{Remember that $U=T+Z$, $V=T-Z$. We leave a prefactor of $1/4\pi$ implicit.}
\begin{equation}
G(P,Q)=\ln \abs{\frac{(T-T')^2-(Z-Z')^2}{(T-T')^2-(Z+Z')^2}} = \ln\abs{\frac{(U-U')(V-V')}{(U-V')(V-U')}} \label{greenslc},
\end{equation}
which is found by including an image charge at $Z\to -Z$ to enforce Dirichlet boundary conditions at $Z=0$ \cite{Spradlin:1999bn}.\footnote{Note that this is different than the standard CFT expression on a half-space which is found by imposing Neumann boundary conditions instead.} The field $\phi$ being a scalar, we write:
\begin{equation}
G(f(p),f(q))=\ln \abs{\frac{(f(u)-f(u'))(f(v)-f(v'))}{(f(v)-f(u'))(f(u)-f(v'))}}.\label{greens1}
\end{equation}
Notice that this is an $SL(2,\mathbb{R})$-invariant quadrulocal operator, as it is just (the log of) a crossratio.\footnote{Interpreted as a quantum operator, the way we will compute it, we will implicitly consider the time-ordered quadrulocal operator, i.e. $\mathcal{T}\ln \hdots$. This is also implicitly done in the complexity computation of \cite{zhenbin} and the entanglement computation of \cite{Mertens:2019bvy}.} An observation is now the following:
\begin{equation}
\int_{v}^{u}d t \, \int_{v'}^{u'}d t' \frac{\dot{f}(t) \dot{f}(t')}{(f(t)-f(t'))^2} =\ln \abs{\frac{(f(u)-f(u'))(f(v)-f(v'))}{(f(v)-f(u'))(f(u)-f(v'))}}.\label{greens2}
\end{equation}
The integrand on the left-hand side is recognized as the Schwarzian $\ell=1$ bilocal operator $\mo_{\ell}(t,t')$, or the boundary-anchored Wilson line. The formula \eqref{greens2} relates through the holographic dictionary $m^2 = \ell (\ell-1)$ a boundary primary with $\ell=1$ to a massless field in the bulk. It is instructive to replace $f(u) \to \tanh \frac{\pi}{\beta} f(u)$. In terms of the new field $f(t)$, the JT path integral is over metrics that satisfy $f(t+i \beta)=f( t)+i \beta$. The relation \eqref{greens2} is then:
\begin{equation}
\int_{v}^{u}d t \, \int_{v'}^{u'}d t' \frac{\dot{f}(t) \dot{f}(t')}{\frac{\beta}{\pi}\sinh \frac{\pi}{\beta}(f(t)-f(t'))^2} =\ln \abs{\frac{\sinh \frac{\pi}{\beta}(f(u)-f(u'))\sinh \frac{\pi}{\beta}(f(v)-f(v'))}{\sinh \frac{\pi}{\beta}(f(v)-f(u'))\sinh \frac{\pi}{\beta}(f(u)-f(v'))}}.\label{greens3}
\end{equation}
Using \eqref{greens2}, one can write the bulk two-point correlator as:
\begin{align}
\average{G_{bb}(u,v,u',v')}&= \int_{v}^{u}d t \, \int_{v'}^{u'}d t'\,\average{G_{\partial\partial} (t-t')}. \\ 
&= \int_{-\infty}^\infty d\tau \,\theta(\tau-t-\abs{z}) \int_{-\infty}^\infty d\tau'\,\theta(\tau'-t'-\abs{z'})\,\average{G_{\partial\partial} (\tau-\tau')}.\label{hkll}
\end{align}
The boundary two-point function is well known: \cite{altland,altland2,schwarzian,kitaevsuh,belo,paper3}:\footnote{As compared to the group-theoretic notation used in \cite{schwarzian,paper3,paper4}, we have replaced the irrep label $k_i$ by energies $E_i=k_i^2$. We have also suppressed the Euclidean regulator $e^{-\epsilon E}$, and left some prefactors and $1/Z$ implicit. }
\begin{align}
&\average{G_{\partial\partial}(t-t')}^\beta = \int \left[\dpi f\right]\, \mathcal{O}_{1}(t,t')  \, e^{-S[f]}, \nonumber \\
&= \int_0^\infty d M \sinh 2 \pi \sqrt{M}\, e^{-\beta M} \int_0^\infty d E \sinh 2\pi \sqrt{E}\, e^{ i(t-t')(E-M)}\, \Gamma( 1 \pm i \sqrt{M}\pm i\sqrt{E} ).\label{bdybdy}
\end{align}
Performing the double integral, we find the bulk two-point function:\footnote{Notice that, as $\sin x /x \rvert_{x=0}=1$, there is no additional IR divergence in the bulk two point function as compared to the boundary two-point function.} 
\begin{align}
\average{G_{bb}(t,z,z')}_\beta= \int_0^\infty d M & \sinh 2 \pi \sqrt{M}\, e^{-\beta M} \int_0^\infty d E \sinh 2\pi \sqrt{E}\, e^{ it(E-M)}\,\nonumber\\&\times \frac{\sin z(E-M)}{E-M}\,\frac{\sin z'(E-M)}{E-M}\,\Gamma( 1 \pm i \sqrt{M}\pm i\sqrt{E} ). \label{bulkbulk}
\end{align}
where one recognizes $\omega^{-1}\sin z\omega = z\, \sinc z\omega$ as the Fourier transform of $\theta(t-\abs{z})$. Notice that the result only depends on the time-difference $t$. The result is plotted in Figure \ref{fig:bbfiniteT} for Euclidean times $\tau=it$.
\\
In the remainder of this subsection we present an in-depth analysis of the exact bulk two-point function \eqref{bulkbulk} in Hamiltonian eigenstates and thermal ensembles.

\subsubsection{Pure States}
\begin{figure}[h]
    \centering
    \includegraphics[width=0.4\textwidth]{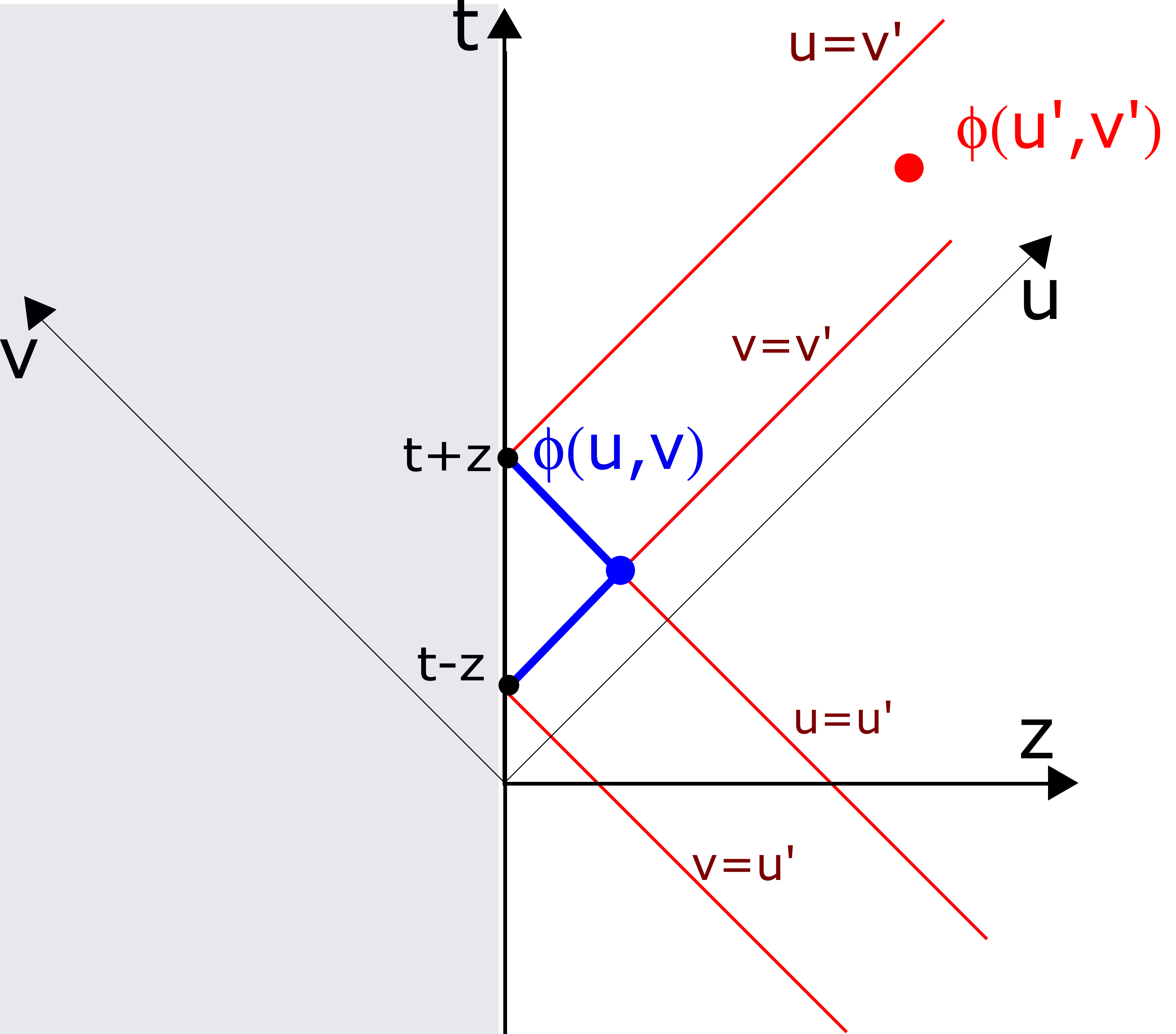}
    \caption{Two scalar operators $\phi$ at $(u,v)$ and $(u',v')$. Singularities in the propagator $\average{\phi(u,v) \phi(u',v')}$ are encountered on the four lines of lightcone separated events: two direct lines, and two indirect lines obtained by reflection on the holographic boundary.}
    \label{bulkreconstr3}
    \end{figure}

\begin{figure}[h]
\centering
\includegraphics[width=0.9\textwidth]{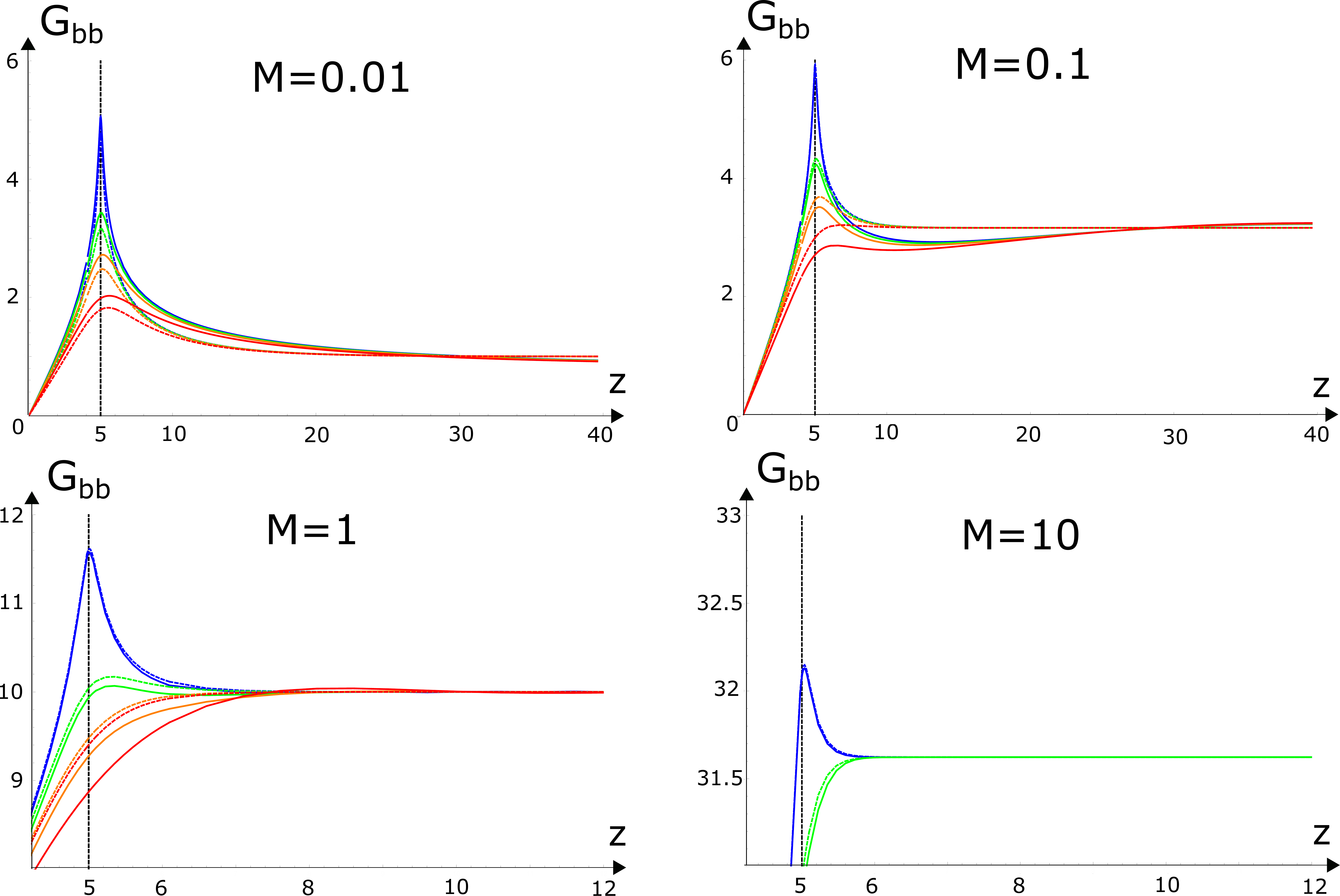}
\caption{Exact (full) and semi-classical (dashed) pure state bulk two-point functions $\average{G_{bb}(t,z,z')}_M$ and $\bar{G}_M(t,z,z')$ for different values of $M$ ($C=1/2$). Evaluated at $z'=5$ (black vertical line) and for a range of $z$ values. Blue: $it=0.1$. Green: $it=0.5$. Orange: $it=1$. Red: $it=2$.}
\label{fig:pureabs}
\end{figure}

Let us first consider pure energy eigenstates $\ket{M}$, and discuss $\average{G}_M$. The density of states is $\rho(M)=\sinh 2\pi \sqrt{M}$, and the thermal two-point function is the Laplace transform of the pure state - or microcanonical - two-point function: 
\begin{equation}
    \average{G_{bb}}_\beta =  \frac{1}{Z} \int_{0}^{+\infty} d M\sinh 2\pi \sqrt{M} e^{-\beta M} \average{G_{bb}}_M. 
\end{equation}
We obtain:\footnote{A direct calculation of $\bra{M}\hat{G}\ket{M}$ is also straightforward via BF techniques. The state $\ket{M}$ turns out to be an eigenstate of the operator $\hat{G}$.}
\begin{align}
\average{G(t,z,z')}_M = \int_0^\infty d E \sinh 2\pi \sqrt{E} &\,e^{-it\frac{(E-M)}{2C}}\,z\sinc \frac{z(E-M)}{2C}\nonumber\\ &\cross  z'\sinc \frac{z'(E-M)}{2C}\,\Gamma (1\pm i(\sqrt{M}\pm \sqrt{E})),\label{pureexact}
\end{align}
where we have reintroduced the appropriate $C$ dependence. Notice immediately that the analytic structure of this formula is that of a two-point function of \emph{local} bulk operators (see also section \ref{s:locality}). In particular the integral is finite and smooth except at the lightcone singularities at $t\pm z \pm z'=0$.\footnote{Notice that the exact answer \eqref{pureexact}, unlike the semiclassical approximation \eqref{pureclas} further on has no singularities \cite{zhenbin,Fitzpatrick:2016ive} at $\tau=\pm i n\pi / \sqrt{M}$.} Let us denote for future reference the smallest absolute value of $t\pm z \pm z'$ by $\tlc$. This is a measure for how far from the lightcone we are (see Figure \ref{bulkreconstr3}).
In Figure \ref{fig:pureabs} we plotted the exact \eqref{pureexact} and the classical \eqref{pureclas} pure state bulk two point functions. 
\\~\\
For the ensuing discussion it is useful to replace $E$ by the energy injection $\omega=E-M$. 
There are several interesting parametric regimes in this formula determined by $\tlc$. We define a macroscopic black hole as a system for which the integral \eqref{thermal} is dominated by its saddle, i.e. a system in the thermodynamic limit. Such states have $\sqrt{M}\sim C / \beta$, in terms of its Hawking temperature $\beta^{-1}$.
\begin{itemize}
    \item Suppose $\tlc\ll C/M$. The integral \eqref{pureexact} is then dominated by $\omega\gg M$. 
    Near the lightcone, all pure state correlators are identical to the zero-temperature correlators. The amplitude \eqref{pureexact} is effectively taken at $M=0$ and one recovers the zero-temperature result where the energy injection of the bilocal is indeed a UV effect. One finds:
    \begin{equation}
    \average{G(t,z,z')}_M\approx  \ln\abs{\frac{(t+z+z')(-t+z+z')}{(t+z-z')(-t+z-z')}}. \label{pureclaszero}
    \end{equation}  
    As we are close to the lightcone this expression reduces to the logarithmic lightcone divergence:
    \begin{equation}
        \average{G(t,z,z')}_M = \ln \frac{\tlc}{C}.
    \end{equation}
    This could have been obtained from a direct series-expansion of the quadrulocal operator \eqref{greens2} around any of the four options $t \pm z \pm z'=0$ in the Schwarzian path integral. Quantum gravitational effects are irrelevant near the lightcones, and the singularity comes entirely from the matter piece. 

    \item Suppose $C/M\ll \tlc \ll C$.\footnote{More precisely we require that the absolute values of all the light cone coordinates $t\pm z\pm z'$ satisfy these constraints. In what follows, for notational purposes we will often not be this explicit.} The integral \eqref{pureexact} is now dominated by $\omega\ll M$. Rescaling $\omega \to 2C \omega$, we obtain:
    \begin{equation}
    2C\sqrt{M}\int_{-\infty}^{\infty} d\omega e^{-i\omega t}(z\sinc z\omega) \,(z'\sinc z'\omega)e^{\frac{2\pi C \omega}{\sqrt{M}}}\Gamma (1\pm 2i\omega C/\sqrt{M}).\label{Fourier}
    \end{equation}
    Doing the Fourier transform we recover the semiclassical answer for a mass $M$ black hole:\footnote{One can check this by inserting $f(\tau)=\tanh \sqrt{M}\tau$ in \eqref{greens1}.}
    \begin{equation}
    \bar{G}_M(t,z,z')=\ln\abs{\frac{\sinh \frac{\sqrt{M}}{2C}(t+z+z')\sinh \frac{\sqrt{M}}{2C}(-t+z+z')}{\sinh \frac{\sqrt{M}}{2C}(t+z-z')\sinh \frac{\sqrt{M}}{2C}(-t+z-z')}}.\label{pureclas}
    \end{equation}
    Physically, the semiclassical result holds here because we are considering small perturbations on top of a black hole state $\omega\ll M$. This means backreaction is negligible.
    \item There is a transient regime $\tlc\sim C/M$ where backreaction does become important: the energy injection $\omega$ is of order the black hole mass $M$. This transient region is clearly visible in Figure \ref{fig:pureabs}. For macroscopic black holes $M\sim C^2$, this regime is absent and there are no UV-modifications of the semiclassical answer.
    \\
    The conclusion is that this type of UV backreaction only becomes important for light black hole states.
\end{itemize}
The regimes of late time $t\gg C$ and $z\gg C$ deserve special attention, they are discussed separately in section \ref{s:quantum}.
\\~\\
As a side remark, note that within the semi-classical regime, one could consider a region far from the singularities: $C/\sqrt{M} \ll \tlc \ll C$ , the Riemann-Lebesgue theorem can be used, which allows a lowest order Taylor expansion. This eliminates the last two terms in \eqref{Fourier}, and we end up with the convolution integral of two Heaviside distributions, resulting in a ramp like-structure that appears in all the plots in Figure \ref{fig:pureabs} as the asymptotes. 
Notice that this is zero for $\abs{t}>z+z'$. As we will discuss in section \ref{s:quantum} this is where IR - quantum gravity - effects take over and the semi-classical approximation fails.

\subsubsection{Thermal Ensemble}
\label{s:thermal}
For the thermal ensemble, we encounter the same parametric regimes as for the pure state, where we again postpone $z,t\gg C$. 
\begin{itemize}
    \item For $\tlc\ll\beta^2/C$ we are in the region close to the lightcones that is effectively zero temperature $\beta=\infty$.
    \item For $\beta^2/C\ll \tlc \ll C $ the semiclassical formula \eqref{thermclas} holds. Notice that this is true regardless of the ratio $\beta/C$. If in addition $\beta/C\ll 1$, we end up with \eqref{pureclas}, which is to be evaluated on the on-shell black hole mass $M_\beta = 4\pi^2 C^2 / \beta^2$. This then holds for all short to medium separations $\tlc\ll C$.   
    \item There is a transient regime dominated by UV effects as in Figure \ref{fig:pureabs} for $\tlc\sim \beta^2/C$. In the thermodynamic limit $\beta/C\ll 1$, the relative error vanishes and there are no UV effects.
\end{itemize}
It has been suggested \cite{vanraamsdonck} that outside of the horizon, pure states and thermal states are geometrically identical, and that they only differ in what is on and behind the horizon. We would like to put this to the test: how and when are observations in pure black hole states different from observations in thermal states?
\\
As we shall discuss further on in section \ref{s:quantum}, the main difference is in late time and near-horizon correlation functions. But clearly both states are also generically different when we back away from the thermodynamic limit $\beta/C\ll 1$. Take thereto $\tlc\ll C$. The thermal state answer is then:
\begin{equation}
\average{G(t,z,z')}_\beta \approx \int d M \sinh 2\pi \sqrt{M} e^{-\beta M} \bar{G}_M(t,z,z'),\label{thermclas}
\end{equation}
that is we approximate each pure state expectation value by its semiclassical answer. The integral is dominated by a saddle $M_\beta = 4\pi^2C^2/\beta^2$ \cite{shockwaves}. Let us include the first correction in $\beta/C$. We find:\footnote{One should use $\bar{G}_k \approx \bar{G}_{k_\beta}+(k-k_\beta)\bar{G}'_{k_\beta}$.}
\begin{equation}
\frac{1}{Z} \int_{0}^{+\infty}dM \left(\sqrt{M} - \frac{2\pi C}{\beta}\right) \sinh(2\pi \sqrt{M}) e^{-\frac{\beta M }{2C}} \approx \frac{1}{2\pi} + \mathcal{O}\left(\sqrt{\frac{C}{\beta}}e^{-\frac{2\pi^2 C}{\beta}}\right)
\end{equation}
The thermal propagator is then the pure state propagator \eqref{pureclas}, but with each of the four terms replaced by:
\begin{equation}
\ln \sinh \frac{\pi}{\beta} \tau  \,\, \to \,\, \ln \sinh \frac{\pi}{\beta}\tau + \frac{1}{4\pi C} \frac{\tau}{\tanh \frac{\pi}{\beta}\tau} \approx \ln \sinh \left[\left(\frac{\pi}{\beta} + \frac{1}{4\pi C}\right)\tau \right]+\mathcal{O}\left(\frac{\beta}{C}\right).\label{thermaltwopoint}
\end{equation}
The thermal two-point function at temperature $\beta^{-1}$ is hence that of a classical black hole with an effective higher temperature:\footnote{In particular, suppose we take $z'>z>t$ and $k_\beta z' \gg 1$, then numerically we indeed find:
\begin{equation}
\average{G(t,z,z')}_\beta=Z(\beta)4\pi\left(\frac{C}{\beta}+\frac{1}{2\pi^2}+\mo \left(\frac{\beta}{C}\right)\right)z'.\label{shift}
\end{equation}}
\begin{equation}
\beta^{-1}\to \beta^{-1}\left(1 + \frac{\beta}{C}\frac{1}{4\pi^2}\right)=\tilde{\beta}^{-1}. \label{shift2}
\end{equation}
The conclusion is that the first order correction in $\beta/C$ - away from the thermodynamic limit - introduces an observable difference between the thermal state $\rho_\beta$ and the pure black hole state $\ket{M_\beta}$: $\rho_\beta\approx \ket{M_{\tilde{\beta}}}\bra{M_{\tilde{\beta}}}$. This is of course as it should be in thermodynamics.

\subsubsection{Massive Bulk Fields}
As a direct extension, one can study correlation functions of a massive bulk field coupled to JT-gravity, with action:
\begin{equation}
S_{\text{mat}}[g,\phi] = \frac{1}{2}\int d^2x \sqrt{-g}\, g^{\mu\nu}\left(\partial_\mu \phi \partial_\nu \phi +m^2 \phi \right).
\end{equation}
We treat the resulting correlators in Appendix \ref{app:massive}.

\subsection{CFT Primary Bulk Two-Point Function}
\label{sect:otherprimaries}
The scalar propagator in 2d CFT is special due to its IR properties. For large spatial separations $\abs{z-z'}$ where we increase $z$, it goes to a constant on the half-plane. Generic 2d CFT correlators on the other hand decay as function of $\abs{z-z'}$. This resonates into the discussion on near-horizon physics in section \ref{s:quantum}. To this end, we investigate the bulk CFT two-point functions for primary fields of generic weights $(h,\bar{h})$. 
\\~\\
The CFT two-point function of weight $(h,\bar{h})$ primaries on the half-plane is:
\begin{equation}
    G_{h,\bar{h}}(u,u',v,v')=\average{\phi_{h,\bar{h}}(u,v)\phi_{h,\bar{h}}(u',v')}_{\text{CFT}}=\frac{1}{(u-u')^{2h}}\frac{1}{(v-v')^{2\bar{h}}}-(u'\leftrightarrow v').
\end{equation}
The second term is a mirror term that imposes Dirichlet boundary conditions at the holographic boundary $z=0$ as for the scalar field in \eqref{greenslc}. We only focus on the disconnected part of the correlator (or we take a free matter sector again).
In a generic off-shell background $f$, this transforms according to:
\begin{equation}
    \left(\frac{\dot{f}(u)\dot{f}(u')}{(f(u)-f(u') )^2}\right)^{h}\left(\frac{\dot{f}(v)\dot{f}(v')}{(f(v)-f(v') )^2}\right)^{\bar{h}}-(u'\leftrightarrow v').\label{primary1}
\end{equation}
This can be viewed as a product of two Schwarzian bilocal operators. In a canonical language, we are led to compute
\begin{equation}
\left\langle \hat{\mathcal{O}}_h (u,u') \hat{\mathcal{O}}_{\bar{h}}(v,v')\right\rangle - (u'\leftrightarrow v').
\end{equation}
It is important to note the operator ordering here. Regardless of the ordering of the four times $u,u',v,v'$, the bilocal operators are applied as a whole one after the other.\footnote{This determines the integration contour for a path integral calculation as one where Wilson lines in the Euclidean disk do not cross, see for example \cite{paper3}. This means in particular that no $6j$-symbols appear.} The bulk two-point function is thus identical to a time-ordered Schwarzian four-point function, computed e.g. in \cite{schwarzian}. For instance, for a pure state $\ket{M}$, one finds:
\begin{align}
    \nonumber\average{G_{h,\bar{h}}(t,z,z')}_M &=\int d E_u \sinh 2\pi \sqrt{E_u}\,\frac{\Gamma(h\pm i\sqrt{M}\pm i \sqrt{E_u})}{\Gamma(2\ell)}e^{i(u-u')(E_u-M)}\\
    &\nonumber\cross\int d E_v \sinh 2\pi \sqrt{E_v}\,\frac{\Gamma(\bar{h}\pm i\sqrt{M}\pm i \sqrt{E_v})}{\Gamma(2\ell)}e^{-i(v-v')(E_v-M)}\\
    &-(u'\leftrightarrow v'),\label{twopointl}
\end{align}
where the second term is the mirrored configuration that implements the Dirichlet boundary condition $\average{G_{h,\bar{h}}(t,z,0)}_M=0$.
\\
Notice that there is holomorphic factorization of the CFT bulk two-point function in a pure state, i.e. it has the same structure as the CFT bulk two point function in all the off-shell geometries \eqref{primary1}. The two chiral sectors of the CFT are only coupled by averaging over energy eigenstates.
\\~\\ 
The properties of \eqref{twopointl} are unsurprising.
\begin{itemize}
    \item The formula \eqref{twopointl} has exclusively lightcone singularities, so it has the analytic structure of propagator of \emph{local} bulk fields.
    \item The other regimes are mainly identical to those discussed around equation \eqref{pureexact}.
\end{itemize}
The main structural difference with the scalar propagator is in its near-horizon behavior, which we will now discuss.   

\subsection{Near-Horizon and Late-Time Physics}
\label{s:quantum}
\begin{figure}[h]
\centering
\includegraphics[width=0.4\textwidth]{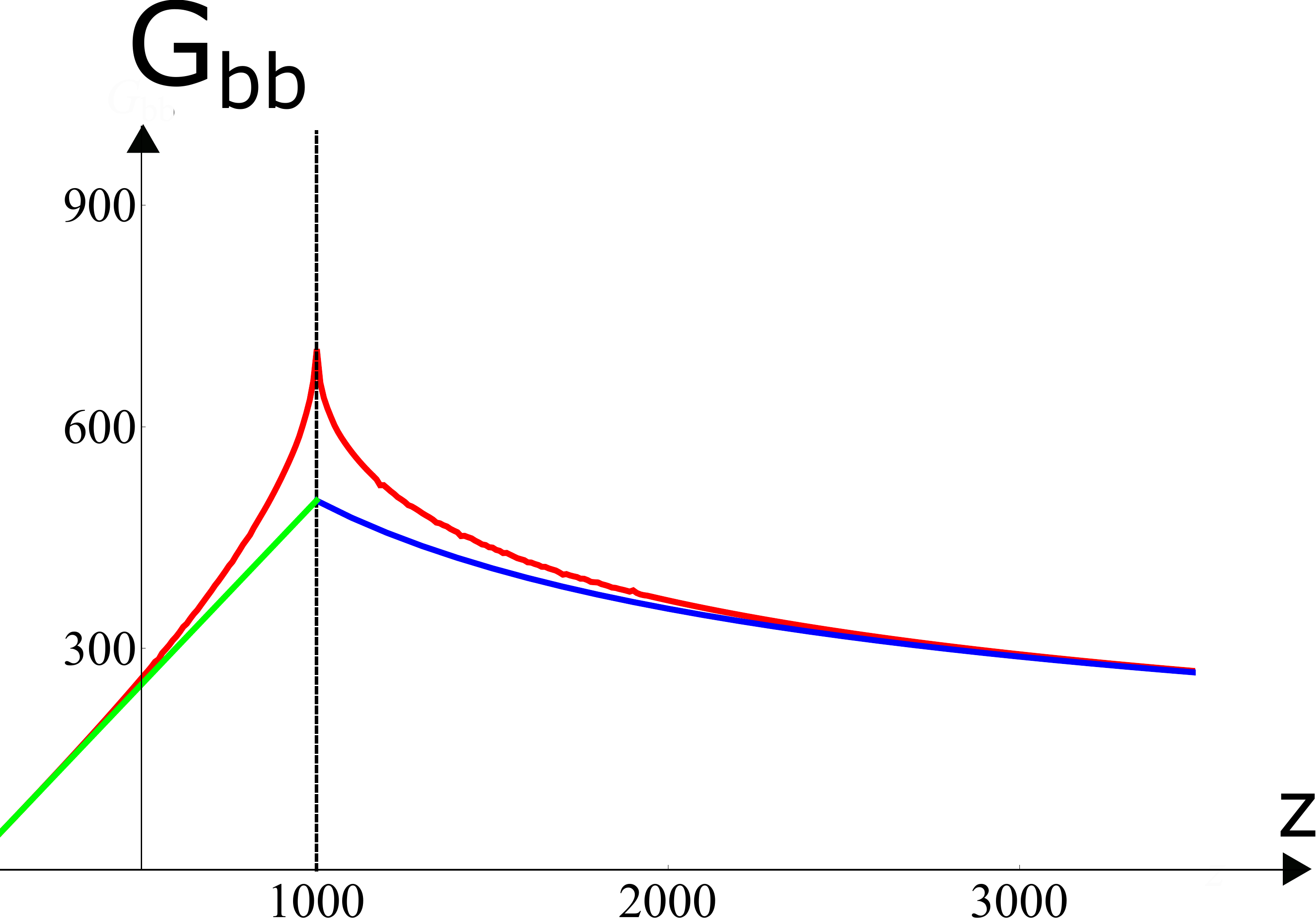}
\caption{Red: Zero-temperature or zero-energy $M=0$ bulk two-point function $G_{bb}^{\infty}$ as a function of $z$ for $z'=1000$, $t=1$ and $C=1/2$. The initial decay close to the lightcone singularity is logarithmic \eqref{pureclaszero}. Blue: power law (inverse square-root) behavior at $z\gg z'$. Green: power law (linear) behavior at $z\ll z'$.}
\label{Bulk2ptyzeroT}
\end{figure}

\begin{figure}[h]
\centering
\includegraphics[width=0.5\textwidth]{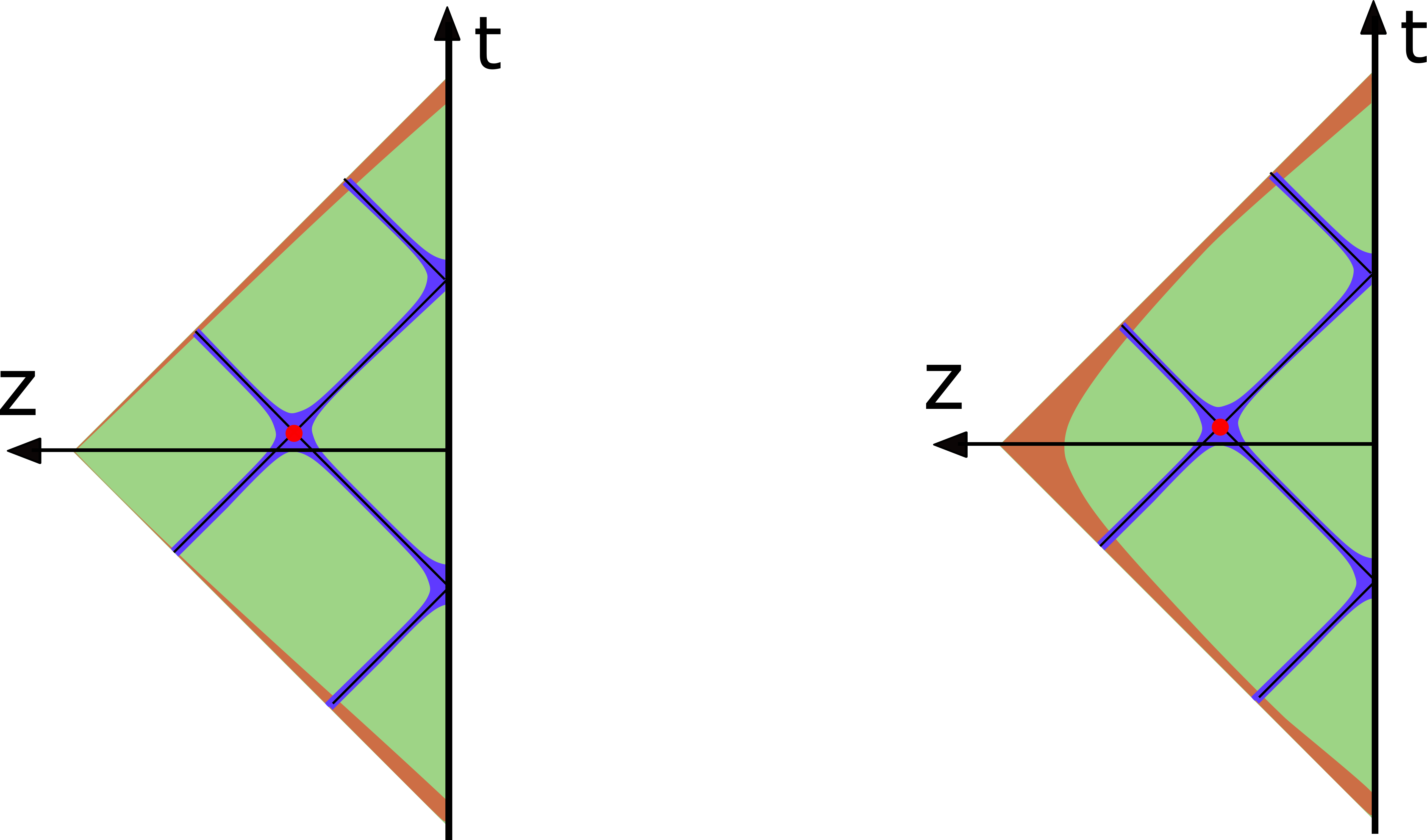}
\caption{Different regimes in the bulk propagator. One of the operators is at the red dot. UV (zero-temperature) effects dominate when the second operator is close to the lightcone $\tlc \ll C/M$ (blue). At larger separation, the semiclassical result is exact. For large separations $\tlc \gg C$, IR quantum gravity effects dominate resulting in power-law decay. We show the scalar \eqref{pureexact} (left) and generic primary two-point function \eqref{twopointl} (right).}
\label{2ptlocationsemi}
\end{figure}

A statement that is generically true is that late-time effects in physics probe IR collective quantum effects. Consider for example the Fourier decomposition of the pure state boundary two point function \eqref{pureexact}:
\begin{equation}
    G_{\partial\partial} (t)=\int_{-\infty}^\infty d \omega\, G_{\partial\partial}(\omega)\,e^{i\omega t}.\label{fourtf}
\end{equation}
From the theory of Fourier analysis we know that large $t$ probes the fine-grained non-analytic properties of $G_{\partial\partial}(\omega)$. It is precisely this fine-grained spectrum of a generic theory that is inherently quantum. For the Schwarzian theory, the function $ G_{\partial\partial} (\omega)$ is sharpest near the vacuum $\omega+M=0$ where it goes like $G_{\partial\partial}(\omega)\sim \theta(\omega+M)\sqrt{\omega+M}$. We can thus safely approximate the late time behavior of the boundary two point function by:\footnote{A phase factor $e^{i M t}$ multiplies this result.}
\begin{equation}
    G_{\partial\partial}(t) \sim \int_0^\infty d E \,\sqrt{E} e^{i E t}\sim t^{-3/2}.\label{powerbdytwo}
\end{equation}
Reintroducing the correct $C$ dependence we see that this approximation holds true whenever $t\gg C$.
\\~\\
Let us return to the pure state scalar propagator $\average{\phi(t,z)\phi(t,z')}_M$ in formula \eqref{pureexact}. For late times the above argument holds and there is a transition from the semiclassical exponential decay at $t\ll C$:
\begin{equation}
\ln\abs{\frac{\sinh \frac{\sqrt{M}}{2C}(t+z+z')\sinh \frac{\sqrt{M}}{2C}(-t+z+z')}{\sinh \frac{\sqrt{M}}{2C}(t+z-z')\sinh \frac{\sqrt{M}}{2C}(-t+z-z')}}\sim e^{-2\frac{\sqrt{M}}{2C}t}\sinh{2\frac{\sqrt{M}}{2C} z}\sinh{2\frac{\sqrt{M}}{2C} z'},\label{missedsaddle}
\end{equation}
to power law behavior at $t\gg C$:
\begin{equation}
\average{G(t,z,z')}_M\sim \sqrt{t+z-z'}+\sqrt{t-z+z'}-\sqrt{t+z+z'}-\sqrt{t-z-z'}.\label{power1}
\end{equation}
In the case of additionally $z' \ll z$, we can series expand the above
\begin{equation}
\average{G(t,z,z')}_M\sim \frac{z'}{\sqrt{t+z}} + \frac{z'}{\sqrt{z-t}}.\label{powerextra}
\end{equation}
This yields inverse square-root behavior when $z \gg z'$ and linear behavior when $z \ll z'$, which is indeed visible in the zero-temperature plot of Figure \eqref{Bulk2ptyzeroT}.\footnote{It is also visible in the small $M$ plots of Figure \ref{fig:pureabs}, for intermediate time ranges, but is eventually defeated by the late-time semi-classical constant result.} 
in the asymptotic regions.
\\
Similar formulas can be obtained for the thermal ensemble.
\\~\\
As a rule of thumb, we can understand that in general a transition from semiclassical behavior to power-law behavior will take place whenever the semiclassical answer gets exponentially small with some large ever increasing parameter in the exponent. Both regimes claim a certain region of $\omega$ dominates for integrals such as \eqref{fourtf}. One should then calculate the approximate answers for either of the assumptions, and the largest one wins.
\\~\\
Consider for example the primary two-point functions \eqref{twopointl}. A transition to power-law behavior is observed whenever either of the light cone separations becomes large, say $u-u'\gg C$.\footnote{Explicit formulas are easily obtained.} In particular this means we will start to see IR quantum gravity effects if one of the operators approaches the horizon closer than the Planck length $z\gg C$. 
\\
It is clear that the same should in fact be true for a generic class of correlation functions in JT gravity. This is in fact precisely the same physics as the universal appearance of power-law decay in boundary correlation functions at order $C$ time separations. Close to the horizon, IR quantum gravity effects take over from semiclassical physics.
This can be thought of as a manifestation of the generic fact that the horizon is a probe for ultra-low energy physics, and hence for quantum effects.\footnote{For related comments in gauge theories, see \cite{paper1,paper2,paper4}.} Any finite-energy excitation at infinity, gets blueshifted to arbitrarily high energies close to the horizon, and is washed out by the Riemann-Lebesgue theorem. The result is dominance by the deep IR spectrum.
\\
The scalar two-point function \eqref{pureexact} is an exception to this mantra: it does not decay at large spatial separations and in fact becomes independent of $z$ for large $z$.\footnote{This is a manifestation of the well-known pathological IR properties of the 2d scalar propagator.} Therefore, it is insensitive to the near-horizon quantum gravity fluctuations.
\\
We summarize the parametric regions where quantum gravity becomes important in Figure \ref{2ptlocationsemi}.

\section{Bulk Reconstruction}
\label{s:localoperator}
\begin{figure}[h]
\centering
\includegraphics[width=0.25\textwidth]{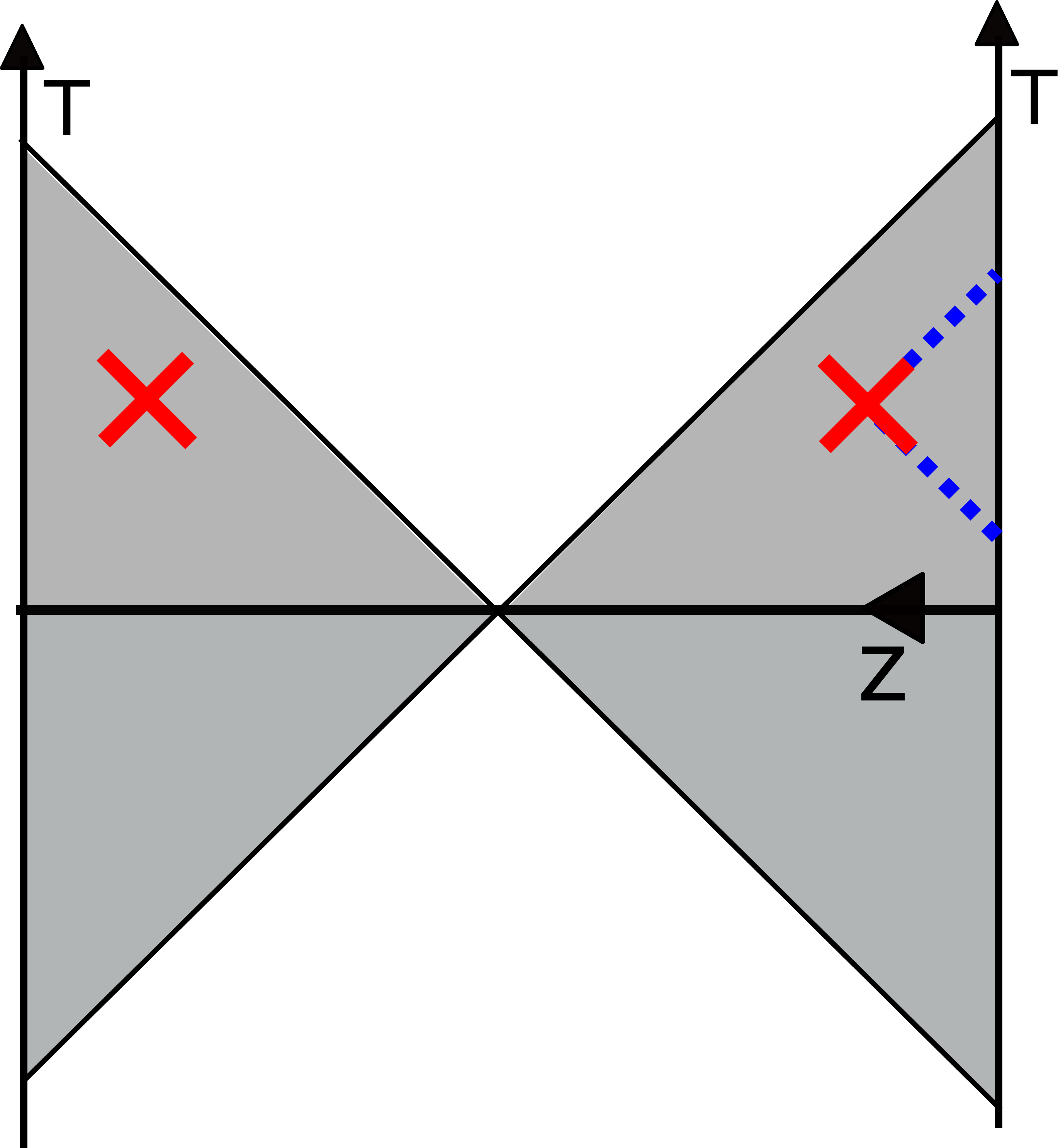}
\caption{Region of spacetime (gray) that can be reconstructed from the boundary using the radar construction of bulk points. Within the thermal ensemble, bulk operators can be taken to the other side by setting $t \to i\frac{\beta}{2} - t$.}
\label{Reconstruction}
\end{figure}

Up to this point, we have directly computed bulk $n$-point correlators. Here we distill the individual bulk operators contained in this definition, and phrase them in the context of bulk reconstruction and microcausality. 
\\~\\
The radar definition of the bulk points allows us to only access the exterior of the bulk black hole. For a pure state, this is it, but for thermal states we can move one of the bulk points to the other side of the TFD state by taking $t \to \frac{i\beta}{2}- t$ (Figure \ref{Reconstruction}).
If we want to define bulk operators behind the horizons (the upper and lower wedge in Figure \ref{Reconstruction}), one way is to utilize the pullback-pushforward procedure of \cite{Almheiri:2017fbd} and use the bulk retarded Green's function to write the bulk operator in terms of a linear combination of bulk operators defined on the $t=0$ slice. The latter can then be reconstructed from the boundary using our radar set-up. \\
An alternative proposal to define interior local operators, is to move an operator as $t\to \frac{i\beta}{4}-t$. This is motivated by the coordinate transformation to map the right Rindler patch into the upper patch. In this case, we define it in terms of the boundary observer's local time, in a similar way as the other side of the TFD. However, it is unclear whether such a computation actually probes interesting physics, as it postulates \emph{a priori} that the bulk interior makes sense in this set-up. We leave a further study of this to future work.

\subsection{Local Bulk Operators}
\begin{figure}[h]
\centering
\includegraphics[width=0.6\textwidth]{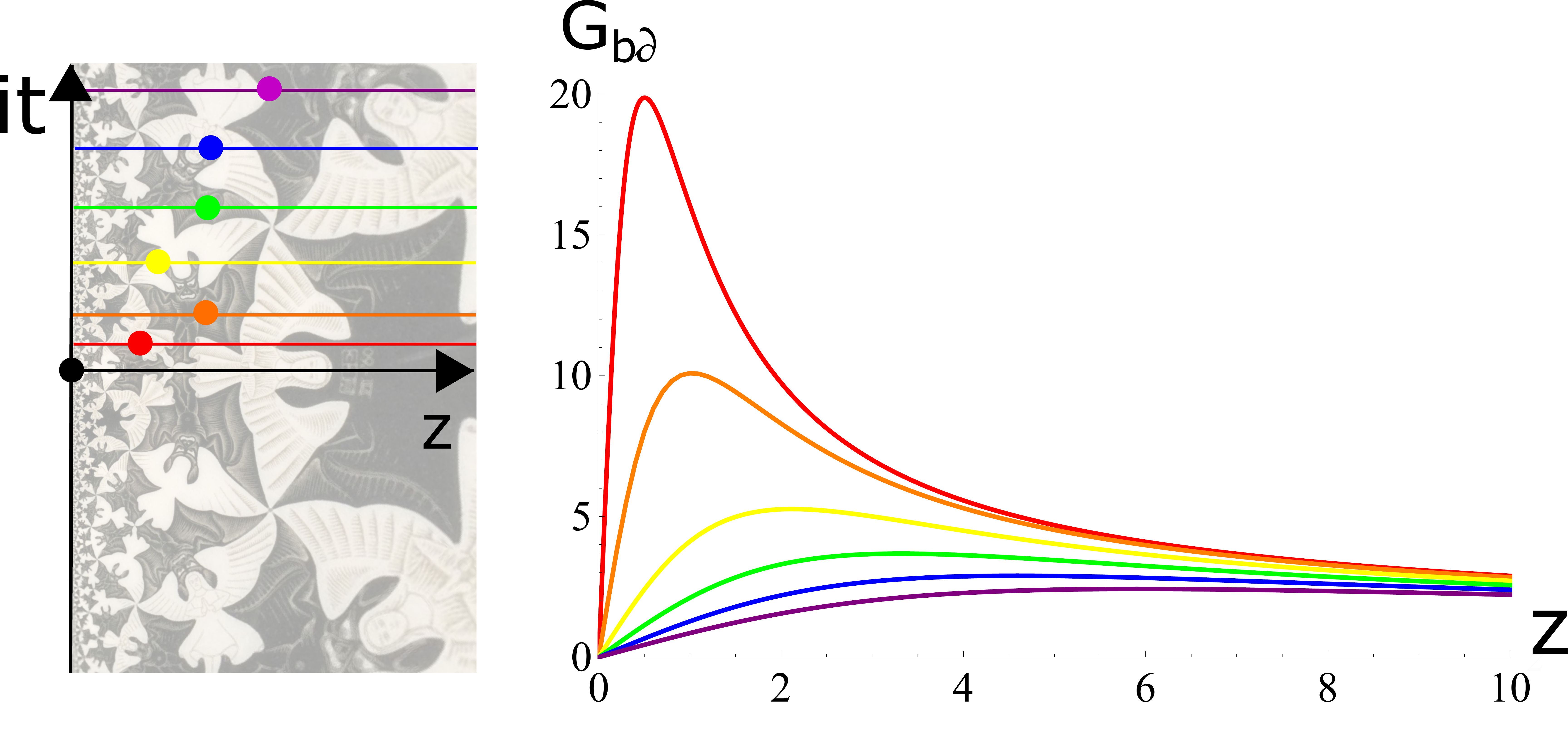}
\caption{Euclidean bulk-to-boundary two point function at zero temperature ($C=1/2$), as a function of radial coordinate $z$ into the bulk. Red: $t=0.5$, Orange: $it=1$, Yellow: $it=2$, Green: $it=3$, Blue: $it=4$, Purple: $it=5$.}
\label{fig:UHPbdelta}
\end{figure}

\begin{figure}[h]
\centering
\includegraphics[width=0.6\textwidth]{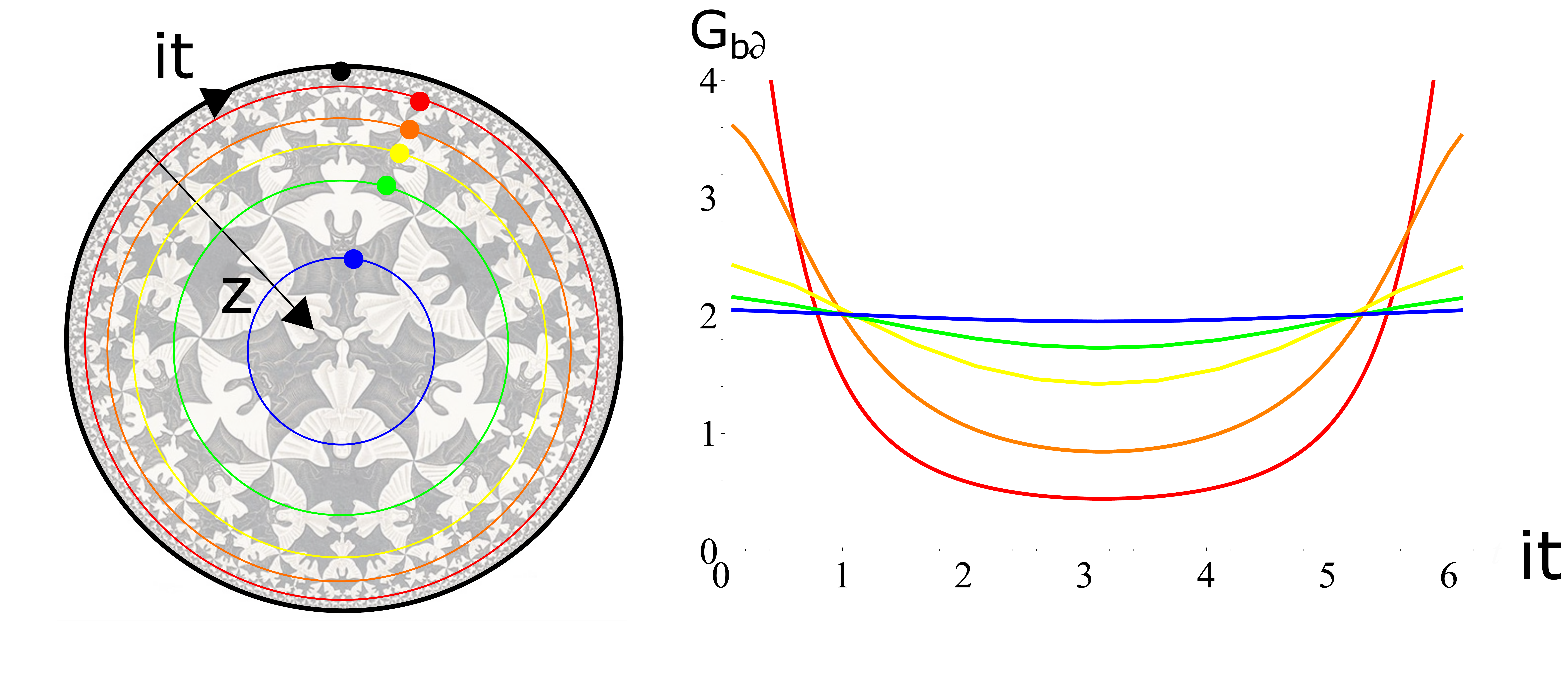}
\caption{Euclidean bulk-to-boundary two point function at finite temperature ($C=1/2$), with the bulk point at ($it,z$), as a function of $\Delta t$. Red: $z=0.5$, Orange: $z=1$, Yellow: $z=2$, Green: $z=3$, Blue: $z=5$.}
\label{fig:bbfiniteT(v2)}
\end{figure}

For a given metric $g$, the HKLL prescription \cite{hkll1,hkll2,kll,kl} associates to a linear combination of boundary operators $\mo_h$:
\begin{equation}
\hat{\phi}_{m^2}(t,z;g) = \int d\tau \, K_{m^2}(t-\tau,z;g) \,\hat{\mo}_h(\tau),
\end{equation}
a local field $\phi_{m^2}$ in the bulk, defined as an operator in the boundary CFT with $m^2 = h(h-1)$. \\
This formula ignores gravitational as well as self-interactions in the bulk.\footnote{Within a holographic CFT, it would be lowest order in $1/N$.} Out of these, gravitational interactions are arguably conceptually the most interesting and puzzling. Our goal here is to fully account for gravitational interactions. We will do this in a similar spirit as \eqref{full}, that is we apply the HKLL procedure for every generically off-shell background and then path-integrate over these backgrounds.
\\~\\
For simplicity let us focus on a $h=1$ primary in the boundary or, via the holographic dictionary, a $m^2=0$ massless scalar in the bulk. This allows checks with formulas from section \ref{s:bulkpropagator}. The generalization to massive fields is included in appendix \ref{app:massive}.
\\
A generic bulk $n$-point correlation function is then:
\begin{equation}
\average{\phi(t_1,z_1) \hdots \phi(t_n,z_n)}=\int \left[\dpi f\right]\left( \prod_i \int d\tau_i \, K(t_i-\tau_i,z_i;f) \average{\mo_1(\tau_1)\hdots \mo_1(\tau_n)}_{\text{CFT}}\right)e^{-S[f]},\label{hkllgrav}
\end{equation}
where we have already carried out the path integral over the dilaton $\Phi$ to localize on the bulk metrics \eqref{bulkmetric}.
For AdS$_2$ in \Poincare coordinates and a primary field $\phi$ of weight $h=1$ the appropriate kernel is just a Heaviside function \cite{Lowe:2008ra,Almheiri:2017fbd}: $K(T-T',Z)=\theta(T-T'-\abs{Z})$ which transforms like a scalar between different frames:
\begin{equation}
\phi(T,Z)=\int_V^U d T' \,\mo_1(T') =\int_v^u d t'\, \mo_1(t').
\end{equation}
So the HKLL kernel is independent of $f$ and just a Heaviside distribution:
\begin{equation}
K(t,z;f)=\theta(t-\abs{z}).\label{kernel}
\end{equation}
This means that the path integral over $f$ can be pulled through the kernels. The massless scalar bulk $n$-point functions in full quantum JT gravity should then from \eqref{hkllgrav} simply be the $n$-fold convolution of the Schwarzian $n$-point functions with Heaviside distributions. This indeed matches with a direct bulk calculation \eqref{hkll}.\footnote{Starting with a non-interacting 1d CFT on the boundary, all bulk and boundary correlators factor into two-point functions.} 
\\~\\
As an example, consider the zero temperature bulk-to-boundary propagator. Starting with the zero-temperature boundary two-point function:
\begin{equation}
\average{G_{\partial \partial}(t)}_\infty = \int d E\sinh 2\pi\sqrt{E}\, e^{ - i E t}\,\Gamma( 1\pm i \sqrt{E})^2,
\end{equation}
and convoluting this once with a Heaviside we obtain the bulk-to-boundary propagator at zero temperature:
\begin{align}
\average{G_{b\partial}(t;z)}_\infty = \int d E\sinh 2\pi\sqrt{E}\, e^{ - i E t}\, z\,\sinc z E \,\Gamma( 1\pm i \sqrt{E})^2.\label{bulkboundary}
\end{align}
This is in sync with the extrapolate holographic dictionary: the $z'\to 0$ limit of the bulk-to-bulk propagator $\average{G_{bb}(t,z,z')}_\infty$ is $z' \average{G_{b\partial}(t;z)}_\infty$. A few profiles of the zero temperature and finite temperature bulk-to-boundary propagators are plotted in respectively Figure \ref{fig:UHPbdelta} and Figure \ref{fig:bbfiniteT(v2)}.

\subsection{Bulk Locality}
\label{s:locality}
Equation \eqref{bulkbulk} is the time-ordered correlation function. The Schwarzian theory has time-reversal invariance.\footnote{Taking $t\to-t$ leaves $\left\{f,t\right\}$ invariant.} Time-reversal is realized in operator language by an anti-unitary (and anti-linear) operator. Time-reversed correlators can then simply be obtained by complex conjugating time-ordered expressions.\footnote{This is indeed true for the actual expressions found in \cite{schwarzian} using a careful Wick-rotation} For $t_1<t_2$, the time-ordered and anti-time-ordered two-point functions are respectively:
\begin{equation}
G^+(t_1,z_1;t_2,z_2) = \average{\phi(t_2,z_2)\phi(t_1,z_1)}, 
\end{equation}
and
\begin{equation}
\qquad G^-(t_1,z_1;t_2,z_2) = \average{\phi(t_1,z_1)\phi(t_2,z_2)} = G^+(t_1,z_1;t_2,z_2)^*.
\end{equation}
Their difference is the gravitational expectation value of the commutator between the bulk operators $\average{\left[\phi(t_1,z_1),\phi(t_2,z_2)\right]}$. At $t=0$, \eqref{bulkbulk} and \eqref{twopointl} are real, hence the equal-time commutators vanish.\footnote{In the case of \eqref{twopointl} this is invariance under the exchange of $E_u$ and $E_v$.}
\\~\\
As pointed out earlier, the bulk-to-bulk propagators \eqref{bulkbulk} and \eqref{twopointl} have (logarithmic) divergences exclusively on the lightcone (including reflections off of the holographic boundary) in the full quantum gravity theory. This analyticity combined with the edge-of-the-wedge theorem proves that the commutator vanishes for all spacelike separations:
\begin{equation}
\label{microcaus}
\boxed{\average{\left[\phi(t_1,z_1),\phi(t_2,z_2)\right]} = 0 , \qquad (t_1,z_1) \text{ and } (t_2,z_2) \text{ spacelike}.}
\end{equation}
Thus the local bulk operators $\phi(t,z)$ satisfy microcausality in the full JT quantum gravity. As mentioned in the introduction, we do not really expect local bulk operators in quantum gravity in spacetime dimensions higher than three, where non-localities at the Planck (or string) scale proliferate \cite{Lowe:1995ac,Susskind:1993aa,Mertens:2015hia}.

\section{Geometric Observables}
\label{sect:geometry}
In this section we exactly calculate the expectation value of the metric operator in JT quantum gravity. As a warm-up, we consider (a variant of) the geodesic distance.
\\~\\
The goal is to find out if and how quantum geometry, interpreted as expectation values of geometrical observables, is different from classical geometry. We should stress that it is not obvious whether this is how one would want to characterize the quantum generalization of classical geometry, but from a quantum mechanical point of view at these these are well defined observable quantities.
\\
As a byproduct we investigate how pure states are geometrically different from thermal states. It has been suggested \cite{vanraamsdonck} that outside of the horizon, pure states and thermal states are geometrically identical, and that they only differ in what is on and behind the horizon. We will emphasize that there are significant differences between pure and thermal states in the very near-horizon geometry. 

\subsection{Geodesic Distance}
\label{s:geodesicdistance}
In appendix \ref{s:geodesics} we review the construction of bulk geodesics in \Poincare AdS$_2$. Consider now the following function of the isometric invariant \eqref{invariant}:
\begin{equation}
d(P,Q)=\ln \abs{1+ \frac{(Z-Z')^2-T^2}{4ZZ'}}=\ln \abs{1+\delta^2(P,Q)} .\label{distancepoinc}
\end{equation}
Let us highlight some of the features of this quantity.
\begin{itemize}
    \item It is related to the geodesic distance $D(P,Q)$ \eqref{distance} between $P$ and $Q$ by 
    \begin{equation}
    \exp \frac{d(P,Q)}{2}=\cosh \frac{D(P,Q)}{2}.\label{dfromD}
    \end{equation}
    There are two interesting regimes here. For $d(P,Q) \gg 1$ we have $d(P,Q) = D(P,Q)$, which we will call large distances. On the other hand, for $d(P,Q) \ll 1$ or small distances we have $d(P,Q)= D^2(P,Q)$.
    \\
    A consequence of the relation \eqref{dfromD} is that $d(P,Q)=0$ is identical to $D(P,Q)=0$, and indeed both vanish only on the lightcone $T=\pm \Delta Z$. Both quantities diverge only when either one of the points approaches the boundary $Z=0$.\footnote{There is also a divergence on the reflected lightcone $T=\pm (Z+Z')$.} 
    \item We can rewrite \eqref{distancepoinc} as:
    \begin{equation}
    d(P,Q)=\ln\abs{ \frac{(U-V')(U'-V)}{(U-V)(U'-V')} }.\label{geodesic}
    \end{equation}
    Comparing this with \eqref{greenslc} we observe that $d(P,Q)$ is obtained from the scalar bulk two-point function $G(P,Q)$ by exchanging $V$ and $U'$:
    \begin{equation}
    d(U,V,U',V')=G(U,U',V,V').\label{dfromg}
    \end{equation}
    This means we have direct analytical control of this observable.\footnote{We have no knowledge of a similar exact calculation for the actual geodesic distance $D(P,Q)$.} 
\end{itemize}
We are interested in calculating the quantum gravity expectation value of the observable $d(p,q)$. To calculate the path integral over all metrics \eqref{bulkmetric} we require the distances in the off-shell metrics \eqref{bulkmetric}. Since the distance is a scalar we readily obtain:
\begin{equation}
d_f(p,q)=\ln \abs{\frac{(f(u)-f(v'))(f(v)-f(u'))}{(f(u)-f(v))(f(u')-f(v'))} }.\label{df}
\end{equation}
The quantum gravity expectation value of $d(P,Q)$ then follows from the exact bulk propagator \eqref{bulkbulk}:\footnote{Alternatively, one might calculate the geodesic distance from the boundary two-point function:
\begin{equation}
\average{d(p,q)}=\int_{u'}^u d t \int_{v'}^v d t' \average{G_{\partial\partial} (t-t')}.
\end{equation}}
\begin{equation}
\average{d(u,v,u',v')}=\average{G_{bb}(u,u',v,v')}.\label{disanceexact}
\end{equation} 
It is useful to make the mapping \eqref{disanceexact} more explicit. Suppose $z'\geq z$ and $z'-z\geq t$ such that the points are spacelike separated. We find:
\begin{equation}
t_G=z+z',\qquad z_G=\frac{z'-z-t}{2},\qquad z_G'=\frac{z'-z+t}{2},\label{coordrel}
\end{equation}
for the coordinates $(t_G,z_G,z'_G)$ to be used in the bulk two-point function in order to obtain the distance in coordinates $(t,z,z')$. This allows us to directly map conclusions about the bulk two-point function such as differences between classical and exact pure state results, as well as differences between pure state and thermal state results, to similar conclusions about expectation values of the geodesic distance operator. 
\\~\\
We will not discuss this in detail, and only mention the main point: there is a sharp difference between pure states and thermal states when we probe in the near-horizon region. This is also the region where a dramatic breakdown of semi-classical gravity can be observed and the geodesic distance gets severely modified by IR quantum gravity effects.
\\
Consider e.g. $t=0$ and the situation where both endpoints are in the very near-horizon region $z' > z \gg C$ with also $z'-z\gg C$. From \eqref{disanceexact} we infer there has been a transition to power-law behavior. From \eqref{power1} we find the equal-time pure state geodesic distance:
\begin{equation}
    \average{d(z,z')}_M\sim 2\sqrt{z+z'}-\sqrt{2z}-\sqrt{2z'},
\end{equation}
independent of $M$. For thermal states we find a very different power-law behavior:
\begin{equation}
    \average{d(z,z')}_\beta \sim \frac{2}{z+z'}-\frac{1}{2z'}-\frac{1}{2z}.
\end{equation}
\subsection{Metric}
\label{s:metric}
\begin{figure}[h]
\centering
\includegraphics[width=0.2\textwidth]{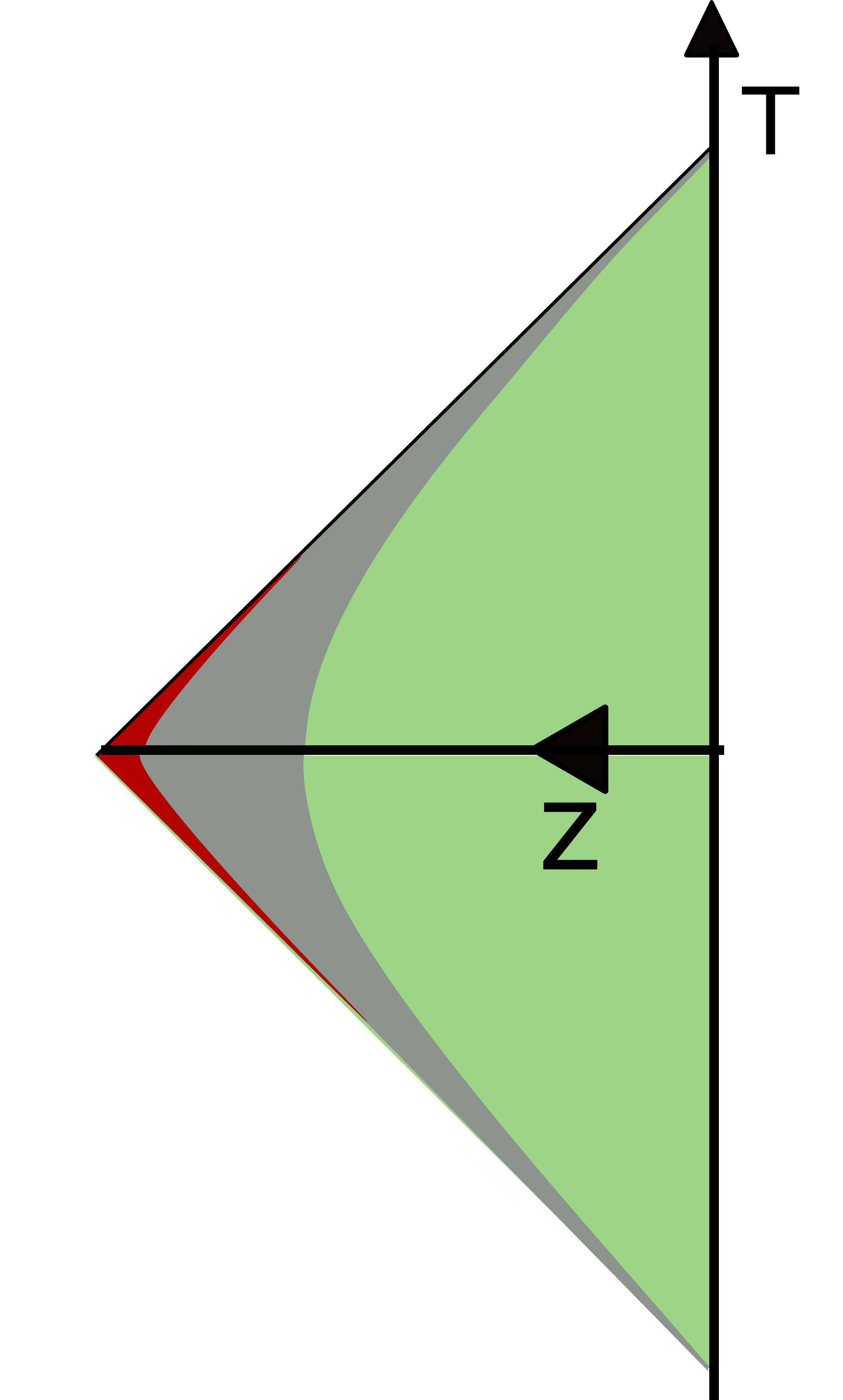}
\caption{Bulk geometry as found by our radar definition. Quantum effects become important and deform the effective geometry at a distance of $z \approx C$ , i.e. at the Planck scale $\ell_P$ from the horizon (red). Outside of this region, there is a band (gray) that extends outwards to the black hole scale $z  \approx C/\sqrt{M}$ that bounds the Rindler approximation to the classical geometry.}
\label{rindlerquantum}
\end{figure}

The quantity $d(p,q)$ can be used to determine the expectation value of the metric operator. The coordinate invariant definition of the metric tensor $g$ is the square of the geodesic distance between points that are infinitesimally separated:
\begin{equation}
ds^2 =g(dx,dx)= D^2(p,p+dx) = d(p,p+dx),
\end{equation}
since for small distances, $d(p,q)=D^2(p,q)$. We wish to emphasize again at this point that this metric observable has a completely boundary-intrinsic, operational definition, so this is an observable that the boundary observer has access to. Remember that bulk coordinates were defined by sending in lightrays from the boundary, the bulk metric is obtained as the geodesic distance between infinitesimally separated such bulk points. 
\\~\\
Differentiating \eqref{df},  we obtain for a generic off-shell $f$:
\begin{equation}
\label{dfu}
d_f(p,p+dx)=\ln\abs{1-\frac{(f(u)-f(u+du))}{(f(u)-f(v))}\frac{(f(v)-f(v+dv))}{(f(u)-f(v))}} = \frac{\dot{f}(u)\dot{f}(v)}{(f(u)-f(v))^2}du\, dv.
\end{equation}
This means we obtain $\average{ds^2(t,z)}$ by computing the boundary two-point function \eqref{bdybdy} with $t\to u$ and $t' \to v$:
\begin{equation}
\label{metrictwopoint}
\average{ds^2(t,z)} = \int \left[\dpi g\right] \left[\dpi \Phi\right]ds^2(t,z)e^{-S[g,\Phi]} =
\average{G_{\partial\partial}(2 z)}(d z^2-d t^2).
\end{equation}
Notice that this is by construction a scalar. Notice also that the causal structure of the exact metric is the same as that of the classical metric (the path of light rays is not affected). 
\\
As before in equation \eqref{primary1}, we should specify an operator ordering, or equivalently a contour for the path integral, whenever we promote a classical object such as \eqref{dfu} to an operator. We want the metric $\hat{ds^2} \equiv \hat{g}$ - as a quantum mechanical operator - to be Hermitian. The bilocal operator $\hat{\mo}(t)\hat{\mo}(t')$ that is used to compute the boundary two-point function, is not Hermitian: $(\hat{\mo}(t)\hat{\mo}(t'))^\dagger=\hat{\mo}(t')\hat{\mo}(t)$.\footnote{The local operators $\hat{\mathcal{O}}$ are Hermitian.} A suitable Hermitian metric operator is obtained as the average of both time-orderings $\hat{\mo}(t)\hat{\mo}(t')+\hat{\mo}(t')\hat{\mo}(t)$, and this is how one should read the two-point function in \eqref{metrictwopoint}. The anti-time-ordered bilocal was considered in section \ref{s:locality}, and is the complex conjugate of the time-ordered one. So the factor $\average{G_{\partial\partial}(2z)}$ in \eqref{metrictwopoint} is to be read as the real part of \eqref{bdybdy}.
\\~\\
For example for a pure state the result is:
\begin{equation}
\boxed{\average{ds^2}_M=(d z^2-d t^2)\int d E \sinh{(2 \pi  \sqrt{E})} \, \cos 2 z(E-M)\,\Gamma( 1 \pm i \sqrt{M}\pm \sqrt{E}).}\label{quantummetric}
\end{equation}
This observable \eqref{quantummetric} has the following features.
\begin{itemize}
    \item Near the boundary the metric reduces to the \Poincare metric $(d z^2-d t^2)/z^2$. Again this result could have been inferred directly from the path integral as every metric $ds^2_f$ reduces to the \Poincare metric for $u \approx v$ via a Taylor series.
    \item In the regime $C/M \ll z \ll C$ and $M \sim C^2$, the above integral is dominated by $E = M+\omega$ with $\omega \ll M$ and we recover the classical metric of a mass $M$ black hole:
    \begin{equation}
    d s^2_M = M \frac{dz^2-d t^2}{\sinh^2 \frac{\sqrt{M}}{2C}2z },\label{dsclassic}
    \end{equation}
	which has a horizon at $z= \infty$. For light black holes there is a transient regime close to the asymptotic boundary where UV effects cause deviations from this semiclassical answer, as in Figure \ref{fig:pureabs}. For regular sized black holes there is no such regime. Let us focus on these from hereon.
	\\
	Closer to the horizon at  $C/\sqrt{M} \ll z$, this classical metric reduces to the Rindler metric:\footnote{This is Rindler space in tortoise coordinates $\rho=e^{-\frac{\sqrt{M}}{C} z},\, \tau=2\sqrt{M}t$, with metric $ds^2 = \rho^2d\rho^2-d\tau^2$.}
    \begin{equation}
    d s^2_M=4 M e^{- \frac{2\sqrt{M}}{C}z}(dz^2-d t^2).\label{rindler}
    \end{equation}
\end{itemize}
As we have come to expect, funny things start to happen when we probe very close to the horizon. There is now a transition in the measured metric \eqref{quantummetric} from the semiclassical Rindler approximation \eqref{rindler}, to power law behavior dictated by nonperturbative quantum gravity effects. The switchover between the Rindler \eqref{rindler} and the quantum gravity regime occurs at $z\approx C$. In the classical Rindler metric \eqref{rindler}, the proper distance $\rho$ from this point to the classical horizon is 
\begin{equation}
	\ell_P = C\, e^{-\sqrt{M}}, \label{planck}
\end{equation}
to be interpreted as the Planck length in this 2d set-up.\footnote{Note that for a macroscopic black hole $M \sim C^2$ the exponential factor is immensely small.}
\\~\\
When we probe much closer than this Planck length to the classical horizon $\rho\ll \ell_P$, the measured metric is nothing like classical Rindler space. For example for a thermal state we find up to a prefactor:
\begin{equation}
\average{ds^2}_\beta=\frac{d z^2-d t^2}{z^3}.\label{quantummetricthermal}
\end{equation}
We illustrate the different regimes of expectation value of the bulk metric in Figure \ref{rindlerquantum}.
For pure states we find that the scaling factor is $z^{-3/2}$ instead of the $z^{-3}$ in \eqref{quantummetricthermal}, as in \eqref{powerbdytwo}. So the near-horizon metric in a pure state is fundamentally different from the thermal one. Let us make some comments.
\begin{itemize}
    \item It is amusing to calculate the curvature tensor associated with this effective metric \eqref{metrictwopoint}. One should think of this curvature tensor as constructed operationally by the boundary observer. He constructs a manifold using our radar definition of bulk points, and endows this effective classical manifold with the effective metric \eqref{quantummetric}. He then measures the effective curvature tensor by parallel transporting a vector around a small loop on the effective manifold. This operational result will be different from the mathematical curvature $R=-2$, because the points along the loop are now defined using our definition of bulk points, and not as fixed \Poincare coordinates. For $z\ll C$, he finds the classical answer $R_{\left\langle g\right\rangle_\beta}=-2$. For $z\gg C$ however, he finds from \eqref{quantummetricthermal}:
    \begin{equation}
         R_{\left\langle g\right\rangle_\beta}= -3 z.\label{ricci}
    \end{equation}
    This means he concludes from his very near-horizon experiment \eqref{quantummetricthermal} that there is a true singularity at the location of the semiclassical horizon $z=+\infty$.
		    \item A related feature is the strong fluctuations of the metric near the horizon of the thermal system. We could for example calculate the covariance of the metric:\footnote{We thank Zhuo-Yu Xian for suggesting this.}
\begin{equation}
    \text{Cov}(g(z_1),g(z_2)) \equiv \frac{\langle g(z_1)g(z_2)\rangle_\beta- \average{g(z_1)}_ \beta\average{g(z_2)}_\beta}{\average{g(z_1)}_ \beta\average{g(z_2)}_\beta},
\end{equation}
which probes fluctuations away from the saddle $g_0(z)$. In the parametric regions where semiclassical physics holds, the covariance vanishes. When probing close to the horizon though, this changes. For example, if we take $z_1,\,z_2\gg C$ and take furthermore $z_1\gg z_2$, the covariance blows up as: 
\begin{equation}
    \text{Cov}(g(z_1),g(z_2)) \sim z_2^{3/2}.
\end{equation}
So in this sense, the closer we get to the horizon, the more prominent quantum fluctuations become.
    \item We want to stress that one should not think about this metric expectation value as an effective metric on which matter fields propagate. This calculation should merely be considered as additional evidence that quantum gravity effects will generically dominate the expectation value of any operator inserted in the near-horizon region. The calculations of matter correlators in section \ref{sect:mcjt} should be considered more fundamental than the toy-calculations in this section, and have implications for the information paradox, as discussed below in section \ref{s:information}.
\end{itemize}
With the discussion of section \ref{s:quantum} in mind, we arrive at a significant conclusion: the fine structure of the Schwarzian density of states implies that the very near-horizon bulk geometry, as well as all matter correlation functions probing it, are modified at the Planck length $\ell_P$.\footnote{Moreover, we see that at these scales, most observables in JT quantum gravity have very different expectation values in pure states as compared to thermal states.}
\\
Given that the Schwarzian spectrum implies rather radical changes to our classical ideas on near-horizon physics, this raises the question how near-horizon physics is affected in a theory whose spectrum has even finer structure than the Schwarzian density of states, such as SYK. This is the topic of the next section. 
\\~\\
As a side comment, note that the pure energy eigenstates $\ket{M}$ play a privileged role when it comes to quantum geometry. \\
Indeed, all geometric observables we can calculate in JT gravity are, in one way or another, composed of boundary bilocal operators, essentially due to $\sltr$-invariance.  But since the Schwarzian bilocal operators commute with the Hamiltonian \cite{schwarzian}, the energy eigenstates $\ket{M}$ simultaneously diagonalize all geometric observables:
\begin{equation}
    \bra{M}\mathcal{O}\ket{M'}=\delta(M-M')\bra{M}\mathcal{O}\ket{M}.
\end{equation}
So the states $\ket{M}$ are also geometry eigenstates. One could interpret this as evidence that it might be more natural to think about pure states as being the quantum generalization of a classical black hole rather than attributing geometric properties to mixed states such as the thermal state.\footnote{Note that as a consequence, for a fixed energy state $\ket{M}$ there are no quantum fluctuations of these geometrical variables.}

\section{Metric in Topologically Complete JT Gravity}
\label{sect:topo}
\begin{figure}[h]
\begin{center}
    \includegraphics[width=0.85\textwidth]{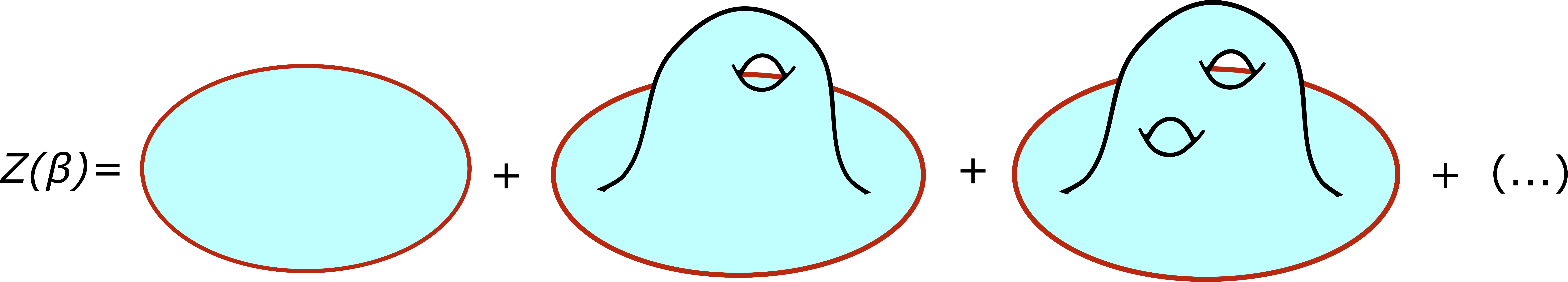}
\end{center}
\caption{A graphical representation of formula \eqref{seriespartitionm}.}
\label{fig:genusexpansion}
\end{figure}

Thus far, we have restricted the JT gravity path integral to Euclidean disk configurations. Let us now lift this restriction. This results in a genus expansion for all observables \cite{sss2}. For example, the partition function has an expansion: 
\begin{equation}
    Z_L(\beta)=\sum_g Z_g(\beta)\, L^{-2g} + (\text{non-perturbative}).\label{seriespartitionm}
\end{equation}
This is shown graphically in Figure \ref{fig:genusexpansion}.
Recently, a possible nonperturbative completion of this genus expansion was proposed to be a specific random matrix theory \cite{sss2,talks}.\footnote{Certain features of this model pop up in a double-scaled limit of SYK where one takes both the number of Majorana fermions $N$ and the effective coupling $\beta\mj$ large, with their ratio held fixed:
\begin{equation}
    \frac{C}{\beta} = \frac{N}{\beta\mj}.
\end{equation}}
This is completely defined by giving its genus zero or $L\to\infty$ density of states:\footnote{In the SYK context we would have $\ln L=N$ up to a prefactor.}
\begin{equation}
\label{linfdos}
\rho_\infty (E) = L \theta(E)\sinh 2\pi \sqrt{E}.    
\end{equation}
We should think of $L$ as some huge parameter $L\gg \beta/C$. We would now like to understand the near horizon geometry in this arguably more realistic theory of quantum gravity. We will find that the conclusions on the metric from the previous paragraph hold up to $z/C\sim L$, at which point a new transition takes place.
\\~\\
In Appendix \ref{app:othermanifolds} we point out an interpretation of higher genus Euclidean contributions to a Lorentzian path integrals. We then derive a formula for the metric expectation value obtained by a boundary observer within this theory. Including the nonperturbative completion we find \eqref{metricRMT}:
\begin{equation}
\label{metriclongt}
    \boxed{\average{ds^2}_\beta = \frac{1}{L}(dz^2-dt^2)Z_L(\beta-2iz)Z_\infty (2iz) + (cc).}
\end{equation}
This is the analogue of formula \eqref{metrictwopoint} for finite $L$. Deep in the bulk $z\gg C$ the bulk metric expectation value thus reduces to a product of late-time partition functions. Notice that in the limit $L\to\infty$ this reproduces \eqref{quantummetric}.
\\~\\
To obtain the very-near-horizon metric we now only require the late-time finite $L$ partition function $Z_L(it)$ of the random matrix model seeded by \eqref{linfdos}. Let us discuss the essential properties. More details are presented and collected in Appendix \ref{app:latetimermt}. 
\\
The finite $L$ partition function is the Fourier transform of the finite $L$ spectral density $\rho_L(E)$:
\begin{equation}
\label{Eupf}
    Z_L(\tau)=\int_\mathcal{C} d E\,\exp(-\tau E)\, \rho_L(E).
\end{equation}
At late times, the saddle in the partition function moves further and further down the energy axis. Structurally the finite $L$ spectral density differs from the Schwarzian one in three ways that are generic in random matrix theory:
\begin{itemize}
    \item The sharp edge $\theta(E)\sqrt{E}$ of the Schwarzian spectrum is replaced by a smooth version. This is shown in Figure \ref{fig:airy}. Roughly speaking, this means there should be some time scale $t$ in the Fourier transform at which we start seeing deviations from the power law behavior, which assumes a sharp edge. We find the relevant time scale to be $t\sim L^{2/3}$ in Appendix \ref{app:latetimermt}. The partition function picks up a phase factor:
    \begin{equation}
    Z_L(it)=t^{-3/2}\exp(i\frac{t^3}{L^2}\frac{1}{12}).\label{airypart}
    \end{equation}
    \item There are periodic wiggles on the curve. As discussed in Appendix \ref{app:latetimermt}, the frequency of these wiggles blow up with increasing energy.\footnote{See also \cite{talks}.}
    \item There is a transition in $\rho_L(E)$ from the smooth edge to a nonperturbative tail that extends all the way along the energy contour $\mathcal{C}$. Its leading behavior is \eqref{nonperturbative}:
    \begin{equation}
    \ln \rho_L(E) = -2L V(E),\label{partgrav}
    \end{equation}
    with the effective potential given in \eqref{potential} and where $E$ can be complex valued.
    The late time partition function can then be obtained using the saddle point method. For $t\ll L$ this results in \eqref{airypart}. For $t\gg L$ we find:
    \begin{equation}
     Z_L(it) = \exp(i\frac{t}{4}),
    \end{equation}
    up to some numerical prefactor.
\end{itemize} 
Plugging these formulas into \eqref{metriclongt} and re-introducing the correct $C$ dependence, we obtain the metric expectation value. 
In particular in the very-near-horizon regime $z/C \gg L$ we find:\footnote{Notice that \eqref{final} is by no means positive definite. This is no surprise: the metric operator defined in section \ref{sect:geometry} is not necessarily positive definite. We view this as evidence that any notion of semiclassical gravity breaks down at $z/C \sim L$.}
\begin{equation}
		\average{ds^2}_\beta =\frac{1} {z^{3/2}}\cos(\frac{1}{2}\frac{z}{C})(dz^2-dt^2).\label{final}
\end{equation}
We note that the metric goes to zero at some point $z\sim C L^{2/3}$, followed by a region where space and time are exchanged.
\\~\\
Rather than focusing on the specific - not particularly enlightening - result, we feel there is a deeper lesson. In the very-near-horizon region, the metric expectation value becomes sensitive to the finite $L$ fine structure of the spectrum. This is the same time scale at which random matrix effects kick in, in the holographic boundary correlators. We touch on this for the two-point function in appendix \eqref{s:bdytwopt}.\footnote{In particular we show that its late-time behavior is identical to that of the spectral form factor discussed in \cite{Cotler:2016fpe}.} Clearly, for $z/C \sim L$ it is safe to say that a spectacular breakdown happens of any classical notion of gravity, and we become sensitive to the discrete nature of the underlying microscopic theory.

\section{Information Paradox}
\label{s:information}
In semi-classical gravity information is lost. This is in contradiction with a holographic theory of quantum gravity, which is unitary. The result is the information paradox \cite{Hawking:1976ra}, which was reincarnated as the firewall paradox \cite{Almheiri:2012rt}. 
\\
A sharp way to state a paradox is as a set of hypotheses that are all assumed to be true, but which are in logical contradiction. The paradox is resolved if one can \emph{prove} one of the hypothesis wrong \cite{compere,info1,info2,info3,info4}.
\\
For the information paradox these hypothesis are basically unitary quantum mechanics, and the assumptions that go into the Hawking calculation \cite{Hawking:1976ra}. In particular, one of these assumptions is that quantum gravity effects are suppressed at microscopic distances from the horizon \cite{compere}.\footnote{One imagines this to be true because classically, spacetime is weakly curved at the horizon.} This is necessary to motivate the use of quantum field theory (for the matter sector) in curved space and in particular in the Rindler geometry, which directly implies the Hawking-Unruh effect.
\\~\\
This hypothesis is not valid in JT quantum gravity. Let us first focus on the JT disk model. We have proven through sections \ref{sect:mcjt} and \ref{sect:geometry} that quantum gravity effects are by no means suppressed in the near-horizon region. On the contrary, they proliferate: near-horizon matter correlation functions are dominated by quantum gravity effects. Therefore, quantum matter on a classical curved spacetime is not a valid approximation of near-horizon quantum JT gravity, invalidating the Hawking calculation in this context. This is how JT gravity evades a version of the information paradox.
\\~\\
Let us make some comments.
\begin{itemize}
    \item This conclusion, like all our results, builds ultimately on our \emph{choice} to use a radar definition of bulk coordinates to construct observables. Intuitively we expect generically that quantum gravity effects will dominate in the near-horizon region, since late time holographic correlation functions seem to demand this. It would still be interesting though, to see how these conclusions hold up for a different construction of bulk frame such as the one discussed in Appendix \ref{app:geodesic}.
    \item The fact that the information paradox is avoided in this model, does not mean we understand how black hole dynamics and evaporation works in JT quantum gravity. A full quantum description of black hole evaporation should be the goal.
    \item In this model, the effective gravitational constant is $1/C$ which has dimensions of energy. Therefore, just by dimensional analysis, probing at distances $z/C\gg 1$ we end up with an effectively strongly coupled theory. One cannot therefore turn off quantum gravitational effects near the horizon.
    \item We should stress that this conclusion is not based on our calculations in section \ref{sect:geometry} and in particular it does not rely on any interpretation of \eqref{metrictwopoint} as a metric. We are building here only on the calculations of section \ref{sect:mcjt}. It is only a posteriori that the metric operator expection value is suggestive for this interpretation.
\end{itemize}
It is amusing to link these conclusions to late-time holographic dual correlation functions and to the fine-structure of the spectrum of the theory, as in section \ref{s:quantum}. \\
A telltale of information loss in semiclassical gravity is that matter correlation functions continue to decay exponentially at ever larger separations. The physics behind this is that information thrown into a classical black hole thermalizes and is eventually lost: the Hawking radiation is uncorrelated with the information thrown into the black hole. Such late time thermal decay is forbidden in any theory that includes quantum gravity \cite{maldacena2001,Dyson:2002nt,Goheer:2002vf,Barbon:2003aq,Fitzpatrick:2016ive}. Usually any microscopic candidate theory of gravity is thought to possess a discrete spectrum, which spoils exponential decay at very late times as erratic fluctuations take over, see for example \cite{Cotler:2016fpe,info2}.
\\
Late-time exponential decay is equally forbidden in a theory of quantum gravity with a continuous spectrum, given some plausible assumptions. Indeed, suppose that the theory has a lowest energy state (say at $\omega=0$), and is Laurent expandable around that as $\rho(\omega) \sim  \theta(\omega) \omega^\alpha$.\footnote{We require $\alpha > -1$ for integrability. This would seem to hold rather generically in quantum physics. It holds for example for the free Bose and Fermi gas, the bosonic Schwarzian ($\alpha = 1/2$) and the $\mathcal{N}=1,2$ super-Schwarzian ($\alpha=-1/2$).} Consider then for example the two point function, where we assume $G(\omega)\sim \omega^\gamma$ is Taylor expandable:\footnote{In the Schwarzian theories, the function $G$ is the product of the vertex functions, which are indeed Taylor-expandable.}$^,$\footnote{It is interesting to see how the semi-classical gravity theory ($t\ll C$) evades this argument. Expanding the integrands of any correlation function around the classical black hole mass: $E_i = M-\omega_i$, where $\omega_i \ll M$, one integrates the leading order result over the entire real $\omega_i$ axis. In terms of the density of states this means we effectively replace
\begin{equation}
\rho_M(\omega) = \theta(M-\omega)e^{2\pi \sqrt{M-\omega}} \to e^{-\frac{2\pi}{\sqrt{M}} \omega},
\end{equation}
which has no sharp feature and hence does not yield a power-law.}
\begin{equation}
\int d \omega \rho(\omega) \, e^{i\omega t} \, G(\omega) \sim 1/t^{\alpha+\gamma + 1}. 
\end{equation}
It is not hard to imagine that such power-law behavior will generically occur for any correlation function provided one of the time separations is larger than some dimensionful coupling $C$. It is for example observed in late-time CFT$_2$ correlation functions which are then dual to (parts of) AdS$_3$ quantum gravity \cite{Fitzpatrick:2016ive,kaplan1606,kaplan1609,kaplan1703num,anoushartmansonner,dyergurari}.
\\~\\
We are hence led to the following lesson: quantum gravity effects will generically dominate near-horizon physics, because late-time boundary correlation functions transition from semiclassical quasi-normal mode decay to power-law decay. JT gravity is special at this point because we are able to exactly compute and prove this quantitatively. This has not yet been achieved for example in AdS$_3$ quantum gravity, and it would be interesting to make this more explicit in that case.
\\~\\
An equivalent statement is that the horizon probes the fine-structure of the spectrum of the quantum gravity theory, much like late-time.
\\~\\
As a further example of this, consider the nonperturbative topological completion of JT gravity discussed in the previous section and Appendices \ref{app:othermanifolds} and \ref{app:latetimermt}. In the holographic random matrix theory, there is a second transition at times of order $L$ (or sooner) from power law-behavior to random matrix behavior. An example of this is the plateau structure observed in SYK correlation functions \cite{Cotler:2016fpe}. This transition originates from fine-grained wiggles on top of the Schwarzian density of states in this model \cite{sss2}. In the previous section, we found that similar fine-grained corrections on the Schwarzian density of states result in a transition in the very-near-horizon metric at distances of order $L$, away from power law behavior to random matrix behavior.

\section{Concluding Remarks}
\label{sect:concl}
Let us conclude with some general lessons about quantum gravity.
\begin{itemize}
    \item To start with, one needs to define a bulk frame to construct diff-invariant observables. Indeed, since the diff-symmetry is a gauge redundancy in gravity, the only physical results are expectation values of diff-invariant operators. To define such a bulk frame, generically in quantum gravity one needs a platform as a reference beacon \cite{Donnelly:2015hta}. In a holographic theory, this is naturally the holographic boundary itself. 
    \item We defined a Hermitian metric operator in quantum gravity and calculated its expectation value. In doing so, we have to be careful about operator ordering ambiguities when going from a classical expression to an operator in Hilbert space. In particular, we choose to define the metric operator as Hermitian, but not necessarily positive definite.\footnote{A positive definite operator whose classical expression is the metric \eqref{dfu} could be obtained as $\hat{\mo}_{1/2}(-z)\hat{\mo}_{1/2}(z)\hat{\mo}_{1/2}(z)\hat{\mo}_{1/2}(-z)$, that is we use two $\ell=1/2$ bilocals. As in \eqref{primary1} and implied by the operator ordering, one inserts the bilocals as a whole one after another. The calculations that follow are very analogous as the one we presented in this work. For example, up to normalization we find for $z \gg C$:
\begin{equation}
\average{ds^2}_M=\average{ds^2}_\beta=\frac{1}{z^3}(dz^2-dt^2),
\end{equation}
and similarly for the finite $L$ version.
} 
We believe that disallowing signature jumps between $(+,-)$ and $(-,+)$ (crossing horizons) is not something you should impose on a theory of quantum gravity by hand.\footnote{Doing the opposite would seem to remove the possibility to create and destroy black hole horizons dynamically.} The local signature should rather follow from an exact calculation.
\\
We learned that in the very-near-horizon region, there are strong quantum fluctuations of the metric, and the expectation value of the metric operator deviates strongly from the semiclassical answer. One would expect generically that such phenomena occur in any theory of quantum gravity.
\item Close to the classical black hole horizon, we found that quantum gravitational effects dominate matter correlators. This represents severe backreaction. On hindsight, this is not surprising. In fact, based on a mapping between late-time holographic boundary correlation functions and near-horizon physics we expect this to be true in any sensible theory of quantum gravity. Some evidence that effective field theory breaks down has also been obtained in string theory \cite{Giveon:2012kp,Mertens:2013pza,Mertens:2013zya,Mertens:2014saa,Dodelson:2015toa}. It would be interesting to make this explicit in 3d gravity for example.
\\
An important point that is manifest in JT gravity, is that this backreaction cannot be avoided in the following sense. No matter how light one chooses a certain operator, or how tiny one makes $G_N$, there is always a region close to the horizon where quantum gravitational fluctuations will become important. In this region, the Hawking calculation can then not be trusted for that particular field. 
\item JT gravity on a disk (dual to the Schwarzian) has a remarkable degree of locality present all the way to the Planck scale. This is embodied by the perfectly local analytic structure of bulk matter correlation functions. It would seem that no such conclusion can hold in more than three dimensions, where gravity is non-renormalizable and we would need e.g. strings which are inherently nonlocal.
\\
The full SYK model certainly has some degree of bulk non-locality \cite{Maldacena:2016hyu}. This suggest that bulk locality in JT gravity may be lost when we lift its constraints and consider instead the topologically complete model dual to a random matrix theory. We leave this to future work.
\end{itemize}
The big questions regarding black holes are thus left unanswered. Can we obtain a solvable quantum mechanical model of black hole evaporation? What happens to an infalling observer? It seems that JT gravity presents our best chance of answering these questions at the moment. 
\\~\\
Even though our computations are explicitly done in 1+1d, there are embeddings of the JT model within higher-dimensional gravity. For instance, the authors of \cite{Nayak:2018qej,Sachdev:2019bjn,Moitra:2018jqs,Moitra:2019bub} argue that certain charged or rotating black holes in higher dimensions are approximated in the near-horizon, near-extremal regime by JT dynamics. In this configuration, the curve separating the near-horizon region from the far region can analogously be used to set up our bulk frame \eqref{bulkmetric}. Whereas there are dynamical gravitons in higher dimensions, requiring an embedding within e.g. string theory, the s-wave sector is still described by the renormalizable 2d JT theory, and we hence expect the lesson on the breakdown of the Rindler geometry and the information paradox to carry through also in these higher-dimensional situations.
\\~\\
Let us end with a comment on the equivalence principle in quantum gravity, and some speculation on the infalling observer.
\\
At several points in this work, it was illustrated that in JT quantum gravity, since all quantities have to be defined operationally, physics in the frame of the boundary observer can be very different from physics in other frames, even for coordinate-invariant quantities such as the infinitesimal distance function \eqref{quantummetricthermal}, or the bulk two-point function \eqref{bulkbulk}. This is not in contradiction with the coordinate invariance from general relativity. For each off-shell metric, there is a relation $(u,v) \leftrightarrow (f(u),f(v))$ between different frames. This relation however, does not exist after the path integral over bulk metrics $f$. As a consequence, one should not a priori expect general coordinate invariance of observables to hold in full quantum gravity, i.e. \emph{after} performing the gravitational path integral.
\\
Our definition of bulk frame \eqref{bulkmetric} seems to be natural for stationary observers such as the boundary observer.\footnote{Note that one still has access to $f$-independent transformations of the type $u\to u(u',v')$, $v\to v(u',v')$ which do preserve the physics, but they also reparametrize the boundary time $t$.} Another frame, for which we have no classical bulk intuition at the moment, is defined in Appendix \ref{app:geodesic}. On the other hand, it is not obvious what is a natural definition of frame for an infalling observer. In particular, it is not even obvious whether or not this should be anchored to the asymptotic boundary. It could be that a natural family of infalling frames is for example the \Poincare family $(T,Z)$ and not the boundary family of frames $(t,z)$. In this scenario there would be no drama in an infalling frame. Because there is no generalized coordinate invariance in quantum gravity, this is not a paradox \cite{Almheiri:2012rt} when combined with our present discussion on stationary observers which do find drama near the horizon. The description of infalling frames in quantum gravity certainly deserves further study.

\section*{Acknowledgements}
We thank J. Cotler, K. Jensen, S. Shenker. D. Stanford, G.J. Turiaci and Z-Y. Xian for discussions. AB and TM gratefully acknowledge financial support from Research Foundation Flanders (FWO Vlaanderen).

\appendix

\section{Geodesic Localizing}
\label{app:geodesic}
\begin{figure}[h]
\centering
\includegraphics[width=0.15\textwidth]{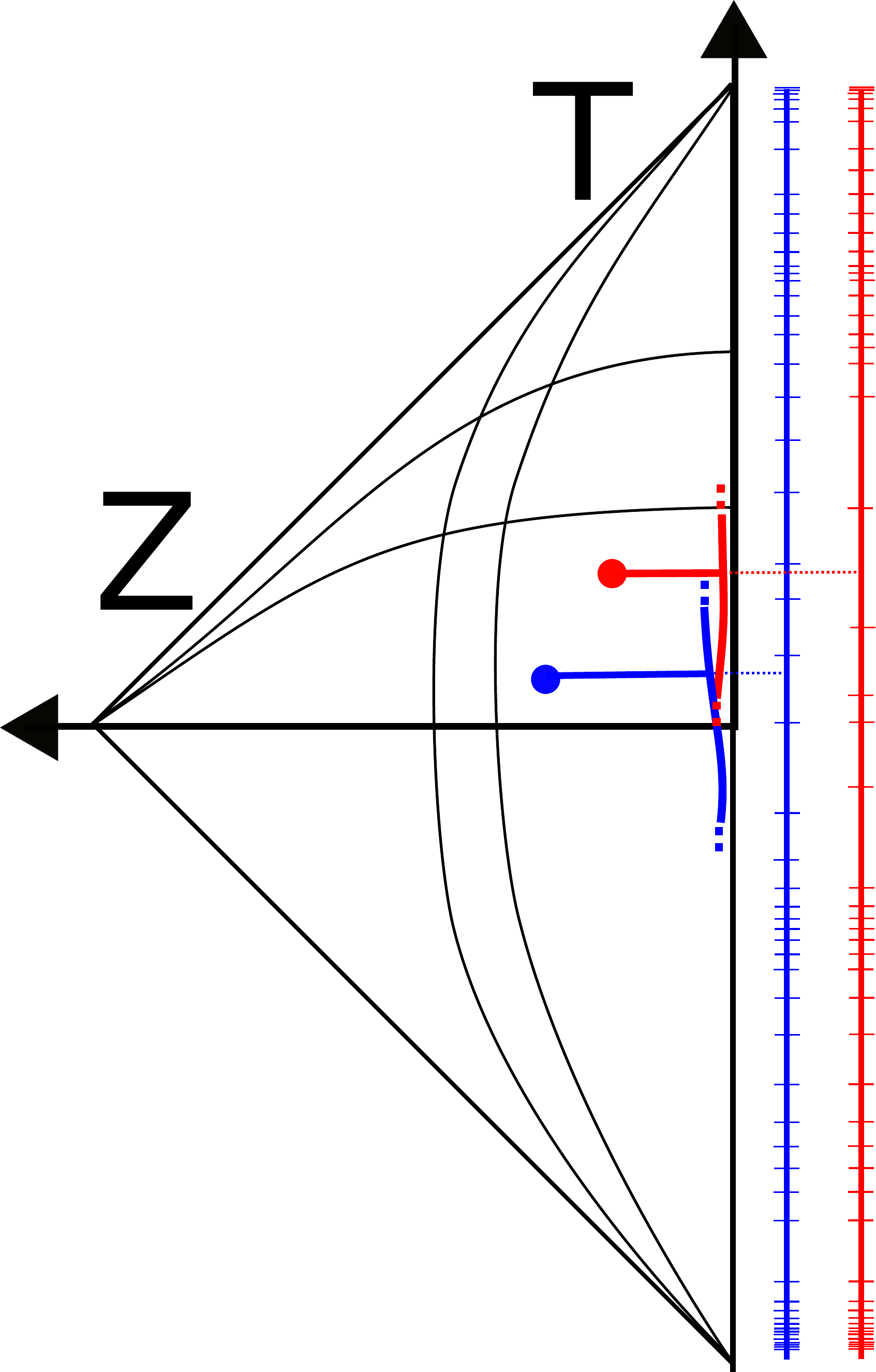}
\caption{Set-up of bulk frame using proper time $t$ and the geodesic length $L$. Both points are the ``the same point'', constructed with the same values of $t$ and $L$, but using a different (off-shell) reparametrization map $f$.}
\label{bulkframeGEO}
\end{figure}

As an alternative definition of a bulk point, we can consider using the geodesic distance measured from the holographic boundary as the radial coordinate, combined with the dynamical time variable $f(t)$. The geodesic length is given by\footnote{One readily checks explicitly that the $T= \text{const}$ lines are spacelike geodesics.} 
\begin{equation}
dL^2 = \frac{dZ^2}{Z^2} \quad \Rightarrow \quad L = \log \frac{Z}{\epsilon f'}.
\end{equation}
Renormalizing the length by dismissing the $\log \epsilon$ term, we perform the radial coordinate transformation $Z = f' e^{L}$, and define a bulk operator as $\mathcal{O}(T,Z) \equiv \mathcal{O}(f(t), f'(t) e^L)$ in terms of the boundary proper time $t$ and the (renormalized) geodesic length $L$ (Figure \ref{bulkframeGEO}).\footnote{This is a version of Wilson line dressing. Note that, due to a lack of angular coordinates, the more symmetric Coulomb dressing is the same in 1+1d. }
Using this coordinate, the constructed metric becomes
\begin{equation}
\label{metrgeo}
ds^2 = dL^2 + 2 \frac{f''}{f'}dt dL + \left( \frac{f''^2}{f'^2}-e^{-2L}\right) dt^2.
\end{equation}
As a check, close to the boundary $e^L = \epsilon$, the metric indeed respects the Poincar\'e asymptotics. The bulk bilocal operator, introduced in section \ref{sect:mcjt}, is written in this case as
\begin{equation}
G(f(p),f(q)) = \ln \left|\frac{(f(t_1)-f(t_2))^2 + (f'(t_1)e^{L_1}-f'(t_2)e^{L_2})^2}{(f(t_1)-f(t_2))^2 + (f'(t_1)e^{L_1}+ f'(t_2)e^{L_2})^2}\right| , 
\end{equation}
and satisfies the correct $Z_2 \to 0$ boundary conditions as well. Next to this, the limit $e^{L_i} \to \epsilon$ indeed yields the boundary bilocal operator (the integrand of \eqref{greens2}). Notice also that the geometry \eqref{metrgeo} is not in conformal gauge, and is not even time-independent for the thermal solution $f(t)=\tan \frac{\pi}{\beta}t$. \\
Whereas this is a suitable definition of fixing the bulk diff-invariance and defining bulk operators, it is not evident how to compute with this definition. Moreover, it is not immediately clear what the semi-classical bulk content is of this particular choice (e.g. our lightcone choice directly leads to the semi-classical Unruh effect \cite{Mertens:2019bvy}). \\
It would be interesting to see though whether more progress can be made for bulk operators defined in this way, or even more generally for correlation functions in which the bulk operators are each defined in different ways.

\section{Massive Bulk Fields and Global Conformal Blocks}
\label{app:massive}
The story in the main text can in principle be extended to massive bulk fields, which we develop here. 
\\~\\
It is useful to introduce another isometric invariant in Poincar\'e coordinates $(T,Z)$ \cite{hkll1,Lowe:2008ra}:
\begin{equation}
\label{sigdelta}
\sigma(p|q) = \frac{(Z^2+Z'^2)-(T-T')^2}{2ZZ'} = 2\delta^2 + 1,
\end{equation}
in terms of $\delta$ introduced in \eqref{invariant}.\footnote{This is also related to the invariant $d$, introduced in \cite{Spradlin:1999bn}, as $\sigma = 1+ d$. The geodesic distance itself can then be written as
\begin{equation}
D(P,Q) = 2 \text{Arcsinh} \delta = \text{Arccosh} \sigma.
\end{equation}
}
After performing a reparametrization, the distance functions $\delta$ and $\sigma$ are:
\begin{align}
\label{deltabulk}
\delta^2(p|q;f) &= -\frac{(f(u)-f(u'))(f(v)-f(v'))}{(f(u)-f(v))(f(u')-f(v'))}, \\
\label{sigmabulk}
\sigma(p|q; f) &= \frac{(f(u)-f(v))(f(u')-f(v'))-2(f(u)-f(u'))(f(v)-f(v'))}{(f(u)-f(v))(f(u')-f(v'))}.
\end{align}
In terms of these, the HKLL kernel of \eqref{kernel} is generalized in the massive case to:
\begin{equation}
K(t,z|\tau)  \sim  \frac{1}{f'(\tau)^{\Delta-1}}\lim_{z'\to 0} (z' \sigma(t,z|\tau;f))^{\Delta-1} \theta(z-\left|t-\tau\right|),
\end{equation}
where $m^2=\Delta(\Delta-1)$ and the $f'$ accounting for the tensor transformation of the kernel, and where 
\begin{align}
 \lim_{z'\to 0} (z' \sigma(t,z|\tau;f)) &= -2\frac{(f(u)-f(\tau))(f(v)-f(\tau))}{(f(u)-f(v))}.
\end{align}
The diff-invariant (dressed) bulk operator can then be constructed as\footnote{We will be more explicit about the prefactors shortly.}
\begin{align}
\label{HKLLm}
\hat{\Phi}(u,v;f ) &= \int_{v}^u d\tau \left(\frac{(f(u)-f(\tau))(f(v)-f(\tau))}{(f(u)-f(v))f'(\tau)}\right)^{\Delta-1} \mathcal{O}_{\Delta}^f (\tau) \nonumber \\
&= \int_{v}^u d\tau f'(\tau) \left(\frac{(f(u)-f(\tau))(f(v)-f(\tau))}{(f(u)-f(v))}\right)^{\Delta-1} \mathcal{O}_{\Delta} (f(\tau)),
\end{align}
and is expressed fully in terms of operators in the Schwarzian + boundary NCFT$_1$ system. It satisfies $(\Box - m^2) \hat{\Phi} = 0$ by construction for $m^2 = \Delta(\Delta-1)$. Since the last way of writing the formula can be simplified by setting $t=f(\tau)$ into the standard Poincar\'e expression, this bulk operator satisfies the extrapolate dictionary in two ways:
\begin{equation}
\lim_{Z\to 0}\hat{\Phi}(u,v;f)  \sim Z^\Delta  \mathcal{O}_{\Delta} (f(\tau)) = \epsilon^\Delta \mathcal{O}_{\Delta}^f (\tau),
\end{equation}
where the first line is in the Poincar\'e fixed frame, and the second line is in the local time frame $\tau$.\footnote{This makes contact with a question of \cite{Giddings:2018umg}, namely which dressing should be used for the bulk operator to find the extrapolate dictionary.}
\\~\\
For the massive bulk two-point function, one is lead to compute
\begin{align}
\label{hyperdouble}
G(f(p),f(q)) \sim \int_{v_1}^{u_1}&d\tau \int_{v_2}^{u_2} d\tau' \left(\frac{f'(\tau)f'(\tau')}{(f(\tau)-f(\tau'))^2}\right)^{\Delta} \nonumber \\
&\times \left(\frac{(f(u_1)-f(\tau))(f(v_1)-f(\tau))}{(f(u_1)-f(v_1))f'(\tau)}\frac{(f(u_2)-f(\tau'))(f(v_2)-f(\tau'))}{(f(u_2)-f(v_2))f'(\tau')}\right)^{\Delta-1}.
\end{align}
Setting $t=f(\tau)$ and $t'=f(\tau')$, one can write this more easily as:
\begin{align}
\int_{V_1}^{U_1}dt \int_{V_2}^{U_2} dt' \left(\frac{(U_1-t)(V_1-t)}{(U_1-V_1)}\frac{(U_2-t')(V_2-t')}{(U_2-V_2)}\right)^{\Delta-1} \left(\frac{1}{(t-t')^2}\right)^{\Delta},
\end{align}
where $U_i = f(u_i)$ and $V_i=f(v_i)$. \\
These integrals are do-able, and we keep track of all prefactors in the computation that follows. We use the boundary two-point function: 
\begin{align}
\left\langle \mathcal{O}_\Delta(t)\mathcal{O}_\Delta(t')\right\rangle_{\text{bdy}} &= \frac{\Gamma(\Delta)}{\Gamma(\Delta+1/2)}\frac{(-)^{\Delta}}{2\sqrt{\pi}}\frac{1}{(t-t'-i\epsilon)^{2\Delta}},
\end{align}
and the HKLL kernel in the form:\footnote{All ingredients to perform this computation can be found in \cite{Spradlin:1999bn, hkll1}. The cosines of the second line are written as a sum of exponentials, leading to 4 exponentials out of which only a single one contributes, due to $\omega,\omega'>0$ and the Heaviside function in \eqref{heavi}.}
\begin{align}
K(T|T',Z') &= \frac{\Gamma(\Delta+1/2)}{\sqrt{\pi}\Gamma(\Delta)} \left(\frac{Z'^2-(T-T')^2}{Z'}\right)^{\Delta-1} \theta(\text{spacelike}) \\
& = \frac{1}{\pi} 2^{\Delta-1/2} \Gamma(\Delta+1/2) \sqrt{Z'}\int_0^{+\infty} d\omega \frac{1}{\omega^{\Delta-1/2}}J_{\Delta-1/2}(\omega Z') \cos \omega (T-T'),
\end{align}
Finally, using:\footnote{This equation can be found by taking integrals of the functional representation of the Heaviside-function and is used to treat the boundary two-point function contribution.}
\begin{align}
\label{heavi}
&\int_{-\infty}^{+\infty}\int_{-\infty}^{+\infty}  dt dt' e^{i(\omega-\omega')\frac{(t+t')}{2}}\frac{e^{i (\omega+\omega') \frac{(t-t')}{2}}}{(t-t'-i\epsilon)^{2\Delta}}= \delta(\omega-\omega') \frac{4\pi^2}{\Gamma(2\Delta)} (-)^{\Delta}\omega^{2\Delta-1} \Theta(\omega),
\end{align}
we can combine the ingredients, and obtain:
\begin{align}
G(f(p),f(q)) &= \frac{\sqrt{ZZ'}}{2}\int_0^{+\infty} d\omega \exp\left( i \omega (T-T')\right) J_{\Delta-1/2}(\omega Z) J_{\Delta-1/2}(\omega Z') \\
&= \frac{\Gamma(\Delta)}{2^{\Delta+1}\sqrt{\pi} \Gamma(\Delta+1/2)}\frac{1}{\sigma^\Delta} {}_2F_1\left(\frac{\Delta}{2},\frac{\Delta+1}{2}; \frac{2\Delta+1}{2};\frac{1}{\sigma^2}\right).
\end{align}
The second line uses (an extension of) a formula from \cite{GR} and leads to the form of the bulk Wightman two-point function written down in \cite{hkll1}, with $\sigma$ the reparametrized invariant given in \eqref{sigmabulk}. Note that this had to work out, as it is a consistency check on the semi-classical HKLL formulas themselves. \\
Using the Kummer-Goursat quadratic transformation and the first Pfaff transformation on the hypergeometric function, combined with the relation between $\sigma$ and $\delta$ \eqref{sigdelta}, with $\delta^2$ given by \eqref{deltabulk} as the crossratio of the four times $u,v,u',v'$, we can rewrite the answer as:
\begin{equation}
\label{hyperblock}
G(f(p),f(q)) =  \frac{\Gamma(\Delta)^2}{4\pi \Gamma(2\Delta)}\left(\frac{1}{\delta^{2}}\right)^\Delta {}_2 F_1\left(\Delta,\Delta;2\Delta; -\frac{1}{\delta^2}\right).
\end{equation}
This version of the result is in the form written by \cite{Spradlin:1999bn}.\footnote{It is in fact half their result and taking both real and imaginary part. This corresponds to the fact that \cite{Spradlin:1999bn} studied the Hadamard two-point function instead.} Written in this way, the two-point function is the global $\sltr$ conformal block for elastic scattering $h_1=h_2$ and $h_3=h_4$. This is partly explained since both are solutions to the $\sltr$ Casimir equation with eigenvalue (mass$^2$) $\Delta(\Delta-1)$. \\
Note that at $\Delta=1$, using $z \, {}_2 F_1(1,1;2;-z) = \ln ( 1+z)$, one finds $\ln (1+ \frac{1}{\delta^2})$, which is indeed the massless two-point function we computed in the main text, including the normalization factor of $1/4\pi$.
\\~\\ 
Whereas \eqref{hyperblock} seems nearly impossible to compute directly within the Schwarzian path integral, there is hope for \eqref{hyperdouble}. In particular, the integrand can be seen as the product of seven bilocals, four of which have negative conformal weight for $\Delta > 1$. For $\Delta \in \mathbb{N}$, these are negative integers, corresponding to the finite-dimensional irreps of $\sltr$, whose correlation functions were initiated in \cite{thomasjoaquin}. Otherwise, they can be found by a simple analytic continuation of the positive weight bilocals. \\ 
When interpreting \eqref{hyperdouble} as an operator, we have to take care of the operator ordering, or equivalently the choice of integration contour in the path integral. The ordering we want is fixed by the HKLL equation \eqref{HKLLm}, and in particular the extrapolate operator dictionary, where two bulk operators are interpreted in the Schwarzian + NCFT$_1$ quantum system as the product of the four operators, schematically $\int \int \hat{K}_1 \mathcal{O}_1 \hat{K}_2 \mathcal{O}_2$, in this specific order, and without crossed lines within each $\hat{K}$ operator. This means the computation we want is given by the Euclidean diagram:\footnote{There is still an ordering ambiguity for the three bilocals within the KHLL kernel $K$ itself, which we have chosen specifically in this diagram. It would be interesting to study which ordering within $K$ is the most natural for this computation, which we leave to future work.}
\begin{equation}
\begin{tikzpicture}[scale=0.8, baseline={([yshift=0cm]current bounding box.center)}]
\draw[thick] (0,0) circle (1.5);
\draw[thick,dashed] (1.4,0.539) arc (250:217:3.1);
\draw[thick,dashed] (-1.4,0.539) arc (290:323:3.1);
\draw[thick,dashed] (1.4,-0.539) arc (110:143:3.1);
\draw[thick,dashed] (-1.4,-0.539) arc (70:37:3.1);
\draw[thick] (-1.4,0.539)--(1.4,0.539);
\draw[thick] (-1.4,-0.539)--(1.4,-0.539);
\draw[thick] (1.5,0)--(-1.5,0);
\draw[fill,black] (1.5,0) circle (0.1);
\draw[fill,black] (-1.5,0) circle (0.1);
\draw[fill,black] (0,1.5) circle (0.1);
\draw (1.75,0) node {\small \color{red} $\tau$};
\draw (-1.75,0) node {\small \color{red} $\tau'$};
\draw (1.75,0.7) node {\small \color{red} $v_1$};
\draw (-1.75,0.7) node {\small \color{red} $u_1$};
\draw (1.75,-0.7) node {\small \color{red} $u_2$};
\draw (-1.75,-0.7) node {\small \color{red} $v_2$};
\draw (0,1.75) node {\small \color{red} $\tau$};
\draw (0,-1.75) node {\small \color{red} $\tau'$};
\draw[fill,black] (-1.4,-0.539) circle (0.1);
\draw[fill,black] (-1.4,0.539) circle (0.1);
\draw[fill,black] (1.4,-0.539) circle (0.1);
\draw[fill,black] (1.4,0.539) circle (0.1);
\draw[fill,black] (0,-1.5) circle (0.1);
\end{tikzpicture}
\end{equation}
where the dashed lines represent the negative weight bilocal lines. The Lorentzian expression is then found by directly setting $\tau_i \to it_i$ everywhere in the resulting expression.\footnote{The above discussion on the correct operator ordering is equivalent to analytically continuing the time-ordered Euclidean expression by going to different sheets of the multi-valued function. By choosing the correct ordering beforehand, we are already on the correct sheet and only require the simple substitution $\tau_i \to i t_i$ to obtain the correct Lorentzian answer.}
Following the diagrammatic rules of \cite{schwarzian,Mertens:2018fds,paper3}, it is hence possible to arrive at an analytic answer for the correlation function in this way, though it does not seem to be very illuminating to write down. \\
This logic is readily generalizable to higher-point functions, where the HKLL kernel is associated to the nested diagrammatic 3-point function that we drew above. Placing the bulk operators in OTO-configurations, the computation reduces to placing the boundary operators $\mathcal{O}$ out-of-time ordered and hence leads to 6j-symbols, whose appearance is hence a direct extension to how they appear for boundary correlators.
\\~\\
As mentioned above, we remark that this computation would also be relevant when computing the Schwarzian integral of a global conformal block. This is needed when considering the connected 4-point function for an interacting boundary NCFT$_1$.

\section{Geodesics in AdS$_2$}
\label{s:geodesics}
\begin{figure}[h]
\centering
\includegraphics[width=0.7\textwidth]{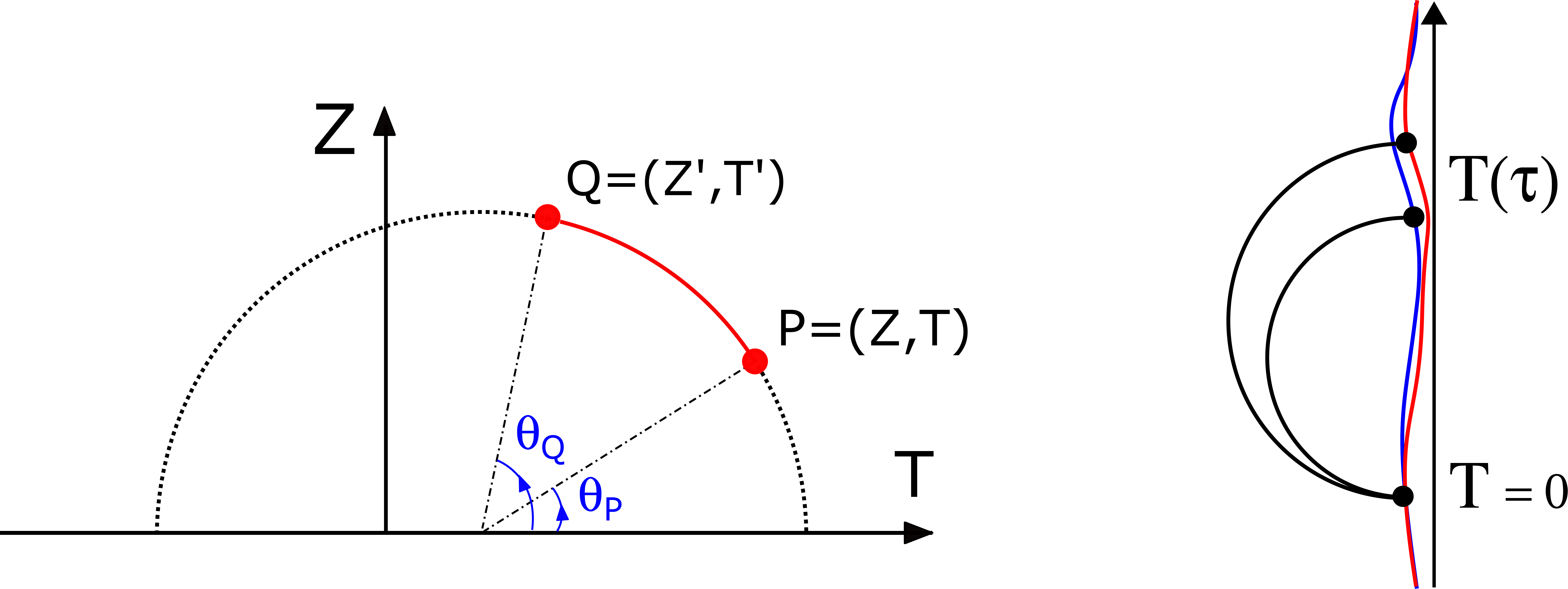}
\caption{Left: Euclidean geodesics are circle segments. Right: Geodesic with endpoints on the holographic boundary at proper time $\tau$ apart. In Poincar\'e coordinates, these endpoints are at $T= 0$ and $T(\tau)$, and the geodesic changes when changing off-shell time reparametrization $T(\tau)$ (blue and red).}
\label{geodistance}
\end{figure}

Within the AdS$_2$ \Poincare patch:
\begin{equation}
ds^2=\frac{dZ^2-dT^2}{Z^2}.
\end{equation}
the geodesic equations for $(Z(s),T(s))$, $s$ the arclength along the geodesic, are:
\begin{equation}
Z T''-2Z'T'=0,\qquad Z Z''-{Z'}^2-{T'}^2=0,
\end{equation}
Using the first of these one proves that 
\begin{equation}
\left(\frac{ZZ'}{T'}-T\right)'=0,
\end{equation}
such that the geodesic satisfies $ZZ'-TT'=CT'$ for some constant $C$. 
This can be integrated again to obtain
\begin{equation}
Z^2-(T-C)^2=\pm R^2\label{geodesic},
\end{equation}
with the $+$ sign applicable to timelike geodesics and the $-$ sign for spacelike geodesics. This is a family of hyperbolas centered around $(0,C)$. Notice that this is just the Wick-rotation $T\to iT$ of the Euclidean geodesics, which are circle segments centered around $(0,C)$ or radius $R$. Figure \ref{geodistance} (left) shows the Euclidean geodesic between two points $P=(Z,T)$ and $Q=(Z',T')$. 
In the limit $R\to 0$ we obtain the lightcone.
\\~\\
Geodesic distances can now readily be calculated: given two points $P=(Z,T)$ and $Q=(Z',T')$ one first determines $C$ and $R$ in function of the parameters $(Z,Z',T,T')$ and then simply integrates along the appropriate geodesic. We demonstrate the Euclidean calculation, it is easy to Wick-rotate everywhere. We parameterize the circle fragment by $Z=R\sin \theta$, $T-C=R\cos \theta$ such that $ds = d\theta / \sin \theta$. The geodesic distance is then:
\begin{equation}
D(P,Q)=\int_P^Q ds =\int_{\theta_P}^{\theta_Q}\frac{d\theta}{\sin\theta}=\ln \frac{1+\cos \theta_P}{\sin \theta_P}\frac{1-\cos \theta_Q}{\sin \theta_Q}.
\end{equation}
Using $\ln \left( x+ \sqrt{x^2+1}\right) = \text{Arcsinh}(x)$ and the relation between $R$ and the coordinates, we can rewrite this in terms of the isometric invariant of AdS$_2$: 
\begin{equation}
\delta (Z,Z',T)=\sqrt{\frac{(Z-Z')^2+T^2}{4ZZ'}},\label{invariant}
\end{equation}
as
\begin{equation}
D(P,Q)=2 \, \text{Arcsinh}\, \delta (P,Q).\label{distance}
\end{equation}
\\
In the special case that both points are close to the $Z=0$ boundary, one can approximate the Arcsinh by a logarithm. For a $\tau$-dependent $Z$-coordinate: $Z(\tau) = \epsilon \dot{T}(\tau)$, one can then write:
\begin{equation}
\label{logdist}
D(P,Q) \approx 2 \, \ln\, \delta (P,Q) \approx \, \ln \frac{(T_1-T_2)^2}{4\dot{T}_1\dot{T}_2} \, - \, \ln \epsilon^2
\end{equation}
The UV-piece can be removed by holographic renormalization techniques. The situation is sketched in Figure \ref{geodistance} (right). This result can be interpreted as the bulk geodesic length between two boundary events separated by observer time $\tau$, given some time reparametrization $T(\tau)$.\footnote{When inserting the operator \eqref{logdist} into the Schwarzian path integral, we have to decide on how to order the operators. Writing the bilocal schematically as $\mathcal{O}_1\mathcal{O}_2$, one can compute the log of this bilocal by computing
\begin{equation}
\partial_\ell \left.\mathcal{O}_1^\ell \mathcal{O}_2^\ell \right|_{\ell=0} = \ln \mathcal{O}_1 + \ln \mathcal{O}_2 = \mathcal{T}\ln \mathcal{O}_1\mathcal{O}_2,
\end{equation}
which corresponds to computing the time-ordered logarithm of the operators involved. This is the choice made implicitly in \cite{zhenbin} and \cite{Mertens:2019bvy} when computing complexity, respectively matter entanglement entropy.
}

\section{Observables in Topologically Complete JT Gravity}
\label{app:othermanifolds}
\begin{figure}[h]
\begin{center}
    \includegraphics[width=0.4\textwidth]{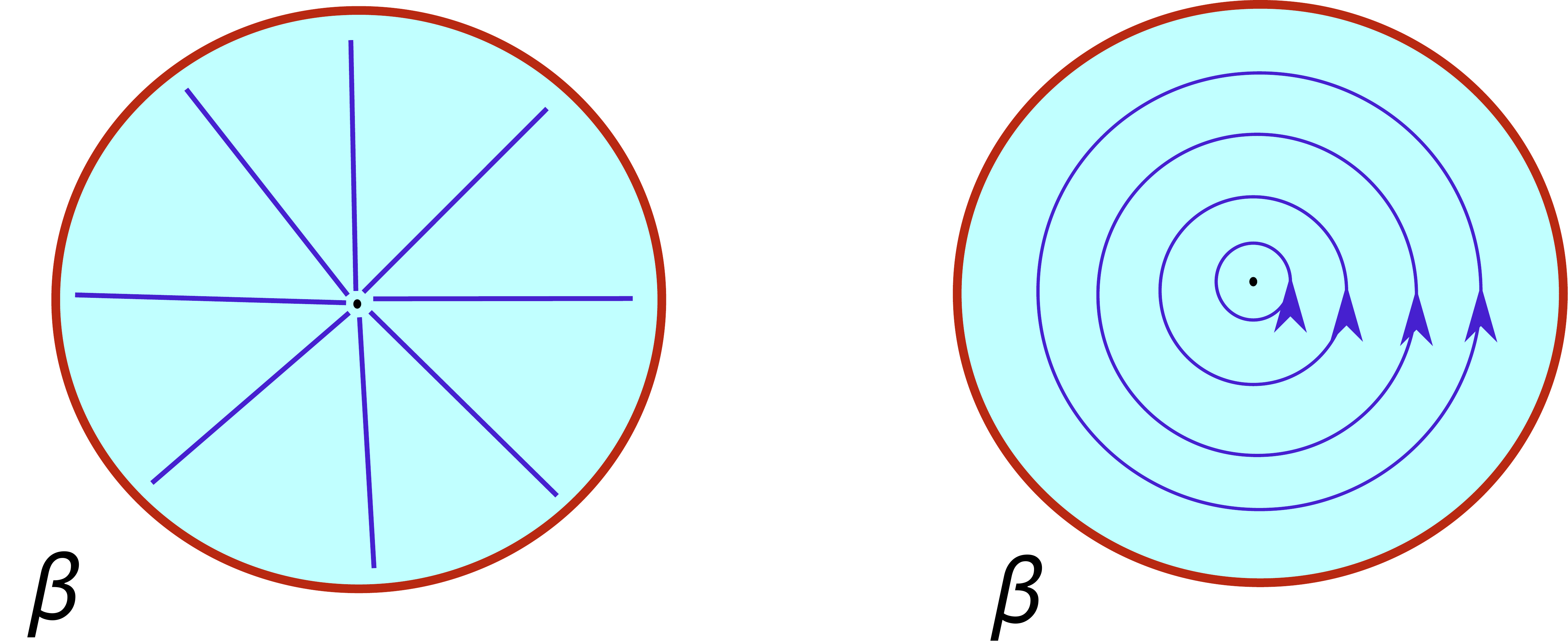}
\end{center}
\caption{Cauchy surfaces (left) and time flow (right) on a Euclidean hyperbolic disk.}
\label{fig:diskflow}
\end{figure}

\begin{figure}[h]
\begin{center}
  \includegraphics[width=0.6\textwidth]{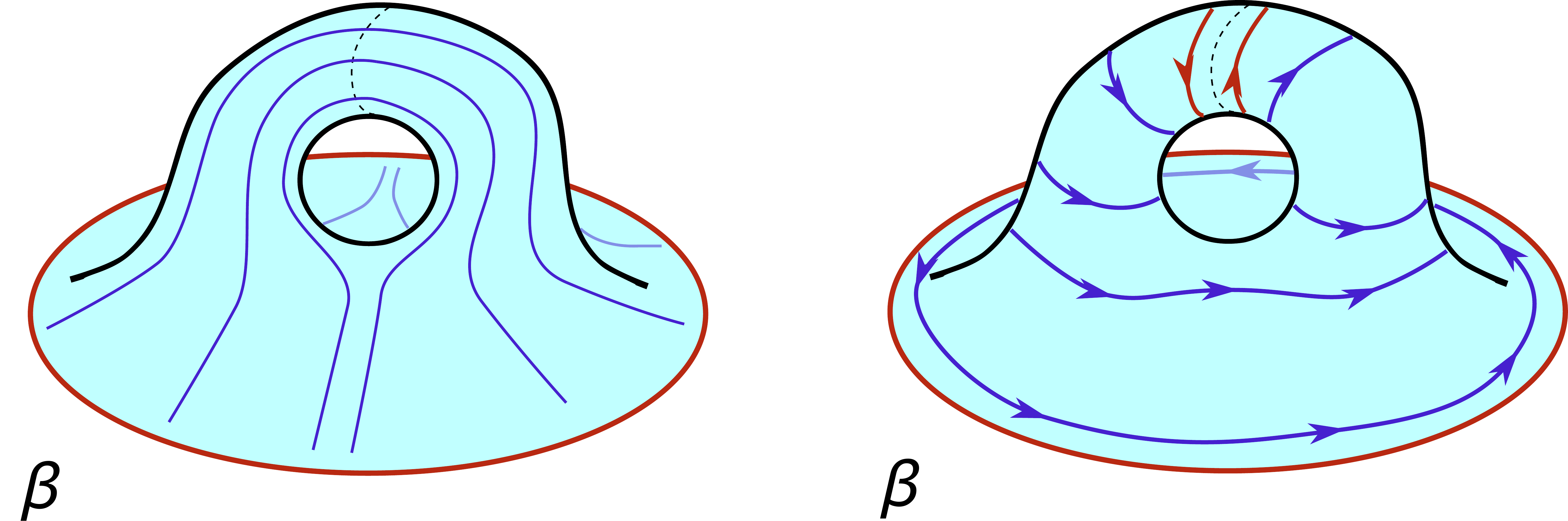}
\end{center}
\caption{An attempt to define a family of oriented Cauchy surfaces (left) or equivalently a global Killing vector field that asymptotes the boundary time flow (right) on a handlebody. This is impossible: we obtain a contradiction on the dotted line (right). }
\label{fig:handleflow}
\end{figure}

As explained in the Introduction, when defining the JT quantum gravity path integral one has to specify whether or not to allow Euclidean disks with handles ending on the asymptotic boundary. In the main text we have studied the Schwarzian theory dual to just the disk. Here we lift this restriction. The JT path integral now has a genus expansion \cite{sss2}:
\begin{equation}
    Z_L(\beta)=\sum_g Z_g(\beta)\, L^{-2g} + (\text{non-perturbative}).\label{seriespartition}
\end{equation}
A possible nonperturbative completion for this theory was found as a specific random matrix theory in \cite{sss2,talks}. The limit $L\to\infty$ of this model is the Schwarzian. Our goal here is to study the finite $L$ effects on the late-time boundary two-point function and on the near-horizon metric. This requires we first understand the perturbative expansion of these objects.
\\~\\
We first remark that it is impossible to directly identify the Euclidean handlebodies with a Lorentzian spacetime. This would require us to specify a global non-singular killing vector field flow on the handle body, with the additional constraint here that the flow should asymptote to the boundary curve time flow. An equivalent task is to specify a set of oriented Cauchy surfaces on the geometry. For the disk, this is shown in Figure \ref{fig:diskflow}, and is similar to the well-known Rindler time flow in Euclidean signature.
Performing this procedure for the disk with the single handle we end up with the set of Cauchy surfaces shown in Figure \ref{fig:handleflow} (left). The associated time flow is shown in Figure \ref{fig:handleflow} (right).\footnote{An important, yet subtle point is that we do not get to choose the topological class of the Killing vector flow on the hyperbolic Riemann surface. This is completely fixed by specifying a consecutive set of geodesic identifications on the \Poincare disk that result in the desired handlebody. This results in the flow of Figure \ref{fig:handleflow}.}
This timeflow has a topological obstruction. Concequently, it does not allow for Wick rotation (see also \cite{sss}).
\\~\\
How should we then interpret the contributions of Euclidean handlebody configurations to Lorentzian observables in JT gravity? We should treat think of them much like instanton corrections in QFT. Indeed, consider the genus $g$ contribution:
\begin{equation}
    Z_g(\beta) = \int \mu d \mu\, Z(\mu,\beta)\, V_{g,1}(\mu),\label{glue}
\end{equation}
This decomposes into a topological piece $V_{g,1}(\mu)$ and a punctured disk amplitude $Z_1(\mu,\beta)$ with an $\slr$ irrep $\mu$ puncture. The former computes the volume of the moduli space of genus $g$ Riemann surfaces ending on a geodesic of length $\mu$ \cite{dijkgraafwitten,Mirzakhani:2006fta,Mirzakhani:2006eta}. 
\\
The latter is a twisted Schwarzian:
\begin{equation}
    Z(\beta,\mu)=\int d k \, \cos 2\pi \mu k \,e^{-\beta k^2}=\int_{\beta,\mu} [\dpi f]e^{-S[f]},\label{partmu}
\end{equation}
where the variables $\mu,\beta$ constrain the integration space to a single Virasoro hyperbolic coadjoint orbit \cite{thomasjoaquin}. The path integral is written in terms of a time reparametrization $f(t)$ of the observer's proper time $t$ to the \Poincare time $f$.
\begin{figure}[h]
\begin{center}
    \includegraphics[width=0.4\textwidth]{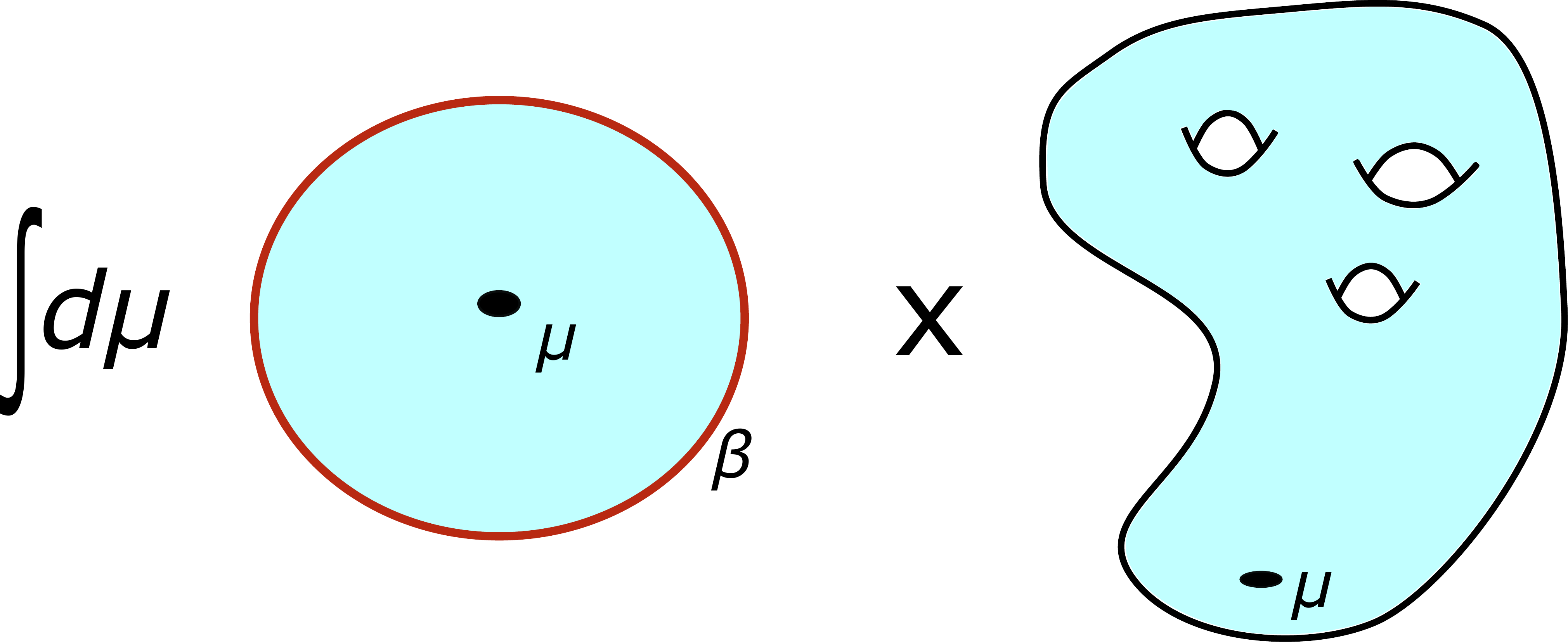}
\end{center}
\caption{A graphical representation of formula \eqref{glue}.}
\label{fig:glue}
\end{figure}
The difference being this is a different class of maps as the one considered in the main text. This path integral has hence a clear Lorentzian interpretation as summing over reparameterizations of hyperbolic orbit geometries. For example, the saddle solution of \eqref{partmu} $f(t)=\tan \frac{\pi\mu}{\beta} t$, results in the spacetime geometry:
\begin{equation}
    d s^2(\mu,\beta)=\frac{4\pi^2\mu^2}{\beta^2}\frac{dz^2-d t^2}{\sin^2\frac{2\pi\mu}{\beta}z}.\label{puncturemetric}
\end{equation} 
In particular, the argument to construct a bulk frame in section \ref{s:building} holds regardless of the specific orbit.
\\
In conclusion, the Lorentzian interpretation of the Euclidean handlebodies is an ensemble of hyperbolic orbit geometries with some weight $P(\mu)$. For example, the perturbative partition function can be rewritten as:
\begin{equation}
\label{pfgen}
		    Z_L(\beta) = Z(\beta)+ \int \mu d\mu\,P(\mu)\, Z(\mu,\beta),
\end{equation}
with: 
\begin{equation}
		    P(\mu)=\sum_{g\geq 1} V_{g,1}(\mu) \, L^{-2g}.\label{pmu}
\end{equation}
This means we should think about measurements of geometric observables by the JT gravity boundary observer as returning a weighted average of geometric observables in all the off-shell metrics $f$. The metric expectation value is then calculated perturbatively as:
\begin{equation}
 \average{ds^2}=\average{ds^2}_\beta + \int \mu d\mu\, P(\mu )\, \average{ds^2}_{\beta,\mu}.\label{metricpertutbativeL}
\end{equation}
We will discuss the near-horizon behavior and nonperturbative completion of this quantity further on.

\subsection{Boundary Two-Point Function }
\label{s:bdytwopt}
\begin{figure}[h]
\begin{center}
    \includegraphics[width=\textwidth]{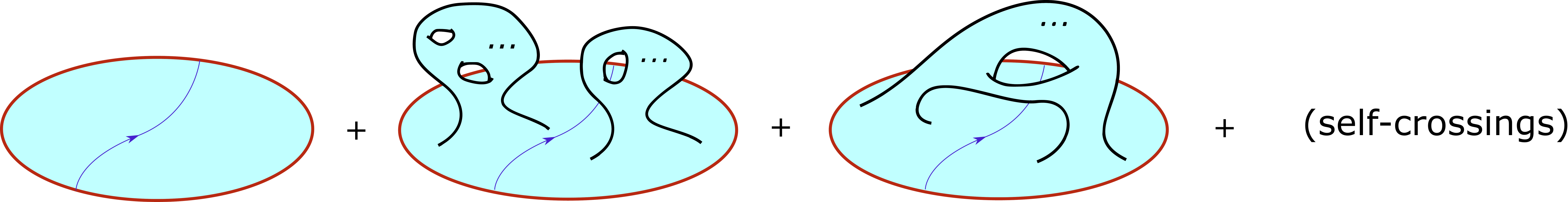}
\end{center}
\caption{Topological corrections to the boundary two-point function. We sum over topological excursions on both sides of the Wilson line, with both connected and disconnected contributions. Also corrections with Wilson lines with self-crossings have to be considered.}
\label{twopttop}
\end{figure}

\begin{figure}[h]
\begin{center}
    \includegraphics[width=0.45\textwidth]{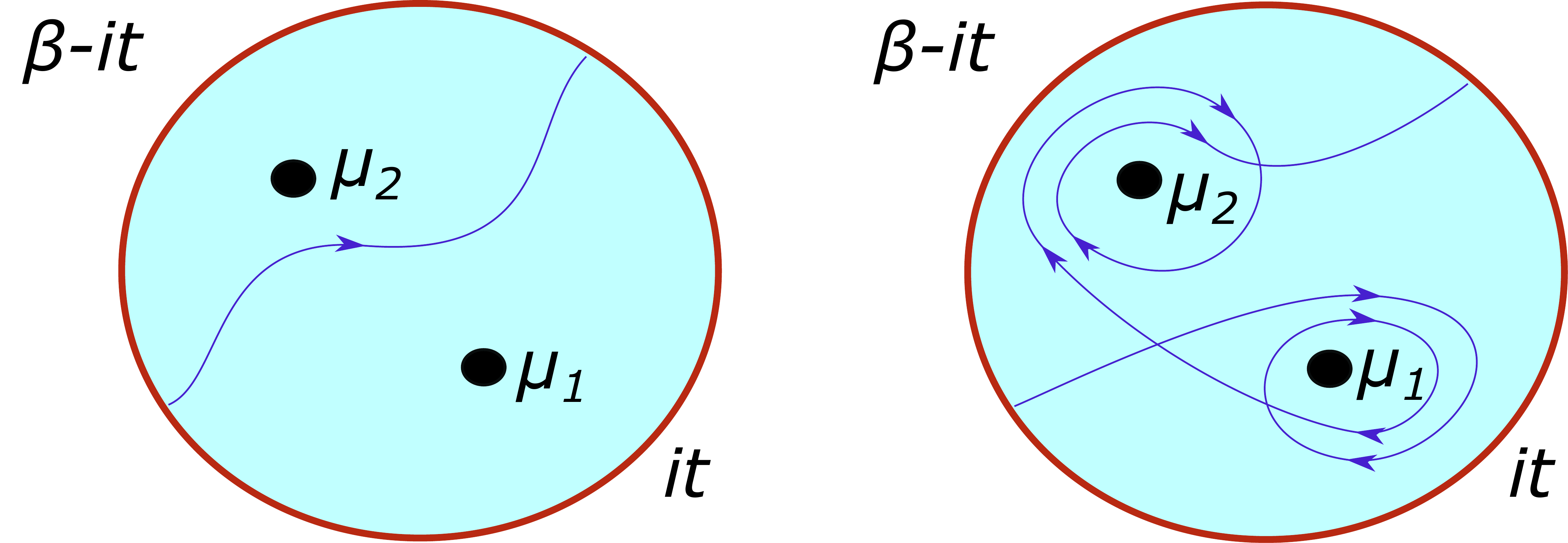}
\end{center}
\caption{Wilson line configuration of the non-self-crossing class (left) and of the self-crossing class (right).}
\label{fig:wilsonconfig}
\end{figure}

\begin{figure}[h]
\begin{center}
    \includegraphics[width=0.75\textwidth]{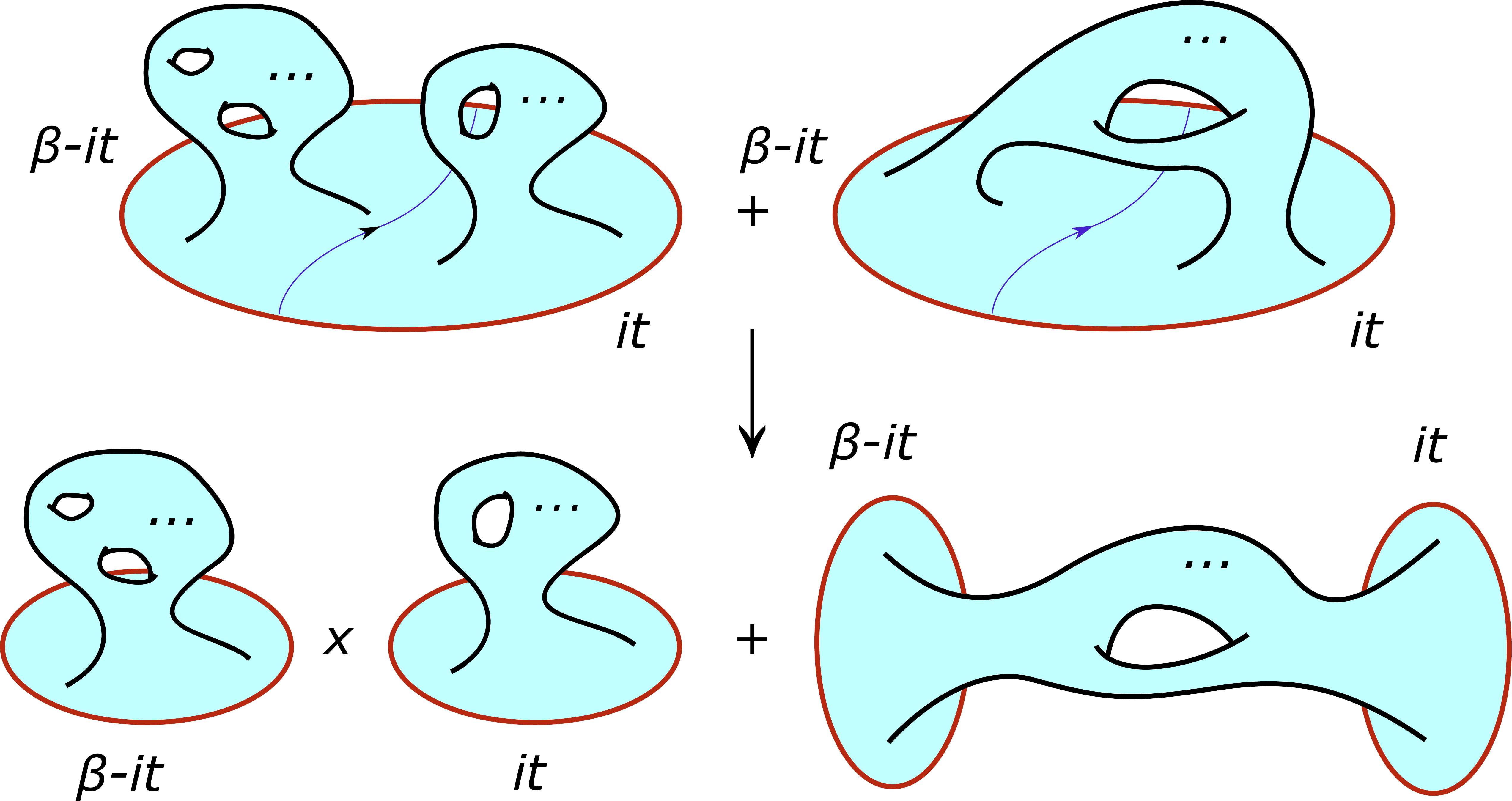}
\end{center}
\caption{Late time reduction of the boundary two point function to the spectral form factor \eqref{twopointRMT}. The Wilson line pinches off distinct regions.}
\label{fig:spectralformfactor}
\end{figure}

First, let us consider the boundary-to-boundary two point function in the topologically completed JT gravity. This is represented by a Wilson line traversing the Riemann surface. Summing over topologies leads to the situation of Figure \ref{twopttop}, where orientable handles are attached to both sides of the Wilson line, both connected and disconnected pieces.
\\
\noindent
The most generic of these topological embedding of a Wilson line can be obtained by introducing several, say $n$, punctures and winding the Wilson lines around it, and then gluing it to some orientable Riemann surface with $n$ geodesic boundaries, which may or may not be disconnected itself. One then sums over $n$ and all possible glued Riemann surfaces. This requires the Weil-Petersson volumes $V_{h,n}$.\footnote{Note that we can fuse together all punctures on the same side of the Wilson line, without loss of generality.}
\\~\\
We can distinguish two classes of configurations: those where the Wilson line crosses itself and those where it does not, see Figure \ref{fig:wilsonconfig}. 
\\\noindent
The self-crossing Wilson line configurations are subdominant at late times. Indeed, follow the Wilson line along its path to its first self-intersection. Out of the four regions that touch this intersection, at least two of them also reach the boundary. Let us denote the associated irrep labels by $k$ and $p$. The late time limit forces us to Taylor expand all functions of $k$ and $p$ to lowest order. The self-crossing comes with a $6j$-symbol \cite{schwarzian,paper3,paper4} which in this Taylor expansion goes to:
\begin{equation}
    \sj{p}{k}{\ell}{q}{r}{\ell} \,\, \to \,\, \sj{0}{0}{\ell}{q}{r}{\ell}\sim \delta_{\ell,0}=0,
\end{equation}
hence the IR contribution is subdominant for these configurations. Therefore, at each level in the perturbative expansion in $1/L^2$, the dominant contribution will come from Wilson lines with no self-crossings.
\\
The most general Wilson line configuration with no self-crossing boils down to Figure \ref{fig:wilsonconfig} (left). We can glue this either to two Riemann surfaces each with a single geodesic boundary, resulting in Weil-Petersson factors of the type $V_{h,1}(\mu_1)V_{g-h,1}(\mu_2)$, or to a single connected Riemann surface with two geodesic boundaries resulting in factors $V_{g,2}(\mu_1,\mu_2)$.
\\~\\
At late times $t \gg C$, the Wilson line pinches off the two regions, as shown in Figure \ref{fig:spectralformfactor}. The vertex functions no longer connects the two pieces of the disk in this regime and is dominated by IR states:\footnote{It is amusing to note that this is essentially identical to what happens when we insert a very heavy Wilson line \cite{pets} in the semi-classical regime:
    \begin{equation}
        \lim_{\ell \gg C}\frac{\Gamma(\ell\pm i\sqrt{E}\pm i\sqrt{M})}{\Gamma(2\ell)}\approx \frac{\Gamma(\ell)^4}{\Gamma(2\ell)}.\label{decouplinggen}
    \end{equation}
		} \footnote{One should be careful in applying this formula to the connected geometries to Taylor expand only this part, and not for example the factors $\cos 2\pi \mu \sqrt{E}$ associated to the geodesic boundaries in \eqref{partmu}. Indeed, the Riemann-Lebesgue inspired argument we are using based on factors such as $\exp(i E t)$ says that for large $t$, the integrals over $E$ are dominates by $E\sim 1/t\ll 1$. This does not necessarily imply that $\mu\sqrt{E}$ is small. In fact, we see that generically the integral over $\mu$ will be dominated at large $t$ by $\mu\sim t^2$. 
		}
    \begin{equation}
        \lim_{E,M \ll 1}\Gamma(\ell\pm i\sqrt{E}\pm i \sqrt{M}) = \Gamma(\ell)^4,\label{decoupling}
    \end{equation}
The result is that the two-point function decomposes into a product of two disk partition functions, and a two-boundary annular partition function:\footnote{The factor $1/L$ arises because the Euler character of the original Riemann surface is different from the Euler character of the pinched Riemann surface. Therefore a relative factor $1/L$ appears when going from from Figure \ref{fig:spectralformfactor} (top) to Figure \ref{fig:spectralformfactor} (bottom). We have dropped some $\ell$-dependent prefactors in writing this formula.}
\begin{equation}
\average{G_\ell(t)}=  \frac{1}{L}Z_L(\beta-it)Z_L(it) + \frac{1}{L} Z_L(\beta-it,it).\label{twopointRMT}
\end{equation}
This quantity is proportional to the spectral form factor \cite{Cotler:2016fpe,sss,sss2}. Its gravitational interpretation as Figure \ref{fig:spectralformfactor} (bottom) will be discussed in \cite{sss2,talks}. Let us just note that, for finite $L$, this perturbative series is asymptotic and we need to use the nonperturbatively exact formula from random matrix theory. The result is a ramp and a plateau. Therefore, at late times $t\gg C$, the boundary two-point function in JT gravity has a ramp and a plateau. Notice that the relative factor $1/L$ between the late two-point function and the spectral form factor matches with the prediction of \cite{sss}.\footnote{We thank Stephen Shenker for bringing this to our attention.}

\subsection{Metric}
\begin{figure}[h]
\begin{center}
  \includegraphics[width=0.45\textwidth]{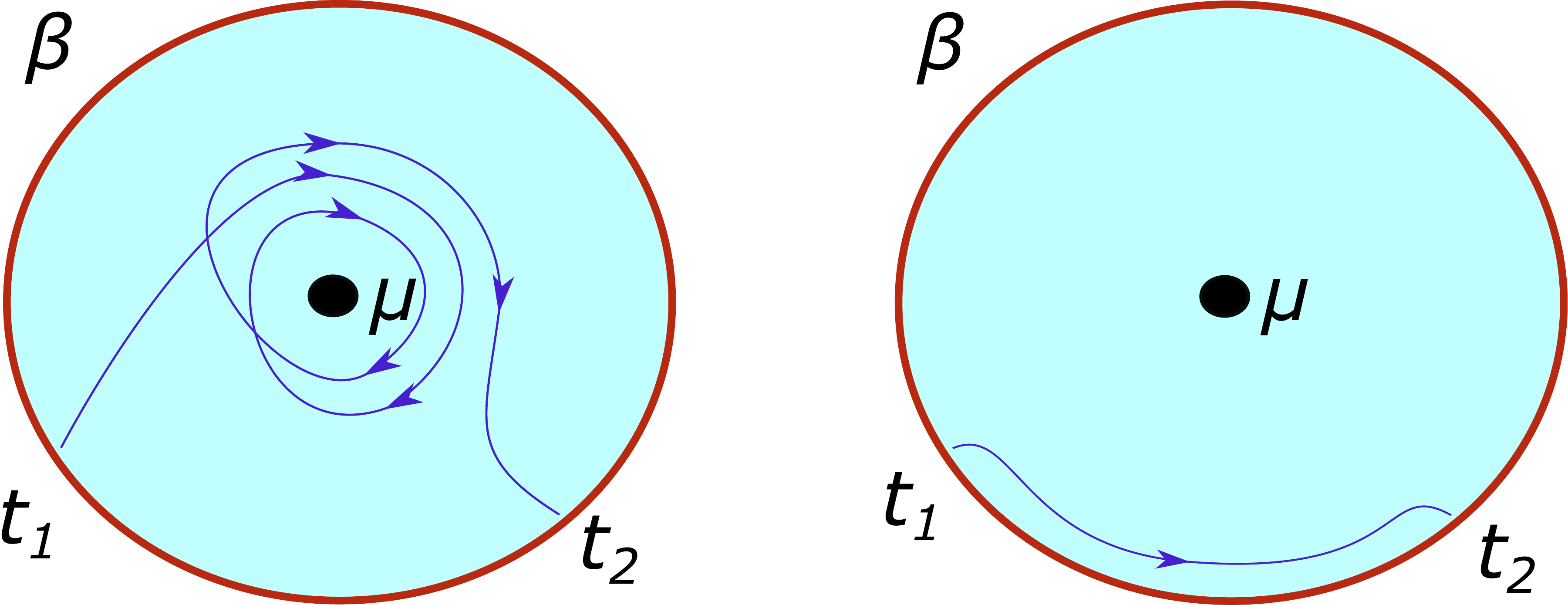}
\end{center}
\caption{Winding boundary-anchored Wilson line amplitudes for the path integral \eqref{winding}: $n=3$ example (left) and $n=0$ example (right). Only the case $n=0$ calculates the metric.}
\label{fig:wind}
\end{figure}

\begin{figure}[h]
\begin{center}
    \includegraphics[width=0.6 \textwidth]{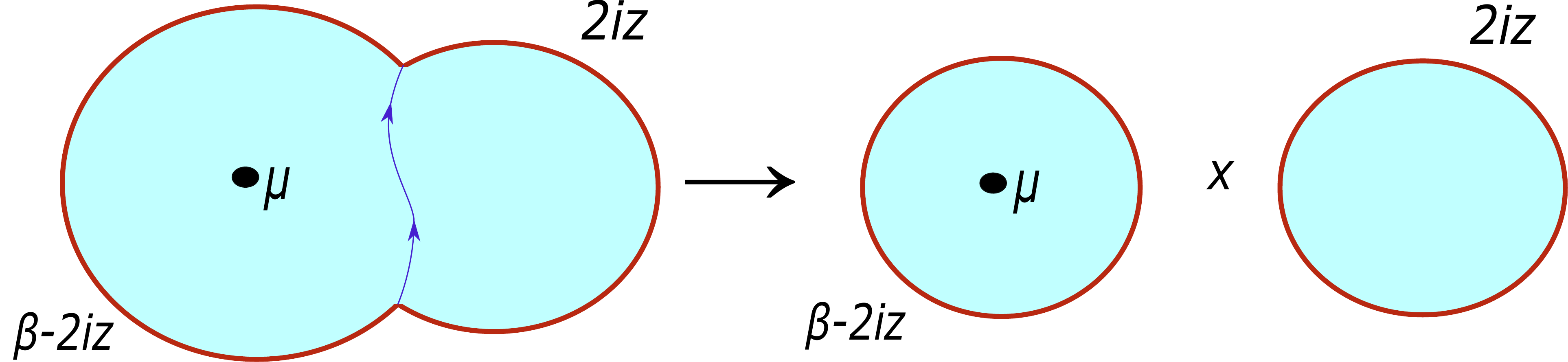}
\end{center}
\caption{Wilson line representation for the metric. For $z$ parametrically large, the Wilson line pinches off two disk-shaped regions which effectively decouple as in \eqref{metricfactor}.}
\label{fig:pinch1}
\end{figure}

\begin{figure}[h]
\begin{center}
    \includegraphics[width=0.65\textwidth]{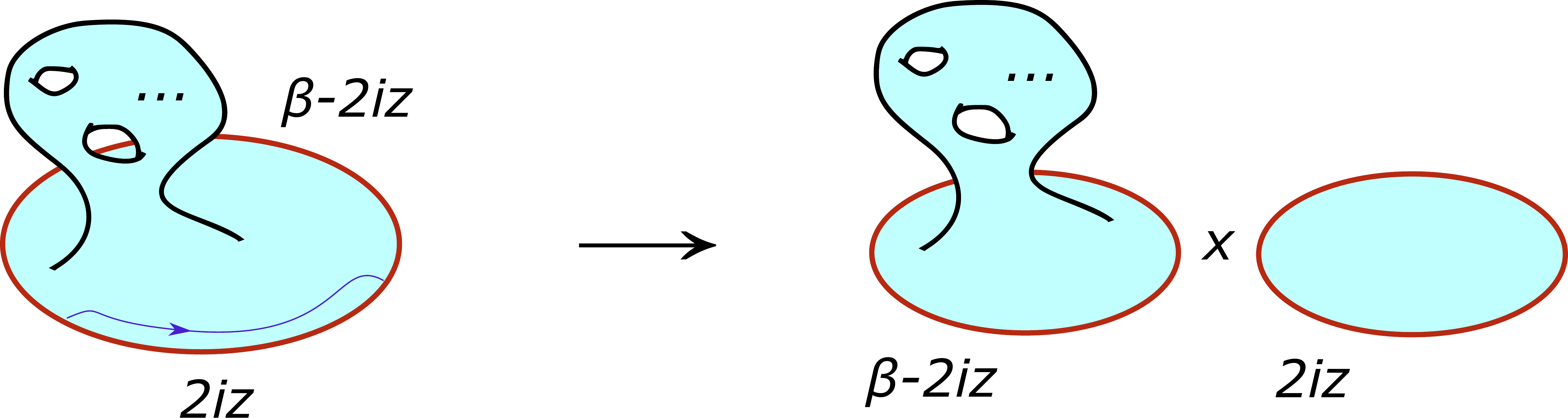}
\end{center}
\caption{A graphical representation of the near horizon metric (left). A sum over the gluing parameter $\mu$ is implicit in the handlebody. For large $z$ the Wilson line pinches off the space (right) and we obtain \eqref{metricRMT}.}
\label{fig:pinch2}
\end{figure}

Let us do a similar geometric analysis of the metric \eqref{metricpertutbativeL}. This means we should find out how the Wilson line is embedded on the Riemann surfaces in order to calculate \eqref{metricpertutbativeL}.
\\~\\
In the twisted Schwarzian theory \eqref{partmu}, via \eqref{metrictwopoint} we find that the Lorentzian bulk metric is a boundary two-point function. This is calculated as an $\ell=1$ boundary anchored Wilson line on a punctured disk.
\\
Notice that for the punctured disk, a boundary-anchored Wilson line can wind around the puncture as in Figure \ref{fig:wind}.
The expectation value of an $\ell=1$ Wilson line between boundary points $\tau_1$ and $\tau_2$ that winds the puncture $n$ times is computed by the twisted Schwarzian path integral:
\begin{equation}
    \int_{\beta,\mu} [\dpi f]\frac{\dot{f}(\tau_1)\dot{f}(\tau_2)}{(f(\tau_1)-f(\tau_2+n\beta))^2}e^{-S[f]}.\label{winding}
\end{equation}
Such winding configurations are included for correlation functions, as we discussed above, but not for the single-valued metric. The metric is only the $n=0$ Wilson line. This follows from the universal formula \eqref{bulkmetric}.
\\
One now immediately applies the rules for calculating amplitudes in JT gravity \cite{paper3,paper4} to obtain for the punctured disk:
\begin{align}
\label{metrmu}
    \average{ds^2}_{\beta,\mu}=(d z^2-d t^2)&\nonumber\int_0^\infty d M\, \frac{\cos 2\pi \mu \sqrt{M}}{\sqrt{M}}e^{-\beta M}\\&\cross \int_0^\infty d E \sinh 2\pi \sqrt{E}\,\cos 2z(E-M)\,\Gamma(1\pm i\sqrt{M}\pm\sqrt{E}).
\end{align}
In the semi-classical regime $z\ll C$, the integral over $M$ develops an imaginary saddle point at $\sqrt{M}=i\lambda(C/\beta)$ and we recover \eqref{puncturemetric}. Upon taking $z \gg C$ in \eqref{metrmu}, we enter the IR-regime where the integrals over $E$ and $M$ are dominated by their vacuum contribution. The coupling vanishes as in \eqref{decoupling} and the Wilson line pinches off the two Euclidean regions as in Figure \ref{fig:pinch1}.
We thus obtain for $z\gg C$:
\begin{equation}
    \average{ds^2}_{\beta,\mu} = \frac{1}{L}(dz^2-dt^2)\, Z\left(\mu,\beta-2iz\right)\, Z\left(2iz\right)+(cc).\label{metricfactor}
\end{equation}
Using the large $z$ behavior $Z(\mu,-2iz) \sim z^{-1/2}$ we find a conformal scaling factor $z^{-2}$ to be compared with \eqref{quantummetric}.
\\~\\
Summing over all Riemann surfaces in \eqref{metricpertutbativeL} results in the metric measured by the JT gravity boundary observer. Graphically, the perturbative formula can be drawn as in Figure \ref{fig:pinch2}: we sum over all Riemann surfaces with an $\ell=1$ boundary-anchored Wilson line included that is contractible to the boundary. For $z\gg C$ this pinches off.
We end up with the near-horizon metric:
\begin{equation}
    \average{ds^2}= \frac{1}{L}(dz^2-d t^2)Z_{L}(\beta-2iz)\,Z_{\infty}(2iz)+(cc).\label{metricRMT}
\end{equation}
This is formula \eqref{metriclongt} in the main text. For the finite $L$ partition function we require again the nonperturbatively exact formula from the matrix integral. This is discussed in Appendix \ref{app:latetimermt}.
\section{Late Time Partition Function in Random Matrix Theory}\label{app:latetimermt}
In this appendix we investigate the late time behavior of the Jackiw-Teitelboim gravity partition function. We will build on results of \cite{sss2}.\footnote{Their paper was yet to appear at the release date of this work, but its content had already partly been made public during a series of talks. These can be found online \cite{talks}.}
\\~\\
From \cite{sss2}, we know that a possible nonperturbative completion of the JT gravity genus expansion is a random matrix theory with $L\to\infty$ spectral density:
\begin{equation}
\label{gravdos}
    \rho_\infty(E)=\theta(E)\sinh 2\pi \sqrt{E}.
\end{equation}
Most of the random matrix theory facts used in this appendix are explained either in the didactic book \cite{mehta} or in the review article \cite{1510.04430}. The partition function is the Fourier transform of the finite $L$ spectral form factor:
\begin{equation}
    Z_L(\tau)=\int_\mathcal{C}  d E\, \exp( -\tau E)\, \rho_L(E), \label{partrmt}
\end{equation}
where we have transitioned to Euclidean times $\tau=it$. We are interested in its behavior for large $\tau$. For larger and larger values of $\tau$, the integral is dominated by smaller and smaller values of $E$.\footnote{We assume the integral converges. This holds true for JT gravity given an appropriate choice of $E$ integration contour $\mathcal{C}$ \cite{sss2}. See in particular Figure 4 in \cite{sss2}.}
\\
For finite $L$, the support of the spectral density is the full real axis. The partition function is then dominated by large negative values of $E$. 
\\~\\
Generically, the spectral density can be written as a trans-series of perturbative and non-perturbative contributions:
\begin{equation}
		\label{spectralde}
\rho_L(E) = \sum_g \rho_g(E)L^{-2g} + \sum_n e^{-L S_n(E)} \sum_m \rho_m^{(n)}(E) L^{-m}.
\end{equation}
The genus zero density of states is found by taking the $L\to \infty$ limit: $\rho_0(E)=\rho_\infty(E)$. Each of the fixed higher genus $g$ contributions can be calculated from equation \eqref{partmu} as:
\begin{equation}
    \rho_g(E)=\theta(E)\int d \mu \frac{\cos 2\pi \mu\sqrt{E}}{\sqrt{E}}\, V_{g,1}(\mu),
\end{equation}
where we recognize the Weil-Petersson volume of genus $g$ Riemann surfaces with a single geodesic boundary. We now see that all perturbative corrections to the spectral density have support only on the positive real axis. This means the only contributions to the spectral density at large negative energies - which dominate the late time partition function - is from nonperturbative contributions to the spectral density.
\\~\\
Nonperturbative corrections to the spectral density were discussed in \cite{1510.04430}.\footnote{In particular see amongst others formulas (1.4.8) and (3.3.9).} For a matrix model for which the support of the spectral density is a single interval (such as gravity), there is only value of $n$ in \eqref{spectralde}. Consequently, the exponential factor of the nonperturbative correction to the spectral density (and hence to the negative energy spectral density) is:
\begin{equation}
\ln \rho_L(E)\approx S_1(E) =- 2 L V(E),\label{nonperturbative}
\end{equation}
where additive subdominant contributions from the prefactor of \eqref{spectralde} are left out. The latter quantity is the effective potential and is specific to each matrix model. It is positive definite and monotonically rising outside the support $\mathcal{S}$ of the perturbative spectral density,\footnote{This ensures $\rho_L(E)\exp -\tau E$ is still a distribution ergo integrals over $E$ of test functions weighed by $\rho_L(E)\exp -\tau E$ converge.} and purely imaginary in $\mathcal{S}$.  For systems with positive real axis support for the $L\to\infty$ spectral density - such as gravity and the Airy model - it is just:
\begin{equation}
    V(E)=i \int_{0}^E d M \rho_\infty (M).\label{potential}
\end{equation}
For negative energies one should use the analytic continuation of the genus zero density of states. This formula allows us to determine the late time behavior of the partition function \eqref{partrmt}, as it is dominated by large negative energies:
\begin{equation}
    Z(\tau)=\int_\mathcal{C} d x \hdots \exp(\tau x +i2 L\int_{0}^x d y \rho_\infty (-y)),\label{beforesaddle}
\end{equation}
where we have introduced $x=-E$. For large $\tau$ and $L$ in a double-scaling regime $\tau \sim L$, we can use the saddle point method to evaluate the integral. For a saddle at $x_0(\tau)$ one then gets the leading behavior:
\begin{equation}
    \ln Z(\tau)\approx \tau x_0(\tau)-2 L V(x_0(\tau)).
\end{equation}
To convince the reader that this works as advertised, we consider per example two matrix models for which the partition function and finite $L$ spectral density are known exactly. These are the Gaussian matrix model with semicircle density of states and the Airy model with density of states $\rho_\infty(E)=\theta(E)\sqrt{E}$. We will check two things: that the negative energy tail is indeed given by the nonperturbative correction \eqref{nonperturbative}, and that the Method of Laplace applied to the resulting density of states agrees with the leading behavior of the exact late time partition function. 

\subsection{Example I: Airy}\label{app:airy}
Consider the Airy model with genus zero density of states $\rho_\infty(E)=\theta(E)\sqrt{E}$. This density of states is structurally of the same type as the gravitational one \eqref{gravdos}: it has support on the positive real axis. At low energies $E \ll 1$ the gravitational density of states \eqref{gravdos} is precisely the Airy density of states with a shift $E\to 4\pi^2 E$.
\\~\\
The finite $L$ density of states of this model is known exactly \cite{airy}:
\begin{equation}
    \frac{L}{\pi}\rho_L(-X)=-X(\airy(X))^2+(\airy ' (X))^2, \qquad X=L^{2/3}E. \label{airyexact}
\end{equation}
Using the asymptotics of the Airy function,\footnote{For $-X\gg 1$:
\begin{equation}
    \sqrt{\pi}\airy(-X) \approx X^{-1/4}\sin \frac{2}{3}X^{3/2}.
\end{equation}
For $X\gg 1$:
\begin{equation}
		\sqrt{\pi}\airy(X) \approx \frac{1}{2}X^{-1/4}\exp \left(-\frac{2}{3}X^{3/2}\right).
\end{equation}}
we recover in the large $L$ regime ($E>0$):
\begin{align}
    \rho_L(E) &\approx \sqrt{E}-\frac{1}{L}\frac{1}{4E}\sin L \frac{4}{3} E^{3/2},\label{rhorightairy} \\
		\rho_L(-E) &\approx \frac{1}{L}\frac{1}{8 E}\exp \left(- L \frac{4}{3} E^{3/2}\right) \label{rholeftairy}.
\end{align}
The high-frequency wiggles \eqref{rhorightairy} and the exponential tail \eqref{rholeftairy} at negative energies are the important qualitatively new features. The finite $L$ spectral density \eqref{airyexact} is shown 
\begin{figure}[h]
\begin{center}
    \includegraphics[width=0.8\textwidth]{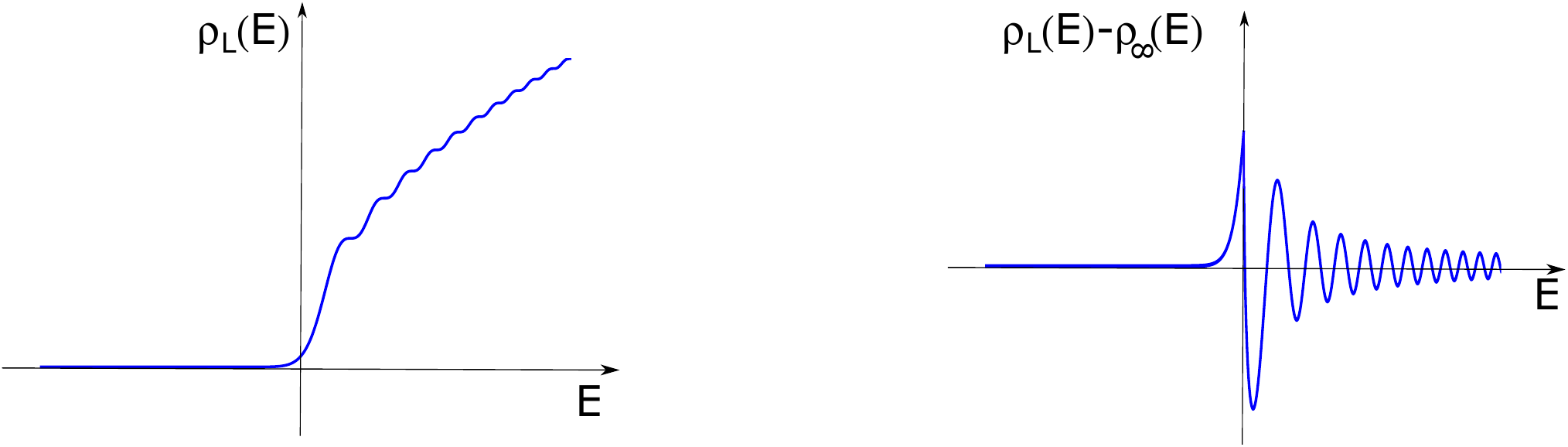}
\end{center}
\caption{Left: Spectral density \eqref{airyexact} for $L=50$. Right: Finite $L$ correction on $\rho_\infty(E)$.}
\label{fig:airy}
\end{figure}
in Figure \ref{fig:airy} (left). In the $L\to\infty$ limit we recover:
\begin{equation}
\label{airyzero}
    \rho_\infty(E)=\theta(E)\sqrt{E}.
\end{equation}
The finite $L$ correction on top of this is shown in Figure \ref{fig:airy} (right). 
\\~\\
The potential is \eqref{potential}:
\begin{equation}
     V(E)=i\int^E_0 d M \rho_\infty (M) = i\frac{2}{3}E^{3/2}.\label{vairy}
\end{equation}
Both the wiggles in \eqref{rhorightairy} and the exponential tail in \eqref{rholeftairy} are immediately observed to be of the form \eqref{nonperturbative}:\footnote{Notice that $V(-E)=\frac{4}{3}E^{3/2}$ which has the desired properties: it is positive definite and monotonically rising.}
\begin{equation}
    \ln \rho_L(E) = -2L V(E),
\end{equation}
predicted by random matrix theory. Performing the saddle approximation on \eqref{beforesaddle} using the seed spectral density \eqref{airyzero}, we obtain:
\begin{equation}
		    \ln Z(it)=\frac{L}{12}\left(\frac{\tau^3}{L^3}\right)=-i\frac{1}{12}\frac{t^3}{L^2}.\label{zairy}
\end{equation}
On the other hand, using the Fourier transform of the exact finite $L$ Airy density of states \eqref{airyexact},
 one obtains, up to numerical prefactors:\footnote{
The Fourier transform of the Airy function is known:
\begin{equation}
    \int d Xe^{i T X}\airy(X)=\exp \left(-\frac{i}{3}T^3\right),
\end{equation}
so the calculation can be done exactly.}
\begin{equation}
    Z(it) \sim t^{-3/2}\exp \left( -i\frac{1}{12}\frac{t^3}{L^2}\right). \label{zairy2}
\end{equation}
This matches precisely with \eqref{zairy}. Notice that formula \eqref{nonperturbative} predicts only the function in the exponential i.e. the leading behavior of $\ln Z(it)$ at late times. A more detailed analyses of the random matrix theory allows to determine the power-law prefactors in $\rho_L(E)$ and $Z_L(it)$ \cite{sss2}. In particular there is a power $t^{-2}$ in \eqref{zairy2} coming from the saddle point evaluation of the $E^{-1}$ power in $\rho_L(E)$ in \eqref{rholeftairy} and a power $t^{1/2}$ from the determinant of fluctuations around the saddle.\footnote{The saddle is $x_0(\tau)\sim \tau^2/L^2$ and $\det(\tau)\sim x_0(\tau)^{-1/2}\sim 1/\tau$.} For the very late time behavior of gravity though, this will turn out not to matter.
\subsection{Example II: Gaussian}\label{app:gaussian}
As a second example, consider the Gaussian Unitary Ensemble (GUE). Its genus zero spectral density is given by the Wigner semi-circle law:
\begin{equation}
    \rho_\infty (E)=\theta(1-E^2)\sqrt{1-E^2}. 
\end{equation}
Close to the each edge $1-\abs{E}\ll 1$, we find a copy of the Airy density of states \eqref{airyzero}, and there are parametric limits in which this model reduces to the Airy model \cite{airy}.
\\~\\
Consider the Hermite functions
\begin{equation}
    \phi_n(\sqrt{2 L}E)=(2^n n!\sqrt{\pi})^{-1/2} H_n(\sqrt{2L}E)\exp \left(-LE^2\right),
\end{equation}
in terms of the Hermite polynomials $H_n(x)$. The finite $L$ spectral density is \cite{mehta}:
\begin{align}
   \frac{\sqrt{2L}}{\pi} \rho_L(E) &= \sum_{n=0}^{L-1}\phi^2_n(\sqrt{2L}E) \nonumber \\
	&=\phi_L^2(\sqrt{2L}E)-\left(\frac{L+1}{L}\right)^{1/2}\phi_{L-1}(\sqrt{2 L}E)\phi_{L+1}(\sqrt{2L}E).\label{densitygaussian}
\end{align}
In the large $L$ limit, this goes indeed to the semicircle:
\begin{equation}
    \rho_\infty(E)=\theta(1-E^2)\sqrt{1-E^2}.
\end{equation}
In the region outside of the semicircle support we find from \eqref{densitygaussian} the tail of the spectral density to be:
\begin{equation}
    \ln \rho_L(E) =- 2LE^2.\label{gaussiantail}
\end{equation}
For the GUE, the effective potential is not just an integral of $\rho_\infty(E)$ but instead we have $V(E)=E^2$, see \cite{1510.04430}. So formula \eqref{gaussiantail} is in sync with \eqref{nonperturbative}. The Fourier transform of \eqref{densitygaussian} can be done exactly \cite{1812rmt}:\footnote{
The Fourier transform of the Hermite function is another Hermite function:
\begin{equation}
    \int d E e^{i E t} \phi_L(\sqrt{2L}E)=\frac{1}{\sqrt{2L}}\phi_L(t/\sqrt{2L}).
\end{equation}
}
\begin{equation}
   Z_L(t) \sim  L_{L-1}^{(1)}\left(\frac{t^2}{4L}\right)\exp \left(-\frac{t^2}{8L}\right),\label{laguerre}
\end{equation}
where $L^{(1)}_n$ is the associated Laguerre polynomial. At times $t\ll\sqrt{L}$, this function is well approximated by:
\begin{equation}
    Z_L(t) \sim t^{-3/2}\cos(t-\frac{3\pi}{4}).
\end{equation}
This makes contact with the Airy result \eqref{zairy2} in the same regime $t\ll\sqrt{L}$.\footnote{The oscillatory piece originates from the fact that there are two spectral edges to the GUE spectral density that both contribute.}
\\~\\
In the late time regime $t\gg L$, the exact GUE partition function \eqref{laguerre} goes like
\begin{equation}
    \ln Z_L(t)=-\frac{t^2}{8L}.
\end{equation}
This matches with the prediction of the saddle point method using the nonperturbative tail \eqref{gaussiantail}. One observes a sharp and sudden drop of in $\abs{Z_L(it)}$ at times of order $t\sim L$. We can understand this from a Fourier transform point of view by observing that the sharpest features in the spectral density \eqref{densitygaussian} are periodic wiggles with an order $1/L$ spacing. Higher frequency components are absent, hence the sudden drop of in the Fourier transform of $\rho_L(E)$ i.e. $\abs{Z_L(it)}$.
\\
We can then also understand why no such sudden drop of is visible in $\abs{Z_L(it)}$ for the Airy model. The wiggles in the Airy model are described by the term
\begin{equation}
    \sim \sin L \frac{4}{3} E^{3/2}
\end{equation}
in \eqref{rhorightairy}. The period of the wiggles decreases with increasing energy as $\sim 1/LE^{1/2}$. This means there is no maximal frequency component in $\rho_L(E)$ for the Airy model.
\subsection{Example III: JT Gravity}
From \eqref{potential} we have:
\begin{equation}
    V(E)=i\int^E_0 d M\sinh 2\pi \sqrt{M}. \label{vgrav}
\end{equation}
As before this will result in high frequency wiggles in $\mathcal{S}$ and an exponential tail outside of $\mathcal{S}$. The frequency of the wiggles as increases with energy as $\sim L\sinh 2\pi \sqrt{E}$ much like it did for the Airy model. By consequence we expect $\emph{no}$ sudden drop of in $\abs{Z_L(it)}$ at late times.
\\~\\
Notice that this potential is not positive definite not monotonically increasing for large negative energies. Therefore the theory would be ill defined for the naive contour $\mathcal{C}=\mathbb{R}$ \cite{sss2}. An appropriate contour is shown in Figure 4 of \cite{sss2}. It follows the real axis up to the saddle point $E=-1/4$ of $V(E)$, and then flows up the path of steepest descent in the complex $E$ plane. With this definition $\rho_L(E)\exp(-\tau E)$ is a good distribution on the contour $\mathcal{C}$.
\\
We arrive at roughly two interesting regimes for $Z_L(it)$:
\begin{itemize}
    \item For $t\ll L$, the saddle point $x_0$ is on the part of the contour where the potential \eqref{vgrav} reduces to the Airy potential \eqref{vairy}. We end up with:
    \begin{equation}
		\ln Z_L(it)=-i\frac{1}{12\pi^2}\frac{t^3}{L^2}.
    \end{equation}
    As in \eqref{zairy2} there is a power $t^{-3/2}$ in $Z_L(it)$ from the determinant and a factor $1/E$ in $\rho_L(E)$.
    \item For $t\gg L$ the saddle point is on the left-most part of the contour which is just the point $E=-1/4$. Indeed, close to this point we may parameterize $\mathcal{C}$ as $ E=-1/4+is$ with $s$ positive. The integral along this path has a saddle at $s=0$ by definition of $\mathcal{C}$. Because the saddle point is independent of $t$ for $t\gg L$, we have that:
    \begin{equation}
        Z_L(it)\approx \exp(\tau x_0(\tau))= \exp(it/4),
    \end{equation}
    up to some constant. Notice that this is a pure phase, confirming our intuition that there should be no sudden drop of in $\abs{Z_L(it)}$.
\end{itemize}
Based on our observations in the main text we may conclude that for a large class of boundary correlation functions, there is a transition from the semiclassical black hole saddle point $E=C^2/\beta^2$ at $t\ll C$ to the nonperturbative saddle point $E=-1/4$ for $t\gg L$ with $t$ any of the time separations.
\\
It is plausible following the logic of section \ref{s:quantum} that we should see a similar transition in bulk matter correlation functions. To verify this one would have to understand bulk matter correlation functions on higher genus surfaces as well as its nonperturbative completion. We leave this to future work.
\\
It is interesting that this nonperturbative saddle point is closely related to the appearance of ZZ branes in JT gravity. This hints at a potential semiclassical interpretation for this late time transition in correlators as being related to the dominant bulk geometry changing from a disk to a disk with an additional ZZ boundary inserted. It would be interesting to understand if there is such a mechanism at play.

\end{document}